%% file: JetPaper.tex
\documentclass[preprintnumbers,reprint,amsmath,amssymb,aps,prd,showpacs,
twocolumn,nofootinbib,floatfix]{revtex4-1}

\pdfoutput=1

\usepackage{graphicx}
\usepackage{dcolumn}
\usepackage{bm}
\usepackage{atlasphysics}
\usepackage{array}
\usepackage{afterpage}
\usepackage{epstopdf}
\usepackage[mathlines]{lineno}
\usepackage{hyperref}
\usepackage{xspace}
\usepackage{mydefs}
\usepackage[parsep]{collref}
\usepackage{subfigure}
\usepackage{amsmath}
\usepackage{amssymb}
\usepackage{helvet}
\usepackage{here}
\usepackage[mathcal]{eucal}
\usepackage{morefloats}

\hypersetup{
  colorlinks=true,
  linkcolor=blue,
  citecolor=blue,
  urlcolor=blue,
  pdftitle={Measurement of inclusive jet and dijet production in $pp$
    collisions at $\sqrt{s}=7$~TeV using the ATLAS detector},
  pdfauthor={The ATLAS Collaboration},
  pdfpagemode={UseOutlines},
  bookmarksopen=true,
  bookmarksnumbered=true,
  pdfstartview={Fit}
 }

\begin{document}

\preprint{CERN-PH-EP-2011-192}
\preprint{Submitted to Physical Review D}
\vspace*{0.2cm}
\title{Measurement of inclusive jet and dijet production in $pp$ collisions \\ at
  $\sqrt{s}=7$~TeV using the ATLAS detector}
\author{The ATLAS Collaboration}
\date{\today}

\begin{abstract}
Inclusive jet and dijet cross sections have been measured in
proton-proton collisions at a centre-of-mass energy of 7 TeV using the
ATLAS detector at the Large Hadron Collider.  The cross sections were
measured using jets clustered with the anti-$k_t$ algorithm with
parameters $R= 0.4$ and $R = 0.6$.  These measurements are based on
the 2010 data sample, consisting of a total integrated luminosity of
37~pb$^{-1}$.  Inclusive jet double-differential cross sections are
presented as a function of jet transverse momentum, in bins of jet
rapidity.  Dijet double-differential cross sections are studied as a
function of the dijet invariant mass, in bins of half the rapidity
separation of the two leading jets.  The measurements are performed in
the jet rapidity range $|y|<4.4$, covering jet transverse momenta from
20~GeV to 1.5~TeV and dijet invariant masses from 70~GeV to 5~TeV.
The data are compared to expectations based on next-to-leading order
QCD calculations corrected for non-perturbative effects, as well as to
next-to-leading order Monte Carlo predictions.  In addition to a test
of the theory in a new kinematic regime, the data also provide
sensitivity to parton distribution functions in a region where they
are currently not well-constrained.
\end{abstract}


\maketitle

\section{Introduction}
\label{sec:intro}
At the Large Hadron Collider (LHC), jet production is the dominant
high transverse-momentum (\pt{}) process.  Jet cross sections serve as
one of the main observables in high-energy particle physics, providing
precise information on the structure of the proton. They are an
important tool for understanding the strong interaction and searching
for physics beyond the Standard Model (see, for example,
Refs.~\cite{Arnison:1983dk, Adeva:1990nu, Alitti:1990aa,
Chekanov:2001bw, Chekanov:2002be, Heister:2002tq, Abdallah:2004uq,
Chekanov:2005nn, Abbiendi:2005vd, Abulencia:2007ez, Abazov:2008hua,
Aaltonen:2008eq, Abazov:2009nc, :2009mh, Aaron:2009he, Aaron:2009vs,
Abramowicz:2010ke, Abazov:2010fr, :2011me, Chatrchyan:2011qt}).

The ATLAS Collaboration has published a first measurement of inclusive
jet and dijet production at $\sqrt{s} = 7$~TeV, using an integrated
luminosity of 17~nb$^{-1}$~\cite{Collaboration:2010wv}.  This
measurement considered only jets with transverse momentum larger than
60 GeV and in a rapidity interval $|y|<2.8$\footnote{ATLAS uses a
right-handed coordinate system with its origin at the nominal
interaction point (IP) in the centre of the detector and the $z$-axis
along the beam pipe. The $x$-axis points from the IP to the centre of
the LHC ring, and the $y$ axis points upward. Cylindrical coordinates
$(r,\phi)$ are used in the transverse plane, $\phi$ being the
azimuthal angle around the beam pipe, referred to the $x$-axis. The
pseudorapidity is defined in terms of the polar angle $\theta$ with
respect to the beamline as $\eta=-\ln\tan(\theta/2)$.  When dealing
with massive jets and particles, the rapidity $y = \frac{1}{2} \ln
\left( \frac{E + p_z}{E - p_z} \right)$ is used, where $E$ is the jet
energy and $p_{z}$ is the $z$-component of the jet momentum.}.

The analysis presented here extends the previous measurement using the
2010 data sample of ($37.3 \pm 1.2$)~$\textrm{pb}^{-1}$, an integrated
luminosity more than 2000 times larger than that of the previous
study.  This more than doubles the kinematic reach at high jet
transverse momentum and large dijet invariant mass.  There are strong
physics reasons to extend the measurement to jets of lower transverse
momentum and larger rapidity as well.  Jets at lower $\pt$ are more
sensitive to non-perturbative effects from hadronisation and the
underlying event, and forward jets may be sensitive to different
dynamics in QCD than central jets.  Moreover, LHC experiments have
much wider rapidity coverage than those at the Tevatron, so forward
jet measurements at the LHC cover a phase space region that has not
been explored before.

\begin{figure}[tb]
\begin{center}
\includegraphics[width=1.0\linewidth]{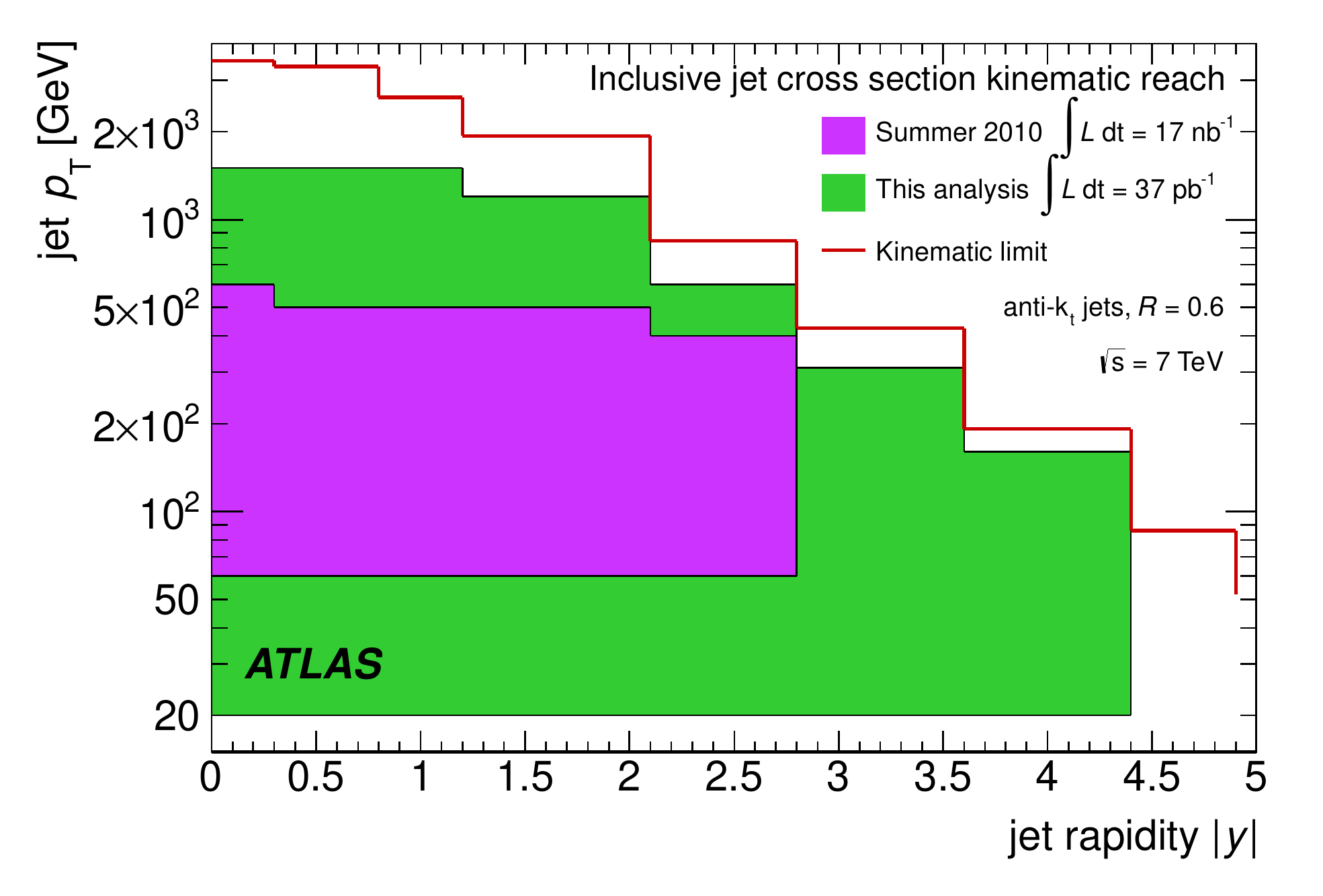}
\caption{Kinematic reach of the inclusive jet cross section measured
in this analysis compared to that of the previous
study~\cite{Collaboration:2010wv} for jets identified using the \AKT
algorithm with $R=0.6$.  The kinematic limit for the center-of-mass
energy of 7~TeV is also shown.}
\label{fig:KineRange}
\end{center}
\end{figure}

The kinematic reach of this analysis is compared to that of the
 previous ATLAS study in Fig.~\ref{fig:KineRange}.  This data sample
 extends the existing inclusive jet $\pt$ measurement from 700~GeV to
 1.5~TeV and the existing dijet mass measurement from 1.8~TeV to
 5~TeV.  Thus this analysis probes next-to-leading order (NLO)
 perturbative QCD (pQCD) and parton distribution functions (PDFs) in a
 new kinematic regime.  The results span approximately $7\times
 10^{-5} < x < 0.9$ in $x$, the fraction of the proton momentum
 carried by each of the partons involved in the hard interaction.

\section{The ATLAS Detector}
\label{sec:detector}

The ATLAS detector is described in detail in Ref.~\cite{:2008zzm}. In
this analysis, the tracking detectors are used to define candidate
collision events by constructing vertices from tracks, and the
calorimeters are used to reconstruct jets.

The inner detector used for tracking and particle identification has
complete azimuthal coverage and spans the region $|\eta|<2.5$.  It
consists of layers of silicon pixel detectors, silicon microstrip
detectors, and transition radiation tracking detectors, surrounded by
a solenoid magnet that provides a uniform field of 2~T.

The electromagnetic calorimetry is provided by the liquid argon (LAr)
calorimeters that are split into three regions: the barrel
($|\eta|<1.475$), the endcap ($1.375<|\eta|<3.2$) and the forward
(FCal: $3.1<|\eta|<4.9$) regions. The hadronic calorimeter is divided
into four distinct regions: the barrel ($|\eta|<0.8$), the extended
barrel ($0.8<|\eta|<1.7$), both of which are scintillator/steel
sampling calorimeters, the hadronic endcap (HEC; $1.5<|\eta|<3.2$),
which has LAr/Cu calorimeter modules, and the hadronic FCal (same
$\eta$-range as for the EM-FCal) which uses LAr/W modules.  The total
calorimeter coverage is $|\eta|<4.9$.

\section{Cross Section Definition}
\label{sec:xsectdef}

Jet cross sections are defined using the \AKT{} jet
algorithm~\cite{Cacciari:2008gp} implemented in the
\fastjet~\cite{Cacciari:2005hq} package.  Two different values are
used for the clustering parameter $R$ (0.4 and 0.6), which can be seen
intuitively as the radius of a circular jet in the plane $(\phi,y)$ of
azimuthal angle and rapidity.  The jet cross section measurements are
corrected for all experimental effects, and thus are defined for the
``particle-level'' final state of a proton-proton
collision~\cite{Buttar:2008jx}.  Particle-level jets in the Monte
Carlo simulation are identified using the $\AKT$ algorithm and are
built from stable particles, which are defined as those with a proper
lifetime longer than 10~ps.  This definition includes muons and
neutrinos from decaying hadrons.

Inclusive jet double-differential cross sections are measured as a
function of jet $\pt$ in bins of $y$, in the region $\pt > 20\gev, |y|
< 4.4$.  The term ``inclusive jets'' is used in this paper to indicate
that all jets in each event are considered in the cross section
measurement.  Dijet double-differential cross sections are measured as
a function of the invariant mass of the two leading (highest $\pt$)
jets, which is given as $\twomass{1}{2} = \sqrt{ (E_{1}+E_{2})^{2} -
(\vec{p}_{1}+\vec{p}_{2})^2}$, where $E_{1,2}$ and $\vec{p}_{1,2}$ are
the energies and momenta of the two leading jets.  The cross sections
are binned in the variable $y^*$, defined as half the rapidity
difference of the two leading jets, $y^* = |y_{1}-y_{2}|/2$.  The
quantity $y^{*}$ is the rapidity in the two-parton centre-of-mass
frame (in the massless particle limit), where it is determined by the
polar scattering angle with respect to the beamline, $\theta^{*}$:
\begin{equation}
y^{*} = \frac{1}{2}
\ln{\left(\frac{1+|\cos{\theta^{*}}|}{1-|\cos{\theta^{*}}|} \right) }
\label{eq:ystar}
\end{equation}
For the dijet measurement, the two leading jets are selected to lie in
the $|y|<4.4$ region, where the leading jet is required to have $\pt >
30\gev$ and the sub-leading jet $\pt > 20\gev$.  Restricting the
leading jet to higher $\pt$ improves the stability of the NLO
calculation~\cite{Frixione:1997ks}.

Theory calculations are used in the same kinematic range as the
measurement.

\section{Monte Carlo Samples}
\label{sec:mc_samples}

The \pythia~6.423 generator~\cite{Sjostrand:2006za} with the MRST LO*
PDF set~\cite{Sherstnev:2008dm} was used to simulate jet events in
proton-proton collisions at a centre-of-mass energy of
$\sqrt{s}=7$~TeV and to correct for detector effects.  This generator
utilizes leading-order perturbative QCD matrix elements (ME) for
$2\rightarrow2$ processes, along with a leading-logarithmic,
$\pt$-ordered parton shower (PS), an underlying event simulation with
multiple parton interactions, and the Lund string model for
hadronisation.  Samples were generated using the ATLAS Minimum Bias
Tune~1 (AMBT1) set of parameters~\cite{ATLAS-CONF-2010-031}, in which
the model of non-diffractive scattering has been tuned to ATLAS
measurements of charged particle production at $\sqrt{s}=900$~GeV and
$\sqrt{s}=7$~TeV.

The particle four-vectors from these generators were passed through a
full simulation~\cite{:2010wqa} of the ATLAS detector and trigger that
is based on GEANT4~\cite{Agostinelli:2002hh}. Finally, the simulated
events were reconstructed and jets were calibrated using the same
reconstruction chain as the data.

\section{Theoretical Predictions}
\subsection{Fixed-Order calculations}
\subsubsection{NLO Predictions}
\label{sec:nlo_predictions}

The measured jet cross sections are compared to fixed-order NLO pQCD
predictions, with corrections for non-perturbative effects applied.
For the hard scattering, both the \nlojet~4.1.2~\cite{Nagy:2003tz}
package and the \powheg generator~\cite{Alioli:2010qp, Nason:2007vt}
were used, the latter in a specific configuration where the parton
shower was switched off and calculations were performed using NLO
matrix elements.  The two programs have been used with the
CT10~\cite{Lai:2010vv} NLO parton distribution functions, and the same
value of normalisation and factorisation scale, corresponding to the
transverse momentum of the leading jet, $\pt^{\rm max}$:
\begin{equation}
  \mu = \mu_{\rm R} = \mu_{\rm F} = \pt^{\rm max}
  \label{eq:ptscale}
\end{equation}
For \powheg, $\pt^{\rm max}$ is evaluated at leading order and is
denoted $\pt^{\rm Born}$.
Using this scale choice, the cross section results of the two NLO
codes are compatible at the few percent level for inclusive jets over
the whole rapidity region.  They are also consistent for dijet events
where both jets are in the central region, while they differ
substantially when the two leading jets are widely separated in
rapidity ($y^*\gtrsim 3$).  In these regions, \nlojet gives an
unstable and much smaller cross section than POWHEG that is even
negative for some rapidity separations.  \powheg remains positive over
the whole region of phase space.  It should be noted that the forward
dijet cross section predicted by \nlojet in this region has a very
strong scale dependence, which however is much reduced for larger
values of scale than that of Eq.~\ref{eq:ptscale}.

\begin{figure}[tb]
\begin{center}
\includegraphics[width=0.5\textwidth]{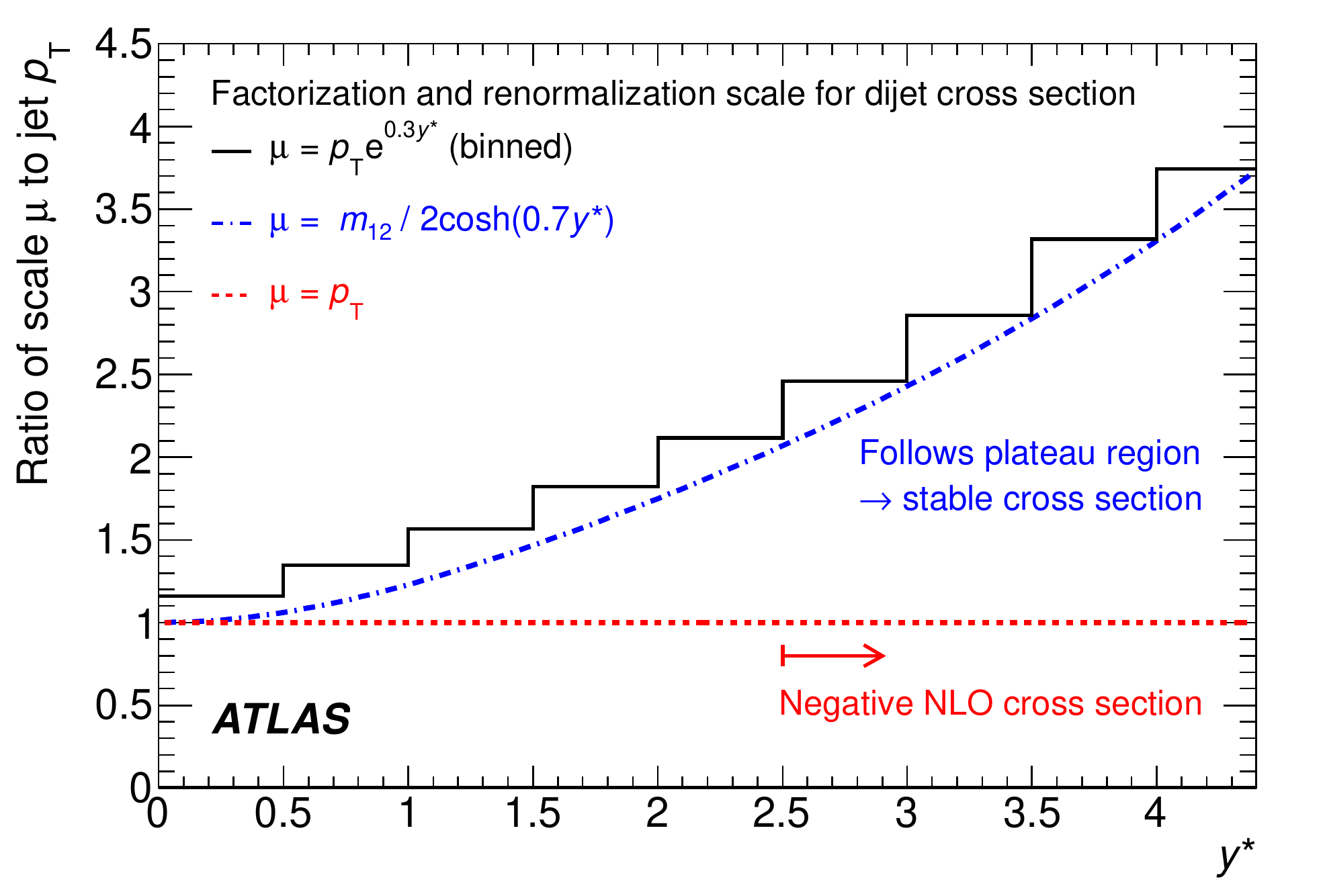}
\caption{The histogram indicates the values of the renormalisation and
  factorisation scales (denoted by $\mu = \mu_{R} = \mu_{F}$) used for
  the dijet predictions obtained using \nlojet, as a function of
  $y^{*}$, half the rapidity separation between the two leading jets.
  This is motivated by the scale choice suggested in
  Ref.~\cite{Ellis:1992en} (dot-dashed line), and is also compared to
  the scale choice used for the inclusive jet predictions (dotted
  line).}
\label{fig:scale}
\end{center}
\end{figure}

The forward dijet cross section for \nlojet is much more stable if
instead of a scale fixed entirely by $\pt$, a scale that depends on
the rapidity separation between the two jets is used.  The values
chosen for each $y^*$-bin follow the formula:
\begin{equation}
\mu = \mu_{\rm R} = \mu_{\rm F} = \pt \, e^{0.3 y^{*}}
  \label{eq:dijetscale}
\end{equation}
and are indicated by the histogram in Fig.~\ref{fig:scale}.  These
values are motivated by the formula (shown by the dot-dashed curve):
\begin{equation}
  \mu = \mu_{\rm R} = \mu_{\rm F} = \frac{m_{\rm 12}}{2\cosh(0.7 y^*)}
  \label{eq:massscale}
\end{equation}
that is suggested in Ref.~\cite{Ellis:1992en}, and are in a region
where the cross section predictions are more stable as a function of
scale (they reach a ``plateau'').  At small $y^*$, the scale in
Eq.~\ref{eq:dijetscale} reduces to the leading jet $\pt$ (dotted
line), which is used for the inclusive jet predictions.  With this
scale choice, \nlojet is again in reasonable agreement with \powheg,
which uses the scale from Eq.~\ref{eq:ptscale}.  The \nlojet
predictions are used as a baseline for both inclusive jet and dijet
calculations, with the scale choice from Eq.~\ref{eq:ptscale} for the
former and that from Eq.~\ref{eq:dijetscale} for the latter. The
\powheg scale used for both inclusive jets and dijets, $\pt^{\rm
Born}$, is given by Eq.~\ref{eq:ptscale} but evaluated at leading
order.  Despite using different scale choices, the dijet theory
predictions from \nlojet and \powheg are stable with respect to
relatively small scale variations and give consistent results.

The results are also compared with predictions obtained
using the MSTW~2008~\cite{Martin:2009iq},
NNPDF~2.1~(100)~\cite{Ball:2010de, Forte:2010ta} and
HERAPDF~1.5~\cite{HERAPDF15} PDF sets.

The main uncertainties on the NLO prediction come from the
uncertainties on the PDFs, the choice of factorisation and
renormalisation scales, and the uncertainty on the value of the strong
coupling constant $\alpha_{\rm s}$.  To allow for fast and flexible
evaluation of PDF and scale uncertainties, the
\applgrid~\cite{Carli:2010rw} software was interfaced with \nlojet in
order to calculate the perturbative coefficients once and store them
in a look-up table.  The PDF uncertainties are defined at 68\% CL and
evaluated following the prescriptions given for each PDF set. They
account for the data uncertainties, tension between input data sets,
parametrisation uncertainties, and various theoretical uncertainties
related to PDF determination.

To estimate the uncertainty on the NLO prediction due to neglected
higher-order terms, each observable was recalculated while varying the
renormalisation scale by a factor of two with respect to the default
choice.  Similarly, to estimate the sensitivity to the choice of scale
where the PDF evolution is separated from the matrix element, the
factorisation scale was separately varied by a factor of two.  Cases
where the two scales are simultaneously varied by a factor 2 in
opposite directions were not considered due to the presence of
logarithmic factors in the theory calculation that become large in
these configurations.  The envelope of the variation of the
observables was taken as a systematic uncertainty.  The effect of the
uncertainty on the value of the strong coupling constant, $\alpha_{\rm
s}$, is evaluated following the recommendation of the CTEQ
group~\cite{Lai:2010nw}, in particular by using different PDF sets
that were derived using the positive and negative variations of the
coupling from its best estimate.

Electro-weak corrections were not included in the theory predictions
and may be non-negligible~\cite{Moretti:2006ea}.

\begin{figure}[tb]
  \begin{center}
    \includegraphics[width=0.45\textwidth]{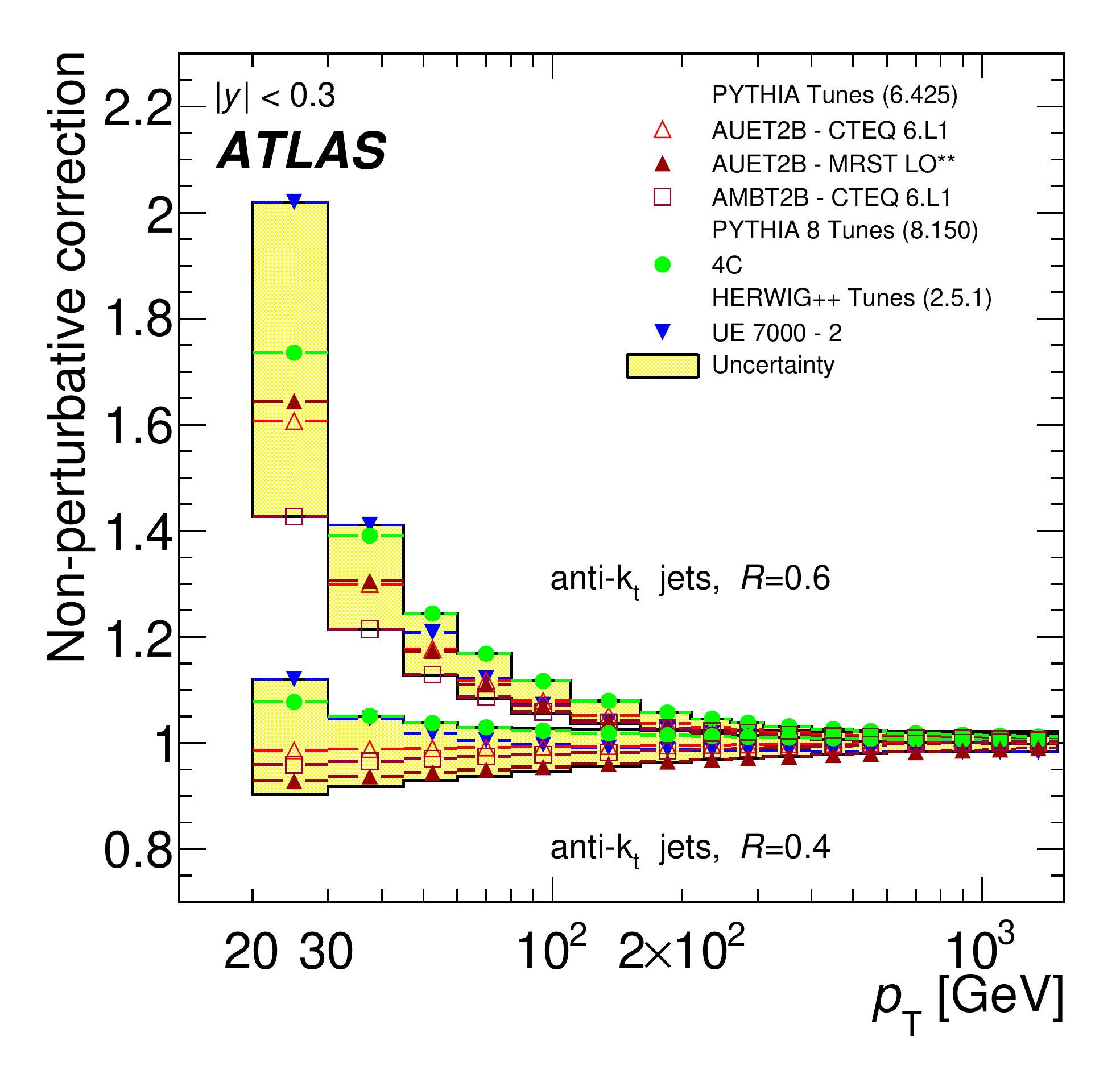}
    \caption{Non-perturbative correction factors for inclusive jets
      identified using the \AKT algorithm with distance parameters
      $R=0.4$ and $R=0.6$ in the rapidity region $|y|<0.3$, derived
      using various Monte Carlo generators. The correction derived
      using \pythia 6.425 with the AUET2B~CTEQ6L1 tune is used for the
      fixed-order NLO calculations presented in this analysis.}
    \label{fig:NP}
  \end{center}
\end{figure}

\subsubsection{Non-Perturbative Corrections}
\label{sec:nonpert}
The fixed-order NLO calculations predict parton-level cross sections,
which must be corrected for non-perturbative effects to be compared
with data.  This is done by using leading-logarithmic parton shower
generators.  The corrections are derived by using \pythia 6.425 with
the AUET2B~CTEQ6L1 tune~\cite{ATL-PHYS-PUB-2011-009} to evaluate the
bin-wise ratio of cross sections with and without hadronisation and
the underlying event.  Each bin of the parton-level cross section is
then multiplied by the corresponding correction.  The uncertainty is
estimated as the maximum spread of the correction factors obtained
from \pythia 6.425 using the AUET2B~LO$^{**}$, AUET2~LO$^{**}$,
AMBT2B~CTEQ6L1, AMBT1, Perugia~2010, and Perugia~2011 tunes
(PYTUNE\_350), and the \pythia~8.150 tune
$4C$~\cite{ATL-PHYS-PUB-2011-008, ATL-PHYS-PUB-2011-009,
Skands:2009zm, Skands:2010ak}, as well as those obtained from the
\herwigpp~2.5.1~\cite{Bahr:2008pv} tune
UE7000-2~\cite{ATL-PHYS-PUB-2011-009}.  The AMBT2B~CTEQ6L1 and AMBT1
tunes, which are based on observables sensitive to the modeling of
minimum bias interactions, are included to provide a systematically
different estimate of the underlying event activity.

The corrections depend strongly on the jet size; therefore separate
sets of corrections and uncertainties were derived for jets with
$R=0.4$ and $R=0.6$.  The correction factors and their uncertainties
depend on the interplay of the hadronisation and the underlying event
for the different jet sizes, and they have a significant influence at
low $\pt$ and low dijet mass.  For $R=0.4$, the correction factors are
dominated by the effect of hadronisation and are approximately 0.95 at
jet $\pt=20$~GeV, increasing closer to unity at higher $\pt$.  For
$R=0.6$, the correction factors are dominated by the underlying event
and are approximately 1.6 at jet $\pt=20$~GeV, decreasing to between
1.0-1.1 for jets above $\pt=100$~GeV.  Fig.~\ref{fig:NP} shows the
non-perturbative corrections for inclusive jets with rapidity in the
interval $|y|<0.3$, for jet clustering parameters $R=0.4$ and
$R=0.6$. The correction factors for the other rapidity bins become
closer to unity as the jet rapidity increases, as can be seen in
Fig.~\ref{fig:NP-other-bins} in Appendix~\ref{sec:nonPert}.

Non-perturbative corrections have been evaluated for the dijet
measurement as well, as a function of the dijet mass and the rapidity
interval $y^*$, for each of the two jet sizes.  These follow a similar
behaviour to those for inclusive jets, with the corrections becoming
smaller for large invariant masses and rapidity differences.

\subsection{NLO Matrix Element + Parton Shower}
\label{sec:powheg}
The measured jet cross sections are also compared to
\powheg~\cite{Alioli:2010xa}, an NLO parton shower Monte Carlo
generator that has only recently become available for inclusive jet
and dijet production.  \powheg uses the \powhegbox
package~\cite{Nason:2004rx, Frixione:2007vw, Alioli:2010xd} and allows
one to use either \pythia or \herwig~\cite{herwig2} +
\jimmy\cite{Butterworth:1996zw} to shower the partons, hadronise them,
and model the underlying event.  The ATLAS underlying event tunes,
AUET2B for \pythia and AUET2~\cite{ATL-PHYS-PUB-2010-014} for \herwig,
are derived from the standalone versions of these event generators,
with no optimisation for the \powheg predictions.  The showering
portion of \powheg uses the PDFs from \pythia or \herwig as part of
the specific tune chosen.

In the \powheg algorithm, each event is built by first producing a QCD
$2 \to 2$ partonic scattering.  The renormalisation and factorisation
scales are set to be equal to the transverse momentum of the outgoing
partons, $\pt^{\rm Born}$, before proceeding to generate the hardest
partonic emission in the event.\footnote{Technical details of the
\powheg generation parameters, which are discussed below, are given in
Refs.~\cite{Alioli:2010qp, Nason:2007vt}.  The folding parameters used
are 5-10-2. A number of different weighting parameters are used to
allow coverage of the complete phase space investigated: 25~GeV,
250~GeV and 400~GeV. The minimum Born $\pt$ is 5~GeV. For all the
samples, the leading jet transverse momentum is required to be no more
than seven times greater than the leading parton's momentum.  The
$\pt$ of any additional partonic interactions arising from the
underlying event is required to be lower than that of the hard scatter
generated by \powheg.  The parameters used in the input file for the
event generation are \texttt{bornktmin} = 5 \GeV,
\texttt{bornsuppfact} = 2, 250, 400 \GeV, \texttt{foldcsi} = 5,
\texttt{foldy} = 10, and \texttt{foldphi} = 2.}  The CT10 NLO PDF set
is used in this step of the simulation.  Then the event is evolved to
the hadron level using a parton shower event generator, where the
radiative emissions in the parton showers are required to be softer
than the hardest partonic emission generated by \powheg.

The coherent simulation of the parton showering, hadronisation, and
the underlying event with the NLO matrix element is expected to
produce a more accurate theoretical prediction.  In particular, the
non-perturbative effects are modeled in the NLO parton shower
simulation itself, rather than being derived separately using a LO
parton shower Monte Carlo generator as described in
Sec.~\ref{sec:nonpert}.

\section{Data Selection and Calibration}
\label{sec:data}

\subsection{Dataset}
\label{sec:dataset}
The inclusive jet and dijet cross section measurements use the full
ATLAS 2010 data sample from proton-proton collisions at
$\sqrt{s}=7$~TeV.

For low-$\pt$ jets, only the first 17~nb$^{-1}$ of data taken are
considered since the instantaneous luminosity of the accelerator was
low enough that a large data sample triggered with a minimum bias
trigger (see Sec.~\ref{sec:trigger}) could be recorded.  This provides
an unbiased sample for reconstructing jets with $\pt$ between
20-60~GeV, below the lowest jet trigger threshold.  In addition,
during this period there were negligible contributions from
``pile-up'' events, in which there are multiple proton-proton
interactions during the same or neighbouring bunch crossings.  Thus
this period provides a well-measured sample of low-$\pt$ jets.  The
first data taking period was not used for forward jets with $|y|>2.8$
and $\pt > 60$~GeV because the forward jet trigger was not yet
commissioned.

For all events considered in this analysis, good
operation status was required for the first-level trigger, the
solenoid magnet, the inner detector, the calorimeters and the
luminosity detectors, as well as for tracking and jet reconstruction.
In addition, stable operation was required for the high-level trigger
during the periods when this system was used for event rejection.

\subsection{Trigger}
\label{sec:trigger}
Three different triggers have been used in this measurement: the
minimum bias trigger scintillators (MBTS); the central jet trigger,
covering $|\eta|<3.2$; and the forward jet trigger, spanning
$3.1<|\eta|<4.9$.  The MBTS trigger requires at least one hit in the
minimum bias scintillators located in front of the endcap cryostats,
covering $2.09<|\eta|<3.84$, and is the primary trigger used to select
minimum-bias events in ATLAS. It has been demonstrated to have
negligible inefficiency for the events of interest for this
analysis~\cite{Aad:2010rd} and is used to select events with jets
having transverse momenta in the range 20-60~GeV.  The central and
forward jet triggers are composed of three consecutive levels: Level 1
(L1), Level 2 (L2) and Event Filter (EF).  In 2010, only L1
information was used to select events in the first 3~pb$^{-1}$ of data
taken, while both the L1 and L2 stages were used for the rest of the
data sample.  The jet trigger did not reject events at the EF stage in
2010.

\begin{figure*}[tb]
  \centering
  \mbox{
    \subfigure[]{\includegraphics[width=.325\textwidth]{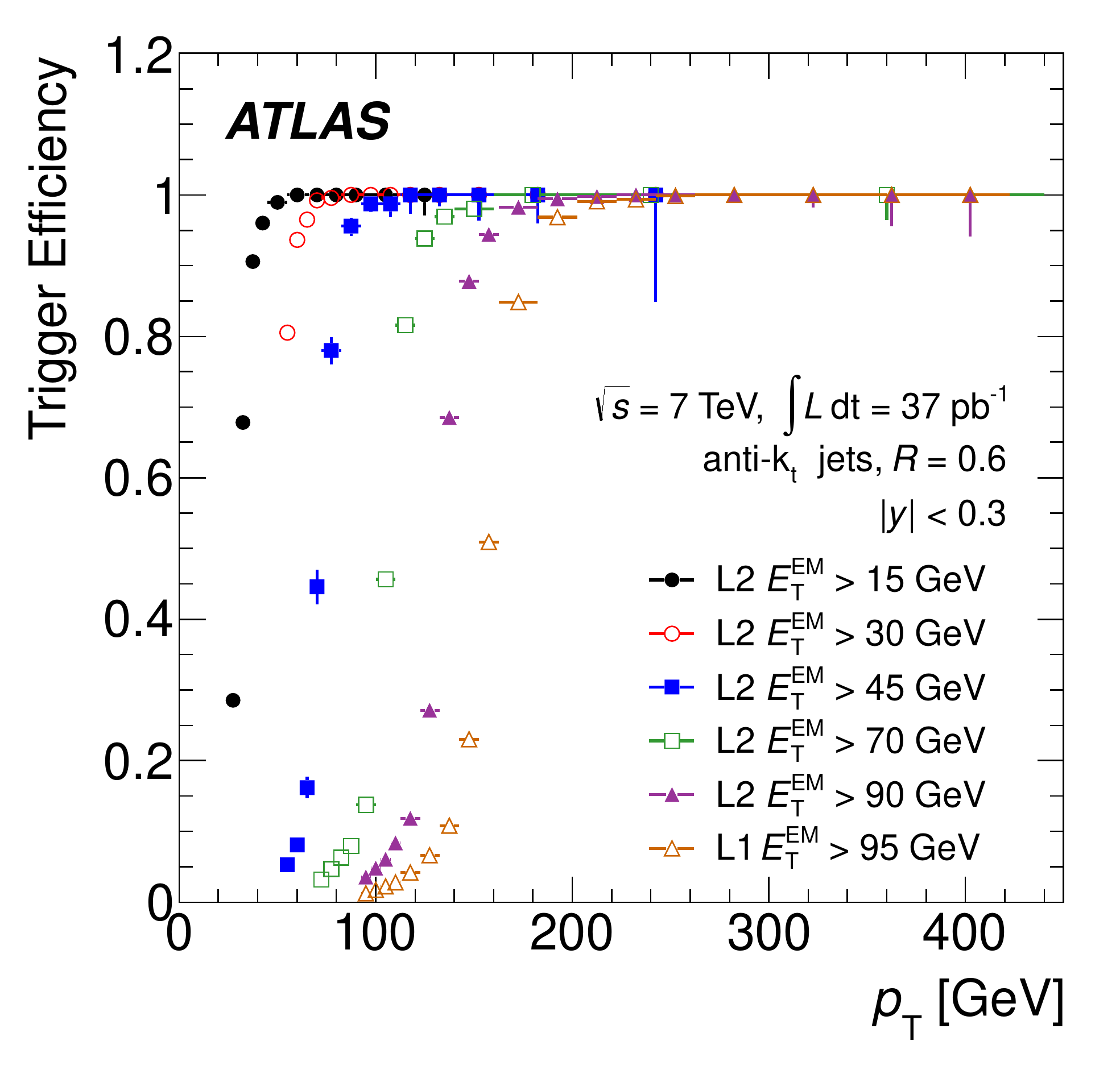}}\quad
    \subfigure[]{\includegraphics[width=.325\textwidth]{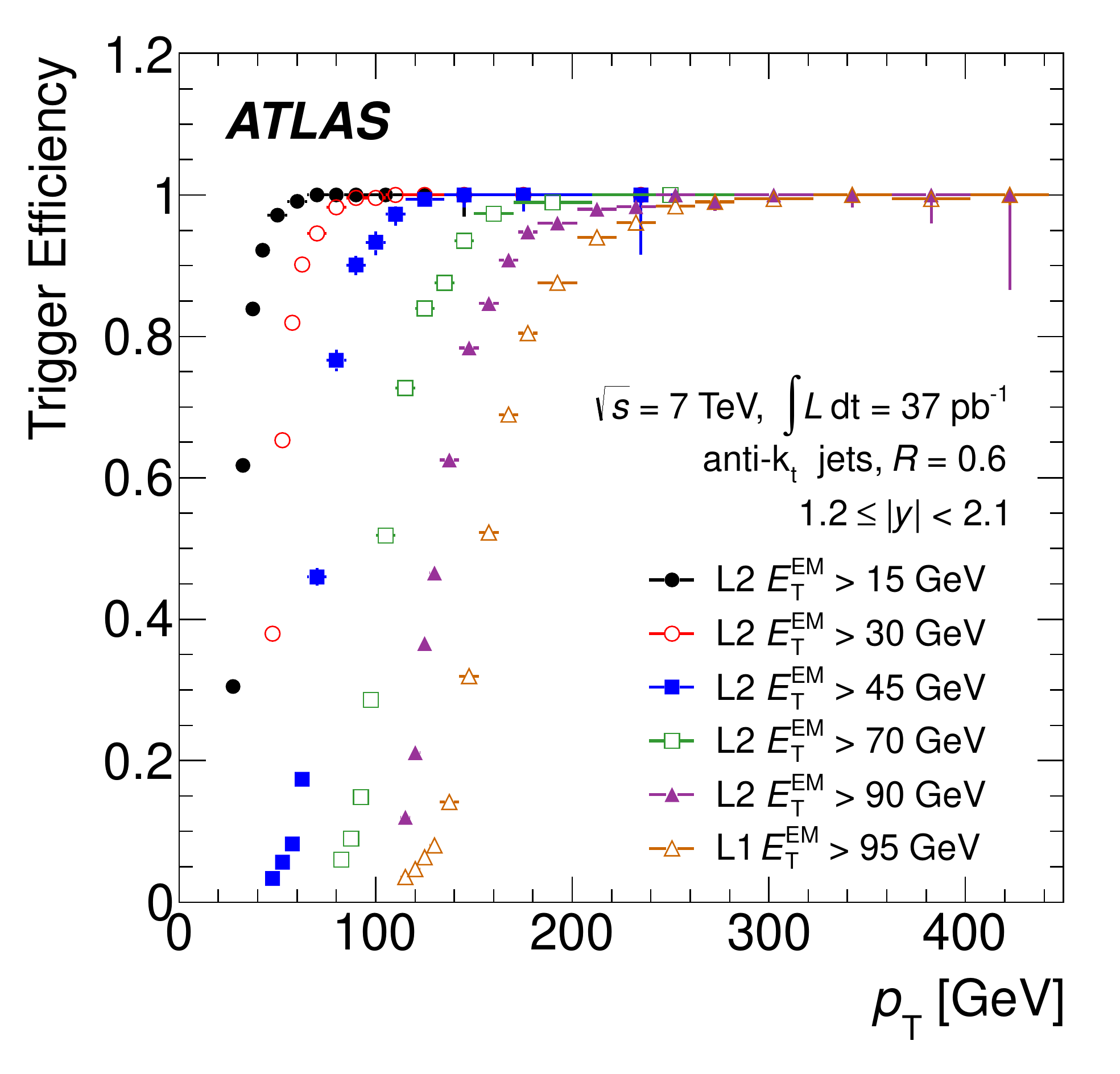}}\quad
    \subfigure[]{\includegraphics[width=.325\textwidth]{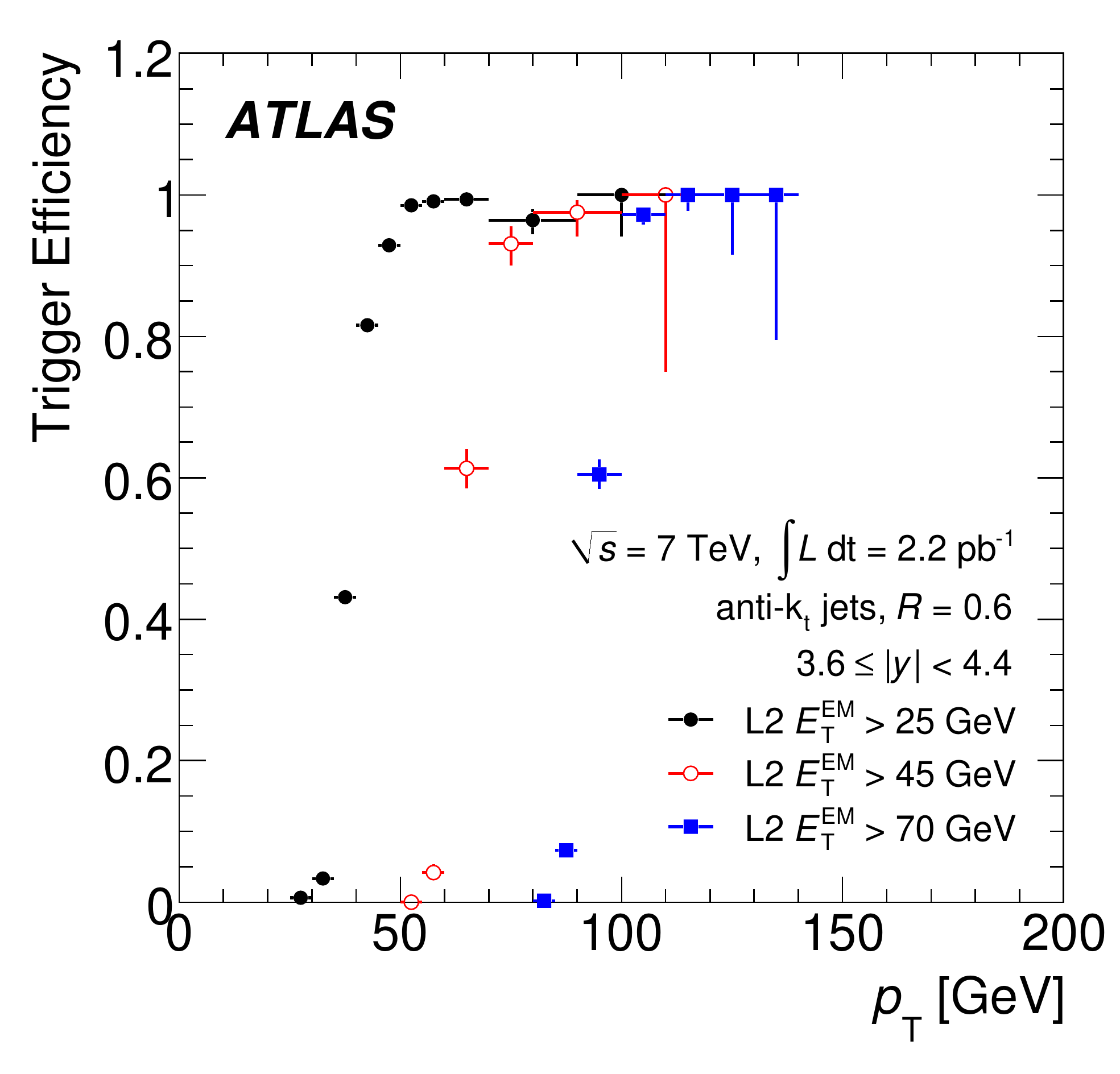}}
  }
  \caption{
    Combined L1+L2 jet trigger efficiency as a function of
    reconstructed jet $\pT$ for \AKT jets with $R = 0.6$ in the
    central region $|y|<0.3$ (a), the barrel-endcap transition region
    $1.2 \le |y|<2.1$ (b) and the FCal region $3.6 \le |y|<4.4$ (c)
    for the different L2 trigger thresholds used in the analysis.  The
    trigger thresholds are at the electromagnetic scale, while the jet
    $\pt$ is at the calibrated scale (see Sec.~\ref{sec:jes}).  The
    highest trigger chain used for $|y|<2.8$ does not apply a
    threshold at L2, so its L1 threshold is listed. The efficiency in
    the $|y|>3.2$ rapidity range is not expected to reach 100\% due to
    the presence of a dead FCal trigger tower that spans 0.9\% of the
    $(\eta,\phi)$-acceptance.  This inefficiency is assigned as a systematic
    uncertainty on the trigger efficiency in the measurement.
  }
  \label{fig:trigger_eff}
\end{figure*}

The central and forward jet triggers independently select data using
several thresholds for the jet transverse energy ($E_{\rm T} \equiv
E\sin\theta$), each of which requires the presence of a jet with
sufficient $E_{\rm T}$ at the electromagnetic (EM) scale.\footnote{The
electromagnetic scale is the basic calorimeter signal scale for the
ATLAS calorimeters.  It has been established using test-beam
measurements for electrons and muons to give the correct response for
the energy deposited in electromagnetic showers, while it does not
correct for the lower response of the calorimeter to hadrons.}  For
each L1 threshold, there is a corresponding L2 threshold that is
generally 15~GeV above the L1 value.  Each such L1+L2 combination is
referred to as an L2 trigger chain.  Fig.~\ref{fig:trigger_eff} shows
the efficiency for L2 jet trigger chains with various thresholds as a
function of the reconstructed jet $\pt$ for $\AKT$ jets with $R=0.6$
for both the central and forward jet triggers.  Similar efficiencies
are found for jets with $R=0.4$, such that the same correspondence
between transverse momentum regions and trigger chains can be used for
the two jet sizes.  The highest trigger chain does not apply a
threshold at L2, so its L1 threshold is listed.

\begin{figure}[tb]
  \centering
  \includegraphics[width=.38\textwidth]{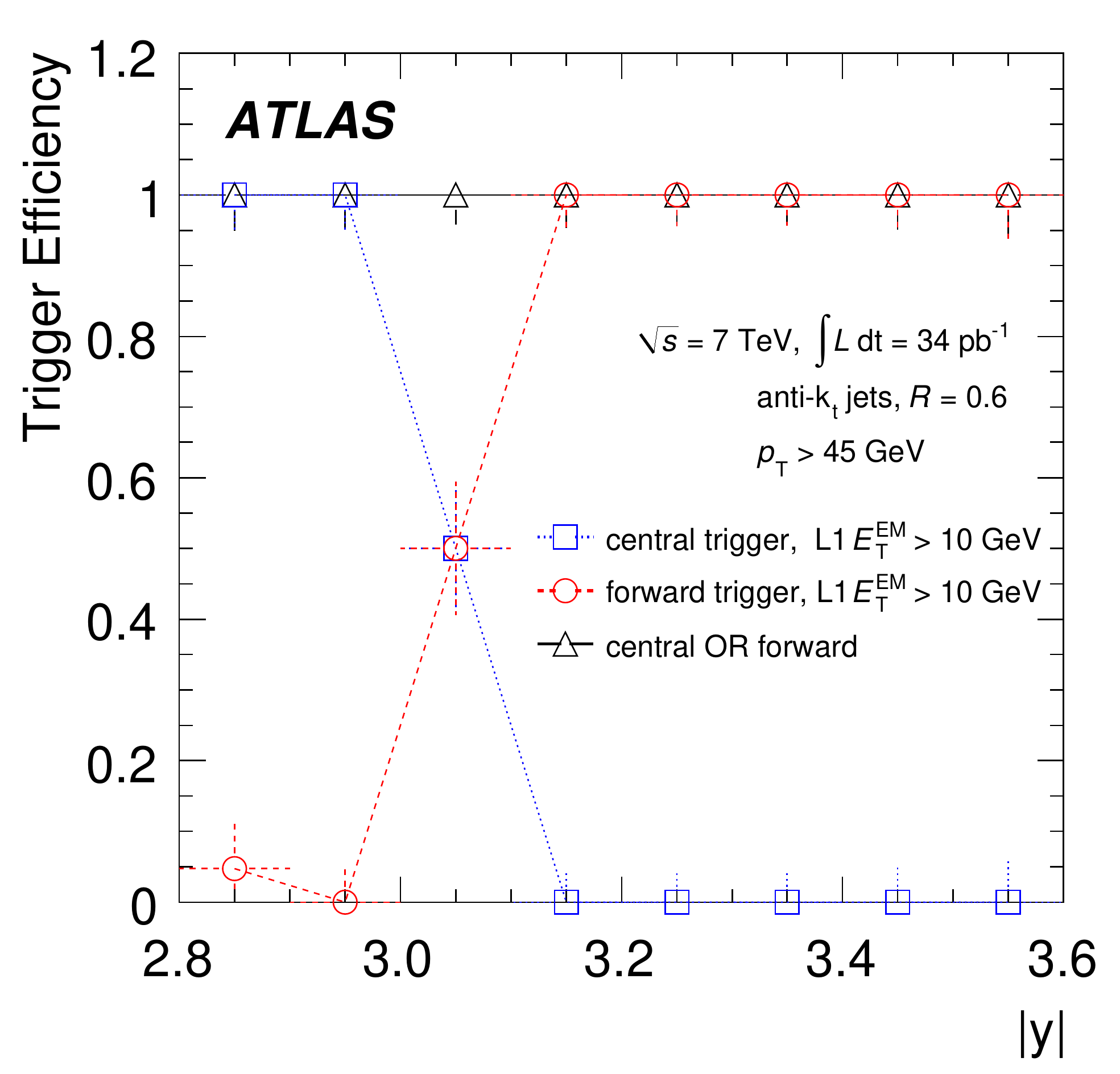}
  \caption{
    Efficiencies for the central and forward jet triggers with a L1
    $E_{\rm T}$ threshold of 10~GeV, and for their logical OR, as a
    function of the rapidity $y$ of the reconstructed jet in the
    transition region between the two trigger systems. The logical OR
    is used for the inclusive jet measurement to collect data in the
    $2.8 \le |y|<3.6$ rapidity slice.
  }
  \label{fig:trigOR}
\end{figure}

As the instantaneous luminosity increased throughout 2010, it was
necessary to prescale triggers with lower $E_{\rm T}$ thresholds,
while the central jet trigger with the highest $E_{\rm T}$ threshold
remained unprescaled.  As a result, the vast majority of the events
where the leading jet has transverse momentum smaller than about 100
GeV have been taken in the first period of data-taking, under
conditions with a low amount of pile-up, while the majority of the
high-$\pt$ events have been taken during the second data-taking
period, with an average of 2-3 interactions per bunch crossing.  For
each $\pt$-bin considered in this analysis, a dedicated trigger chain
is chosen that is fully efficient ($>99\%$) while having as small a
prescale factor as possible. For inclusive jets fully contained in the
central or in the forward trigger region, only events taken by this
fully efficient trigger are considered.  For inclusive jets in the
HEC-FCal transition region $2.8 \le |y|<3.6$, neither the central nor
the forward trigger is fully efficient. Instead, the logical OR of the
triggers is used, which is fully efficient at sufficiently high jet
$\pt$ (see Fig.~\ref{fig:trigOR}).

A specific strategy is used to account for the various prescale
combinations for inclusive jets in the HEC-FCal transition region,
which can be accepted either by the central jet trigger only, by the
forward jet trigger only, or by both.  A similar strategy is used for
dijet events in a given $(m_{12},y^*)$-bin, which can be accepted by
several jet triggers depending on the transverse momenta and
pseudorapidities of the two leading jets.  Events that can be accepted
by more than one trigger chain have been divided into several
categories according to the trigger combination that could have
accepted the events. For inclusive jets in the transition region,
these correspond to central and forward triggers with a similar
threshold; for dijets the trigger combination depends on the position
and transverse momenta of the two leading jets, each of which is
``matched'' to a trigger object using angular criteria.  Corrections
are applied for any trigger inefficiencies, which are generally below
1\%.  The equivalent luminosity of each of the categories of events is
computed based on the prescale values of these triggers throughout the
data-taking periods, and all results from the various trigger
combinations are combined together according to the prescription given
in Ref.~\cite{Lendermann:2009ah}.

\subsection{Jet Reconstruction and Calibration}
\label{sec:jes}
Jets are reconstructed at the electromagnetic scale using the \AKT{}
algorithm.  The input objects to the jet algorithm are
three-dimensional topological clusters~\cite{topoclusters} built from
calorimeter cells.  The four-momentum of the uncalibrated, EM-scale
jet is defined as the sum of the four-momenta of its constituent
calorimeter energy clusters.  Additional energy due to multiple
proton-proton interactions within the same bunch crossing
(``pile-up'') is subtracted by applying a correction derived as a
function of the number of reconstructed vertices in the event using
minimum bias data.  The energy and the position of the jet are next
corrected for instrumental effects such as dead material and
non-compensation.  This jet energy scale (JES) correction is
calculated using isolated jets\footnote{An isolated jet is defined as
a jet that has no other jet within $\Delta R=2.5R$, where $R$ is the
clustering parameter of the jet algorithm.}
in the Monte Carlo simulation as a
function of energy and pseudorapidity of the reconstructed jet.
The JES correction factor ranges from about 2.1 for low-energy jets with
 $\pt=20$~GeV in the central region $|y|<0.3$ to less than 1.2 for
high-energy jets in the most forward region $3.6\le|y|<4.4$.
The corrections are cross-checked using {\it in-situ}
techniques in collision data (see below)~\cite{Aad:2011he}.

\subsection{Uncertainties in Jet Calibration}
\label{sec:JESuncertainty}
The uncertainty on the jet energy scale is the dominant uncertainty
for the inclusive jet and dijet cross section measurements.  Compared
to the previous analysis~\cite{Collaboration:2010wv}, this uncertainty
has been reduced by up to a factor of two, primarily due to the
improved calibration of the calorimeter electromagnetic energy scale
obtained from $Z\to ee$ events~\cite{Aad:2011mk}, as well as an
improved determination of the single particle energy measurement
uncertainties from {\it in-situ} and test-beam
measurements~\cite{testbeam}.  This improvement is confirmed by
independent measurements, including studies of the momenta of tracks
associated to jets, as well as the momentum balance observed in
$\gamma$+jet, dijet, and multijet events~\cite{Aad:2011he}.

In the central barrel region ($|\eta|<0.8$), the dominant source of
the JES uncertainty is the knowledge of the calorimeter response to
hadrons.  This uncertainty is obtained by measuring the response to
single hadrons using proton-proton and test-beam data, and propagating
the uncertainties to the response for jets.  Additional uncertainties
are evaluated by studying the impact on the calorimeter response from
varying settings for the parton shower, hadronization, and underlying
event in the Monte Carlo simulation.  The estimate of the uncertainty
is extended from the central calorimeter region to the endcap and
forward regions, the latter of which lies outside the tracking
acceptance, by exploiting the transverse momentum balance between a
central and a forward jet in events where only two jets are produced.

In the central region ($|\eta|<0.8$), the uncertainty is lower than
4.6\% for all jets with $\pt > 20$~GeV, which decreases to less than
2.5\% uncertainty for jet transverse momenta between 60 and 800 GeV.
The JES uncertainty is the largest for low-\pt{} ($\sim$20~GeV) jets
in the most forward region $|\eta|>3.6$, where it is about 11-12\%.
Details of the JES determination and its uncertainty are given in
Ref.~\cite{Aad:2011he}.

\subsection{Offline Selection}
\label{sec:selection}

\subsubsection{Event Selection}
\label{eventSelection}

To reject events due to cosmic-ray muons and other non-collision
backgrounds, events are required to have at least one primary vertex
that is consistent with the beamspot position and that has at least
five tracks associated to it.  The efficiency for collision events to
pass these vertex requirements, as measured in a sample of events
passing all selections of this analysis, is well over 99\%.

\subsubsection{Jet Selection}
\label{sec:jet_selection}

For the inclusive jet measurements, jets are required to have
$\pt>20$~GeV and to be within $|y|<4.4$.  They must also pass the
specific fully-efficient trigger for each $\pt$- and~$|y|$-bin, as
detailed in Sec.~\ref{sec:trigger}.  For the dijet measurements,
events are selected if they have at least one jet with $\pt > 30$~GeV
and another jet with $\pt > 20$~GeV, both within $|y| < 4.4$.
Corrections are applied for inefficiencies in jet reconstruction,
which are generally less than a few percent.

Jet quality criteria first established with early collision data are
applied to reject jets reconstructed from calorimeter signals that do
not originate from a proton-proton collision, such as those due to
noisy calorimeter cells~\cite{Aad:2011he}.  For this analysis,
various improvements to the jet quality selection have been made due
to increased experience with a larger data set and evolving beam
conditions, including the introduction of new criteria for the forward
region.

The main sources of fake jets were found to be: noise bursts in the
hadronic endcap calorimeter electronics; coherent noise from the
electromagnetic calorimeter; cosmic rays; and beam-related
backgrounds.

Quality selection criteria were developed for each of these categories
by studying jet samples classified as real or fake energy depositions.
This classification was performed by applying criteria on the
magnitude and direction of the missing transverse momentum,
$\vec{E}_{\mathrm{T}}^{\mathrm{miss}}$. Following this, about a dozen
events with $|\vec{E}_{\mathrm{T}}^{\mathrm{miss}}| > 500$~GeV were
found that pass the standard analysis selection.  These events were
visually scanned and were generally found to be collision events with
mostly low $\pt$ jets and a muon escaping at low scattering angle.

\begin{figure}[tb]
  \begin{center}
  \mbox{
    \includegraphics[width=.48\textwidth]{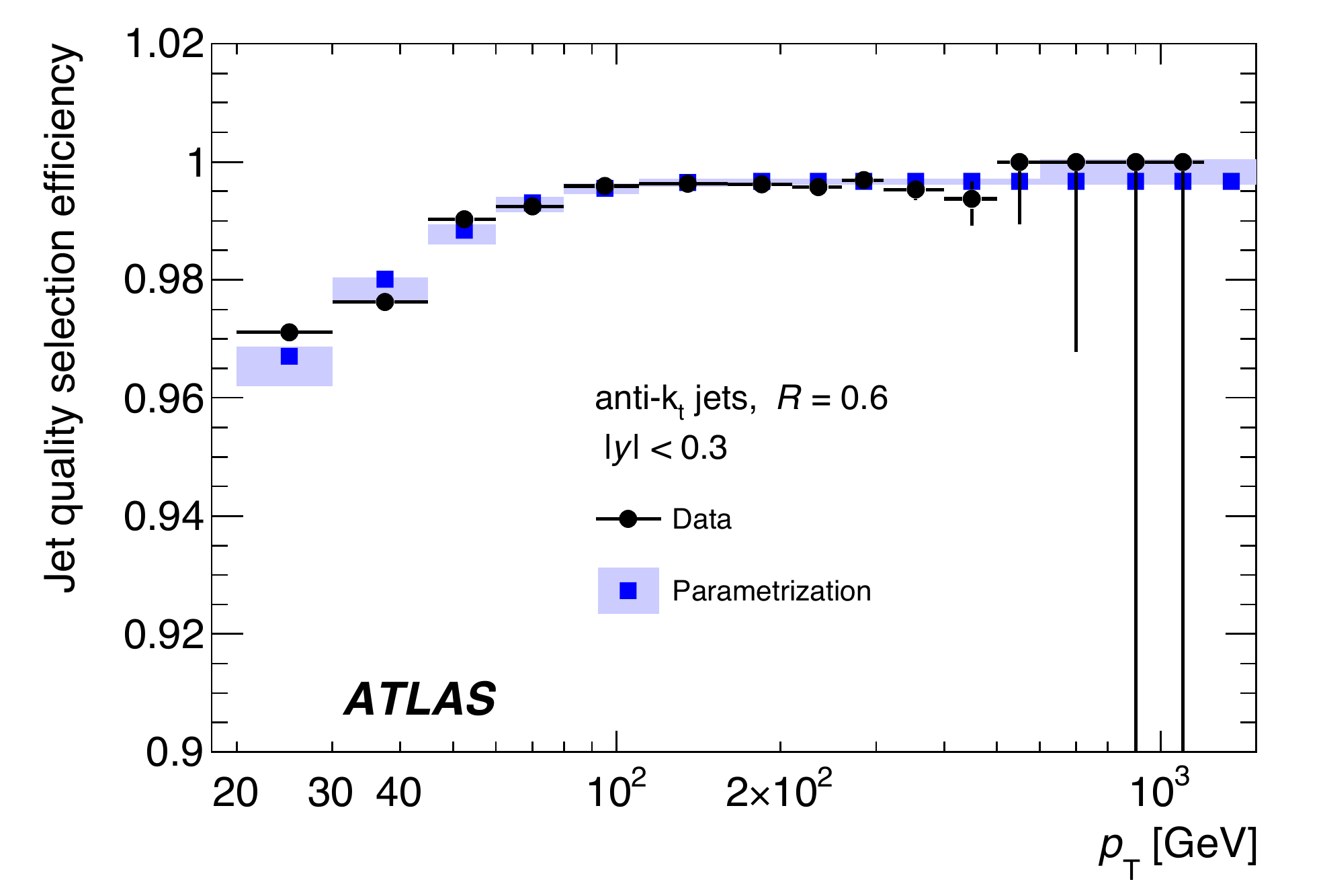}
  }
  \caption{
    Efficiency $\epsilon$ for jet quality selection as a function of
$\pt$ for \AKT jets with $R=0.6$ in the rapidity region $|y|<0.3$.
The black circles indicate the efficiency measured {\it in-situ} using
a tag-probe method.  The blue squares indicate the fit to the
parameterisation $\epsilon(\pt) = A - e^{-(B \pt -C)}$ used in this
analysis, where $A$, $B$, and $C$ are fitted constants, and the shaded
band indicates the systematic uncertainty on the efficiency obtained
by varying the tag jet selection.  The turn-on is due to more
stringent jet quality selection at low jet $\pt$.
  }      
  \label{fig:jetID}
  \end{center}
\end{figure}

The efficiency for identifying real jets was measured using a
tag-and-probe method.  A ``probe jet'' sample was selected by
requiring the presence of a ``tag jet'' that is within $|\eta|<2.0$,
fulfills the jet quality criteria, and is back-to-back ($\Delta \phi
>$ 2.6) and well-balanced with a probe jet
($|{\pt}_{1}-{\pt}_{2}|/{\ptavg} <0.4$, with
${\ptavg}=({\pt}_{1}+{\pt}_{2})/2$ and where ${\pt}_{1,2}$ are the
transverse momenta of the tag and probe jets).  The jet quality
criteria were then applied to the probe jet, measuring as a function
of its $|\eta|$ and $\pt$ the fraction of jets that are not rejected.

The efficiency to select a jet is shown in Fig.~\ref{fig:jetID} for an
example rapidity region, along with the systematic uncertainty on this
efficiency.

The jet quality selection efficiency is greater than 96\% for jets
with $\pt=20$~GeV and quickly increases with jet $\pt$. The efficiency
is above 99\% for jet $\pt>60$~GeV in all rapidity regions.  The
inclusive jet and dijet cross sections are corrected for these
inefficiencies in regions where the efficiency is less than 99\%.  The
systematic uncertainty on the efficiency is taken as a systematic
uncertainty on the cross section.

\subsection{Background, Vertex Position, and Pile-Up}
\label{sec:background}
Background contributions from sources other than proton-proton
collisions were evaluated using events from cosmic-ray runs, as well
as unpaired proton bunches in the accelerator, in which no real
collision candidates are expected.  Based on the duration of the
cosmic-ray runs and the fact that only one event satisfied the
selection criteria, the non-collision background rates across the
entire data period are considered to be negligible.

The primary vertices span the luminous region around the nominal
beamspot.  To determine the systematic uncertainty due to possibly
incorrect modeling of the event vertex position, the jet $\pt$
spectrum was studied as a function of the $|z|$ position of the
primary vertex with the largest $\sum \pt^{2}$ of associated tracks.
The fraction of events with $|z|>200$~mm is 0.06\%, and the difference
in the $\pt$ spectrum compared to events with $|z|<100$~mm is small.
Consequently, the uncertainty from mis-modeling of the vertex position
was taken to be negligible.

The $\pT$ of each jet is corrected for additional energy from soft
pile-up interactions in the event (see Sec.~\ref{sec:jes}).  An
uncertainty associated to this pile-up offset correction is assigned
that is dependent on the number of reconstructed primary vertices, as
described in Sec.~\ref{sec:recoCalib}.  The jet measurements are then
compared to the Monte Carlo simulation without pile-up.

\subsection{Luminosity}
\label{sec:luminosity}
The integrated luminosity is calculated by measuring interaction rates
using several ATLAS devices, where the absolute calibration is derived
using van der Meer scans~\cite{Aad:2011dr}.  The uncertainty on the
luminosity is 3.4\%~\cite{ATLAS-CONF-2011-011}.  The calculation of
the effective luminosity for each bin of the observable for inclusive
jets follows the trigger scheme described in Sec.~\ref{sec:trigger}.
The integrated luminosity for each individual trigger is derived using
separate prescale factors for each luminosity block (an interval of
luminosity with homogeneous data-taking conditions, which is typically
two minutes).  For dijets, each bin receives contributions from
several trigger combinations, for which the luminosity is calculated
independently.  The luminosity that would be obtained without
correction for trigger prescale is (37.3 $\pm$ 1.2)~pb$^{-1}$.  Since
the central jet trigger with the largest transverse momentum threshold
was always unprescaled, this is the effective luminosity taken for
jets with transverse momentum above about 220~GeV.

\section{Unfolding}
\label{sec:unfolding}

\subsection{Technique used}
\label{sec:unfoldingTechnique}

Aside from the jet energy scale correction, all other corrections for
detector inefficiencies and resolutions are performed using an
iterative unfolding, based on a transfer matrix that relates the
particle-level and reconstruction-level observable, with the same
binning as the final distribution.  The unfolding is performed
separately for each bin in rapidity since the migrations across
rapidity bins are negligible compared to those across jet $\pt$ (dijet
mass) bins.  A similar procedure is applied for inclusive jets and
dijets, with the following description applying specifically to the
inclusive jet case.

The Monte Carlo simulation described in Sec.~\ref{sec:mc_samples} is
used to derive the unfolding matrices.  Particle-level and
reconstructed jets are matched together based on geometrical criteria
and used to derive a transfer matrix.  This matrix contains the
expected number of jets within each bin of particle-level and
reconstructed jet $\pt$.  A folding matrix is constructed from the
transfer matrix by normalising row-by-row so that the sum of the
elements corresponding to a given particle-level jet $\pt$ is unity.
Similarly, an unfolding matrix is constructed by normalising
column-by-column so that the sum of the elements corresponding to a
specific reconstructed jet $\pt$ is unity.  Thus each element of the
unfolding matrix reflects the probability for a reconstructed jet in a
particular $\pt$ bin to originate from a specific particle-level $\pt$
bin, given the assumed input particle-level jet $\pt$ spectrum.  The
spectra of unmatched particle-level and reconstructed jets are also
derived from the simulated sample.  The ratio between the number of
matched jets and the total number of jets provides the matching
efficiency both for particle-level jets, $\epsilon^{{\rm ptcl},i}$,
and for reconstructed jets, $\epsilon_{{\rm reco},j}$.

The data are unfolded to particle level using a three-step procedure,
with the final results being given by the equation:
\begin{equation}
  N^{{\rm ptcl},i} = \sum_{j} N_{{\rm reco},j}
  \times \epsilon_{{\rm reco},j}
  A_{{\rm reco},j}^{{\rm ptcl},i}
  / \epsilon^{{\rm ptcl,i}}
\end{equation}
where $i$ and $j$ are the particle-level and reconstructed bin
indices, respectively, and $A_{{\rm reco},j}^{{\rm ptcl},i}$ is an
unfolding matrix refined through iteration, as discussed below.

The first step is to multiply the reconstructed jet spectrum in data
by the matching efficiency $\epsilon_{{\rm reco},j}$, such that it can
be compared to the matched reconstructed spectrum from the Monte Carlo
simulation.  In the second step, the iterated unfolding matrix
$A_{{\rm reco},j}^{{\rm ptcl},i}$ is determined using the Iterative,
Dynamically Stabilised (IDS) method~\cite{Malaescu:2009dm}.  This
procedure improves the transfer matrix through a series of iterations,
where the particle-level distribution is reweighted to the shape of
the corrected data spectrum, while leaving the folding matrix
unchanged.  The main difference with respect to previous iterative
unfolding techniques~\cite{D'Agostini:1994zf} is that, when performing
the corrections, regularisation is provided by the use of the
significance of the data-MC differences in each bin.  The third step
is to divide the spectrum obtained after the iterative unfolding by
the matching efficiency at particle level, thus correcting for the jet
reconstruction inefficiency.

The statistical uncertainties on the spectrum are propagated through
the unfolding by performing pseudo-experiments.  An ensemble of
pseudo-experiments is created in which each bin of the transfer matrix
is varied according to its statistical uncertainty.  A separate set of
pseudo-experiments is performed where the data spectrum is varied
while respecting correlations between jets produced in the same event.
The unfolding is then applied to each pseudo-experiment, and the
resulting ensembles are used to calculate the covariance matrix of the
corrected spectrum.

As a cross-check, the results obtained from the iterative
unfolding have been compared to those using a simpler bin-by-bin
correction procedure, as well as the ``singular value decomposition''
(SVD) method implemented in \texttt{TSVDUnfold}~\cite{Hocker:1995kb,
TSVDUnfold2}. These methods use different regularisation procedures
and rely to different degrees on the Monte Carlo simulation modelling
of the shape of the spectrum.  The unfolding techniques have been
tested using a data-driven closure test~\cite{Malaescu:2009dm}.  In
this test the particle-level spectrum in the Monte Carlo simulation is
reweighted and convolved through the folding matrix such that a
significantly improved agreement between the data and the
reconstructed spectrum from the Monte Carlo simulation is attained.
The reweighted, reconstructed spectrum in the Monte Carlo simulation
is then unfolded using the same procedure as for the data. The
comparison of the result with the reweighted particle-level spectrum
from the Monte Carlo simulation provides the estimation of the bias.

The bin-by-bin method gives results consistent with those obtained
using the IDS technique, but requires the application of an explicit
correction for the NLO k-factor to obtain good agreement.  A somewhat
larger bias is observed for the SVD method.

\begin{figure}[tb]
\begin{center}
\vspace*{-0.7cm}
\includegraphics[width=0.95\linewidth]{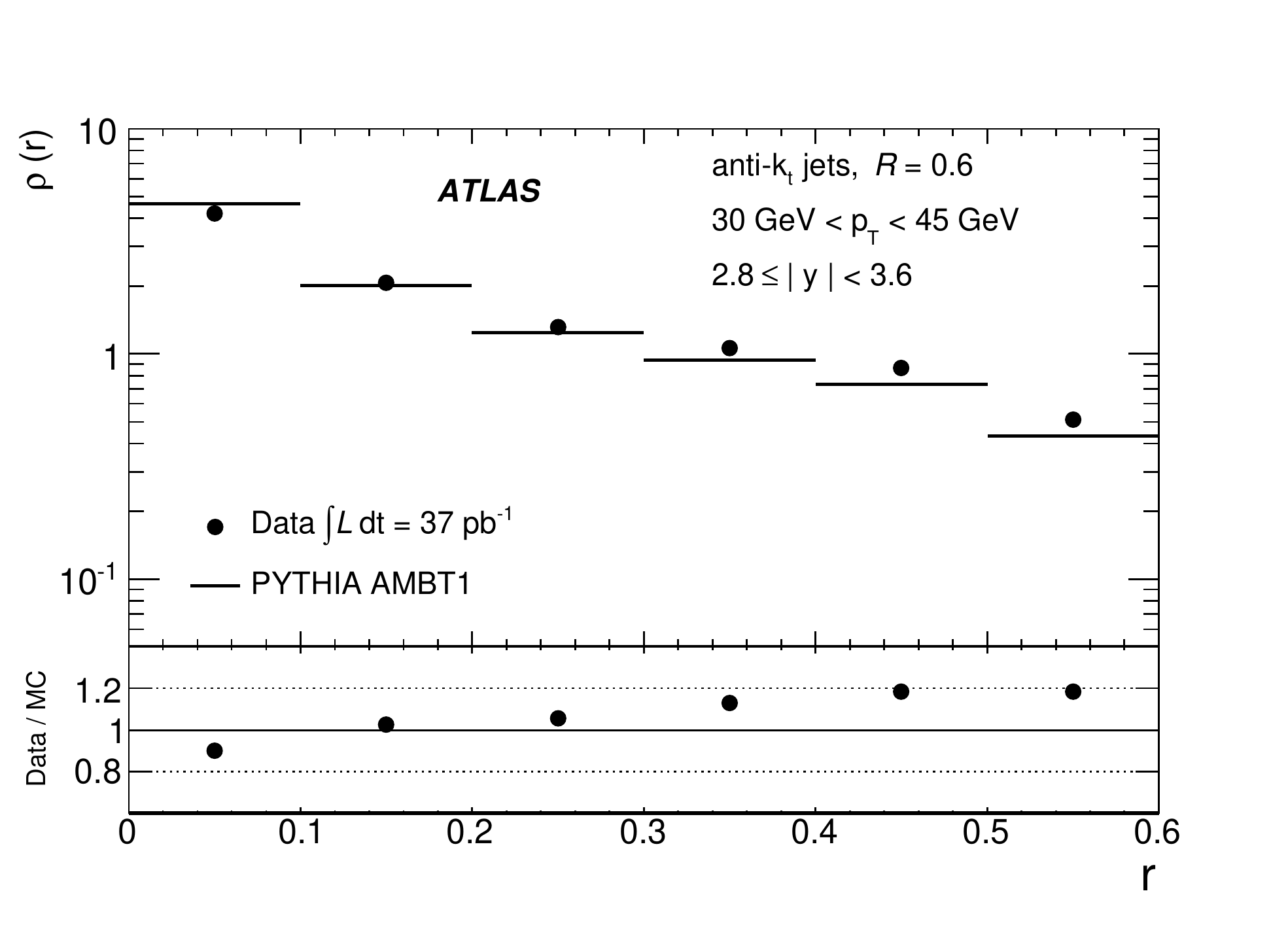}
\caption{
The jet shape $\rho(r)$ measured using calorimeter energy clusters for
\AKT jets with $R=0.6$ in the rapidity interval $2.8 \le |y| < 3.6$,
compared to \pythia with tune AMBT1 (used for unfolding), and for jets
with transverse momenta in the range $30 < \pt < 45\gev$.  The
statistical error bars are smaller than the size of the markers, while
systematic errors are not shown.
}
\label{fig:forwardshapes_28_36__06}
\end{center}
\end{figure}

\subsection{Cross-check with jet shapes}
The use of Monte Carlo simulation to derive the transfer matrix in the
unfolding procedure requires that the simulation models the jet
properties well.  The modelling of the energy flow around the jet core
provides a useful test of this.  The energy and momentum flow within a
jet can be expressed in terms of the differential jet shape, defined
for a jet with radius parameter $R$, as the fraction $\rho(r) =
\frac{1}{\Delta r} \frac{\pt^r}{\pt^R}$, where $\pt^R$ is the
transverse momentum within a radius $R$ of the jet centre, and $\pt^r$
is the transverse momentum contained within a ring of thickness
$\Delta r = 0.1$ at a radius $r = \sqrt{(\Delta y)^2 +
(\Delta\phi)^2}$ from the jet centre.

Jet shape measurements using calorimeter energy clusters and tracks
were performed with 3~pb$^{-1}$ of data~\cite{Aad:2011kq}, and show
good agreement with the \pythia and \herwig+ \jimmy Monte Carlo
simulations in the kinematic region $30~{\rm GeV} < \pT < 600$~GeV and
rapidity $|y| < 2.8$.  Using the same technique, the uncorrected jet
shapes in the forward rapidity region $2.8 \le |y| < 4.4$ have been
studied in the context of the present analysis.  As an example, the
results for the HEC-FCal transition region $2.8 \le |y| < 3.6$, the
most difficult detector region to model, are shown in
Fig.~\ref{fig:forwardshapes_28_36__06}.  The maximum disagreement in
shape between data and the Monte Carlo simulation is approximately
20\%, demonstrating that the distribution of energy within the jets is
reasonably well-modeled even in this worst case.  Any bias from
mis-modeling of the jet shape is included in the unfolding
uncertainties described below, so this jet shape study serves only as
a cross-check.

\section{Systematic uncertainties and correlations}

\subsection{Uncertainty sources from jet reconstruction and calibration}
\label{sec:recoCalib}

The uncertainty on the jet reconstruction efficiency for $|y|<2.1$
(within the tracking acceptance) is evaluated using track jets, which
are used to play the role of ``truth jets''.  In this paper, truth
jets are defined to be jets at the particle level, but excluding muons
and neutrinos.  The efficiency to reconstruct a calorimeter jet given
a track jet nearby is studied in both data and the MC simulation.  The
data versus MC comparison of this efficiency is used to infer the
degree to which the calorimeter jet reconstruction efficiency may be
mis-modeled in the Monte Carlo simulation.  The disagreement was found
to be 2\% for calorimeter jets with $\pt$ of 20~GeV and less than 1\%
for those with $\pt > 30$~GeV.  The disagreement for jets with
$|y|<2.1$ is taken as a systematic uncertainty for all jets in the
rapidity range $|y|<4.4$.  This is expected to be a conservative
estimate in the forward region where the jets have higher energy for a
given $\pt$.

The JES uncertainty was evaluated as described in
Sec.~\ref{sec:JESuncertainty} and in Ref.~\cite{Aad:2011he}.  The
jet energy and angular resolutions are estimated from the Monte Carlo
simulation using truth jets that have each been matched to a
reconstructed calorimeter jet.
The jet energy resolution (JER) in the Monte Carlo simulation is
compared to that obtained in data using two {\it in-situ} techniques,
one based on dijet balance and the other using a bisector method~\cite{ATLAS-CONF-2010-054}.  In general the two resolutions agree within 14\%, and
the full difference is taken as a contribution to the uncertainty on
the unfolding corrections, which propagates to a systematic
uncertainty on the measured cross section as described in
Sec.~\ref{sec:uncProp}.  The angular resolution is estimated from the
angle between each calorimeter jet and its matched truth-level
jet. The associated systematic uncertainty is assessed by varying the
requirement that the jet is isolated.

The JES uncertainty due to pile-up is proportional to $(N_{\rm
PV}-1)/\pt$, where $N_{\rm PV}$ is the number of reconstructed
vertices.  The total pile-up uncertainty for a given $(\pt, y)$-bin is
calculated as the average of the uncertainties for each value of
$N_{\rm PV}$ weighted by the relative frequency of that number of
reconstructed vertices in the bin.

\subsection{Uncertainty propagation}
\label{sec:uncProp}

The uncertainty of the measured cross section due to jet energy scale
and jet energy and angular resolutions has been estimated using the
Monte Carlo simulation by repeating the analysis after systematically
varying these effects.  The jet energy scale applied to the
reconstructed jets in MC is varied separately for each JES uncertainty
source both up and down by one standard deviation.  The resulting
$\pt$ spectra are unfolded using the nominal unfolding matrix, and the
relative shifts with respect to the nominal unfolded spectra are taken
as uncertainties on the cross section.  The effects of the jet energy
and angular resolutions are studied by smearing the reconstructed jets
such that these resolutions are increased by one standard deviation of
their respective uncertainties (see Sec.~\ref{sec:recoCalib}).  For
each such variation, a new transfer matrix is constructed, which is
used to unfold the reconstructed jet spectrum of the nominal MC
sample.  The relative shift of this spectrum with respect to the
nominal unfolded spectra is taken as the uncertainty on the cross
section.

The impact of possible mis-modeling of the cross section shape in the
Monte Carlo simulation is assessed by shape variations of the
particle-level jet spectra introduced to produce reconstructed-level
spectra in agreement with data as discussed in
Sec.~\ref{sec:unfolding}.

The total uncertainty on the unfolding corrections is defined as the
sum in quadrature of the uncertainties on the jet energy resolution,
jet angular resolution, and the simulated shape.  It is approximately
4-5\% at low and high $\pt$ (except for the lowest $\pt$-bin at
20~GeV, where it reaches 20\%), and is smaller at intermediate $\pt$
values.  This uncertainty is dominated by the component from the jet
energy resolution.

\subsection{Summary of the magnitude of the systematic uncertainties}
The largest systematic uncertainty for this measurement arises from
the jet energy scale.  Even with the higher precision achieved
recently as described in Sec.~\ref{sec:JESuncertainty}, the very
steeply falling jet $\pt$ spectrum, especially for large rapidities,
translates even relatively modest uncertainties on the transverse
momentum into large changes for the measured cross section.

As described in Sec.~\ref{sec:luminosity}, the luminosity uncertainty
is 3.4\%.  The detector unfolding uncertainties have been discussed in
the previous subsection.  Various other sources of systematic
uncertainties were considered and were found to have a small impact on
the results.  The jet energy and angular resolutions, as well as the
jet reconstruction efficiency, also contribute to the total
uncertainty through the unfolding corrections.

The dominant systematic uncertainties for the measurement of the
inclusive jet $\pt$ spectrum in representative $\pt$ and $y$ regions
for $\AKT$ jets with $R=0.6$ are shown in Table~\ref{tab:systsumm}.
Similarly, the largest systematic uncertainties for the dijet mass
measurement are given for a few representative $\twomass{1}{2}$ and
$y^{*}$ regions in Table~\ref{tab:systsummDijet}.

\begin{figure*}[p]
  \centering
  \includegraphics[width=.388\textwidth]{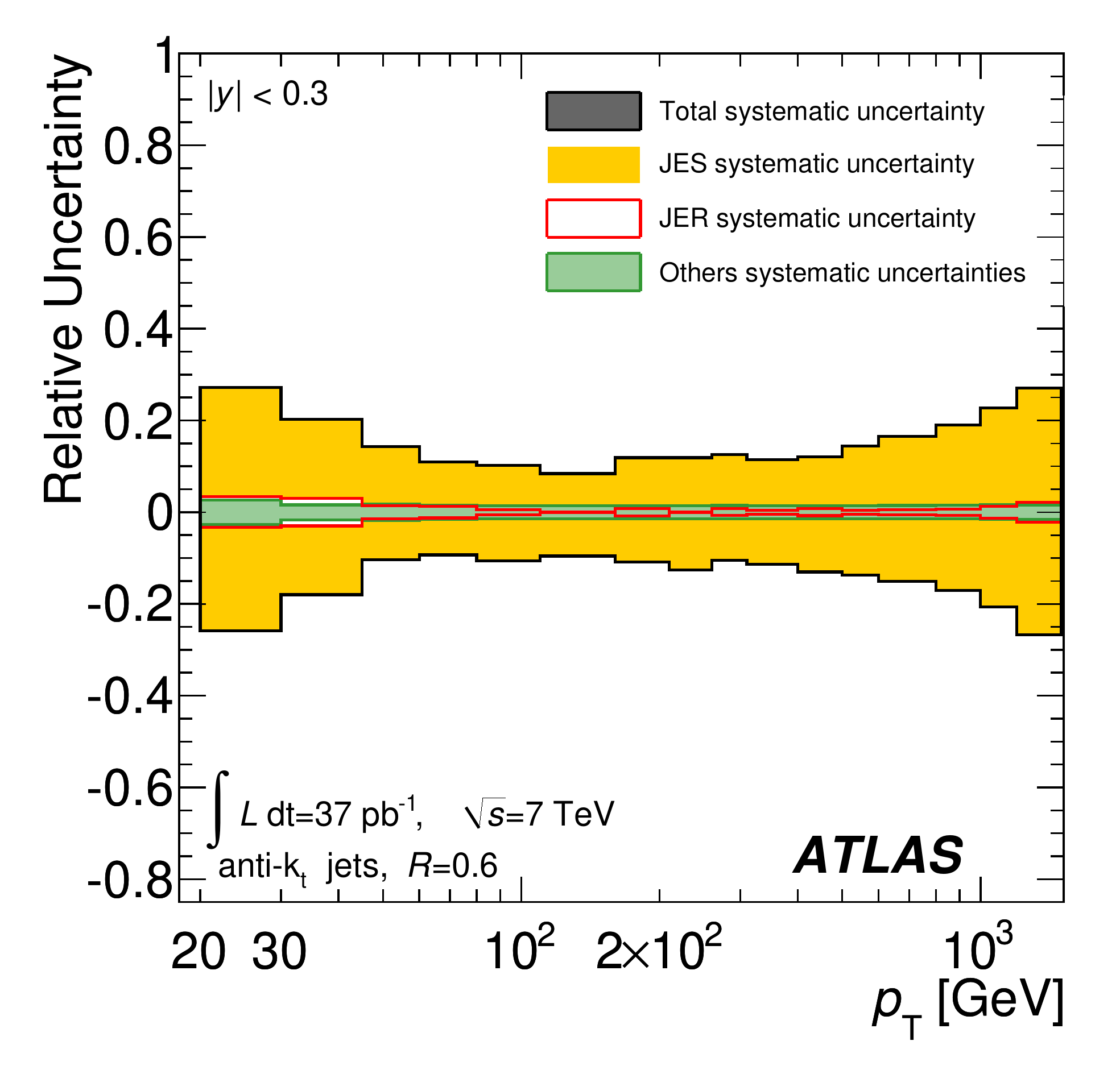}
  \includegraphics[width=.452\textwidth]{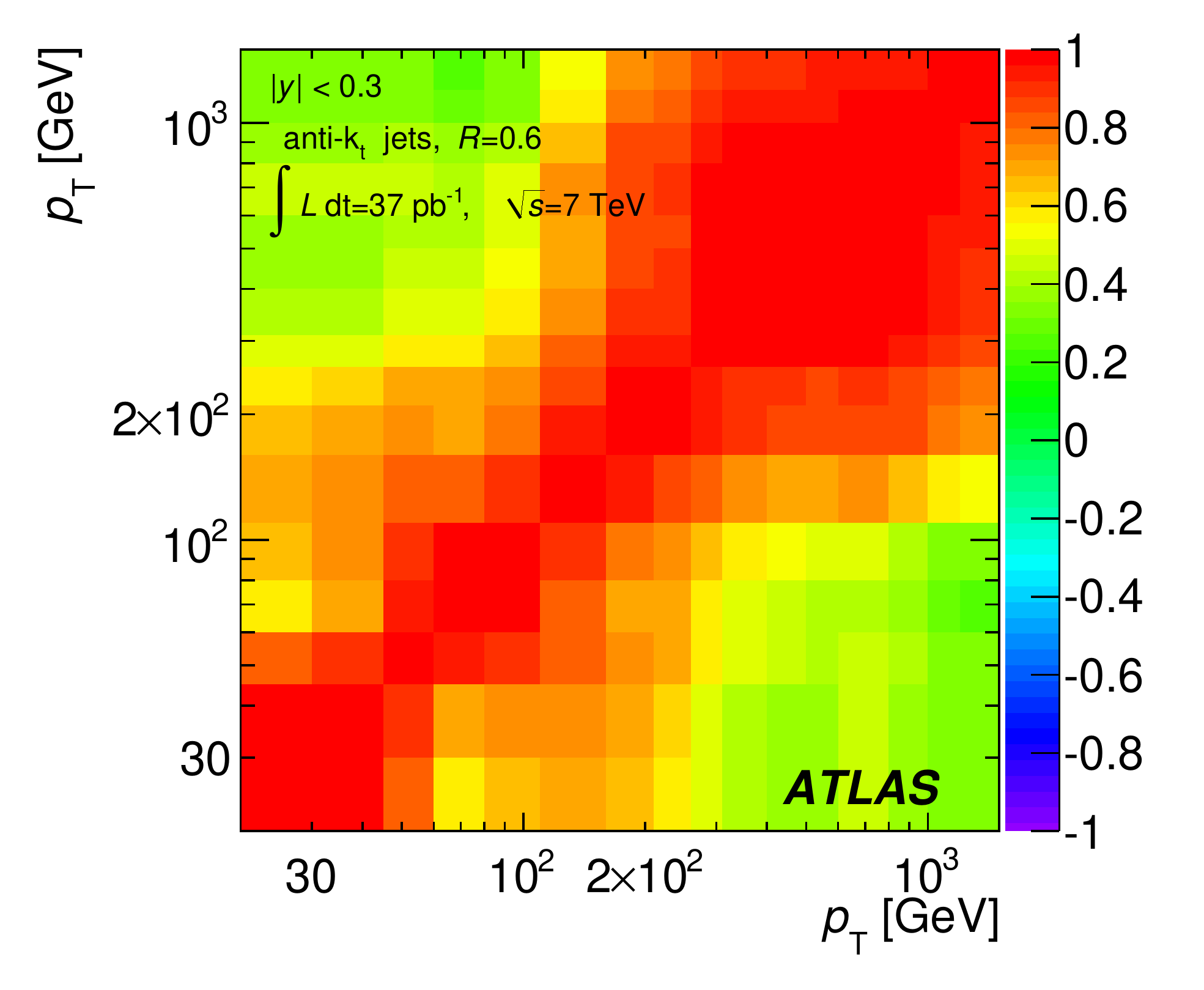}\\
  \includegraphics[width=.388\textwidth]{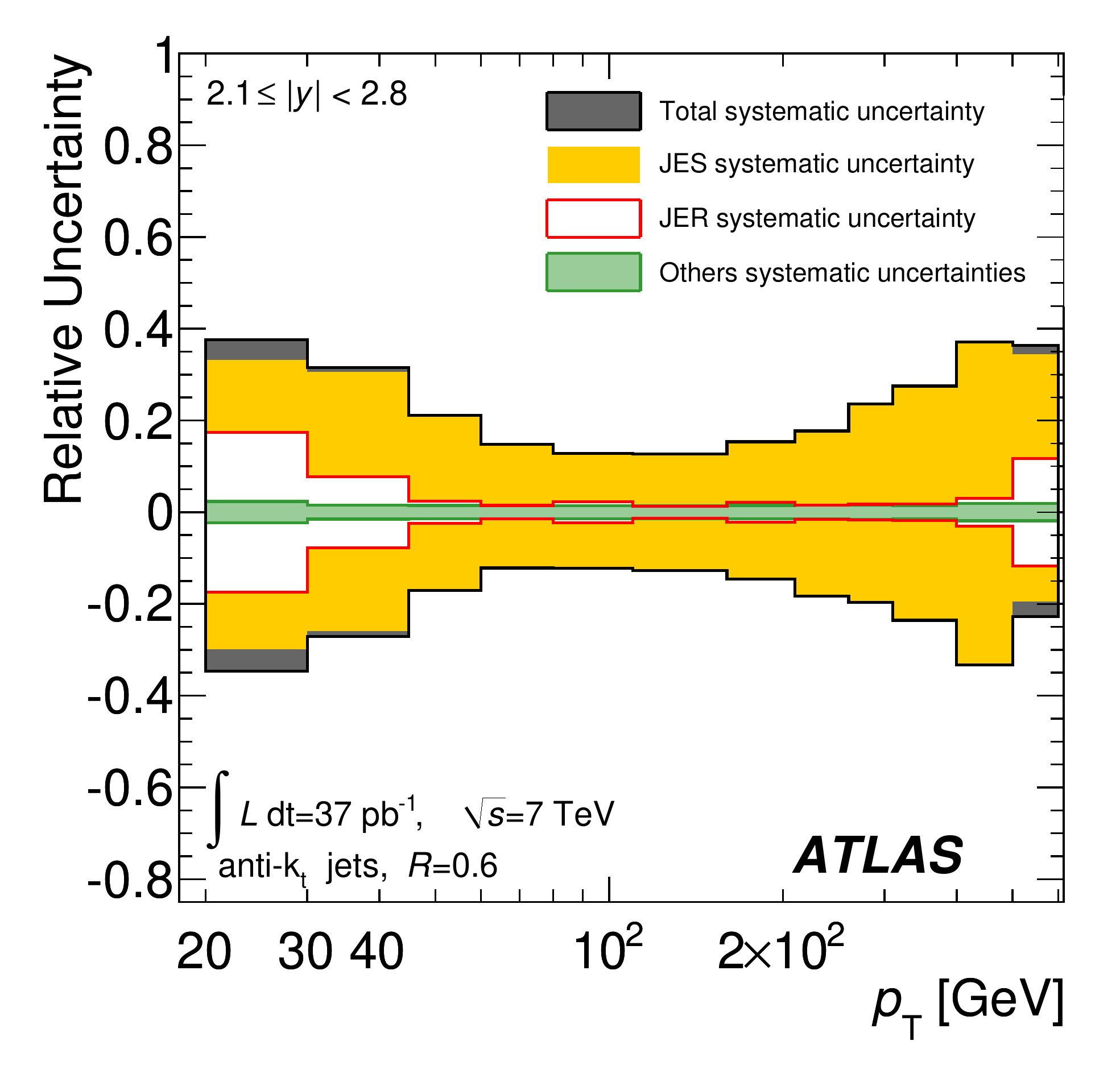}
  \includegraphics[width=.452\textwidth]{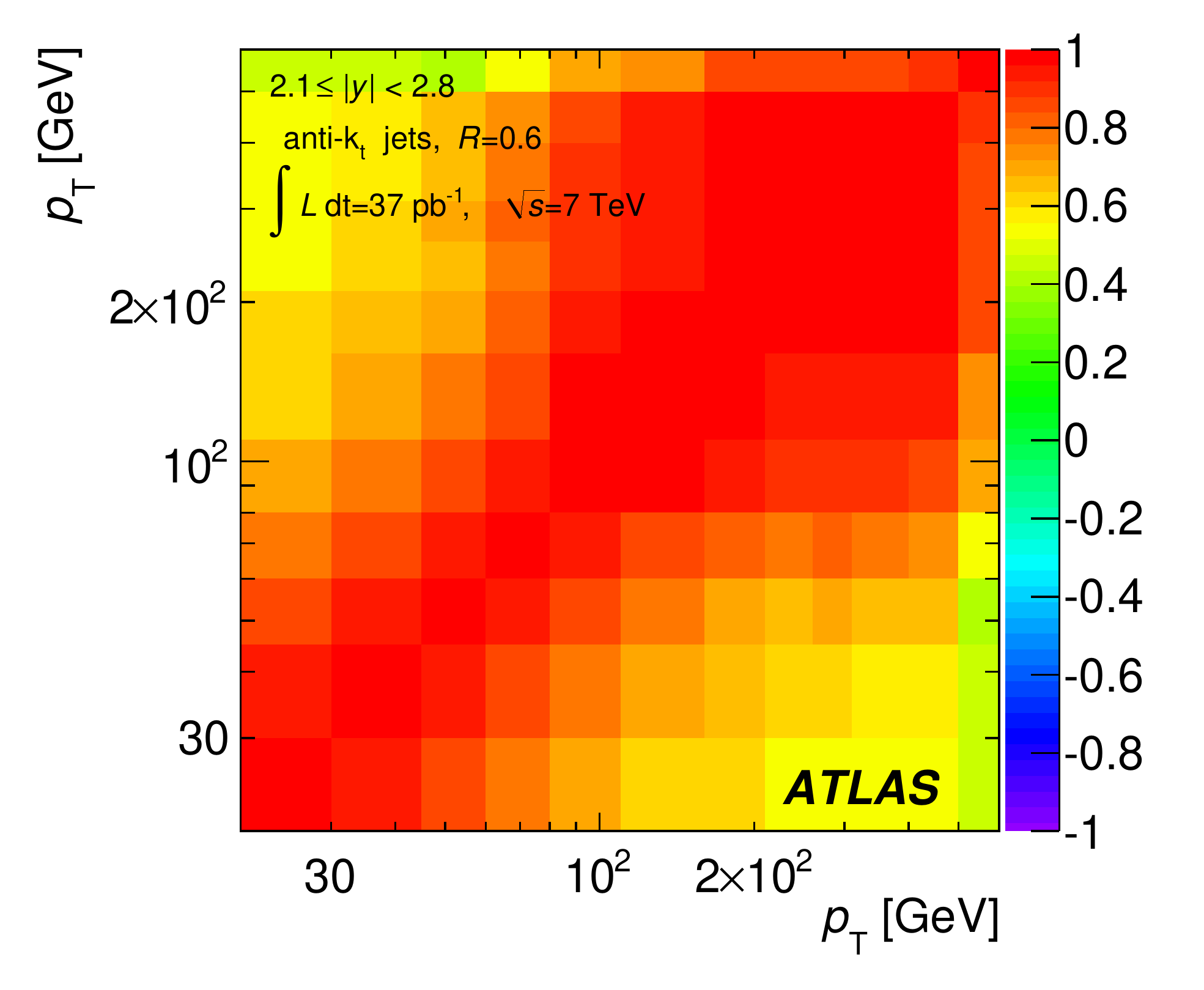}\\
  \includegraphics[width=.388\textwidth]{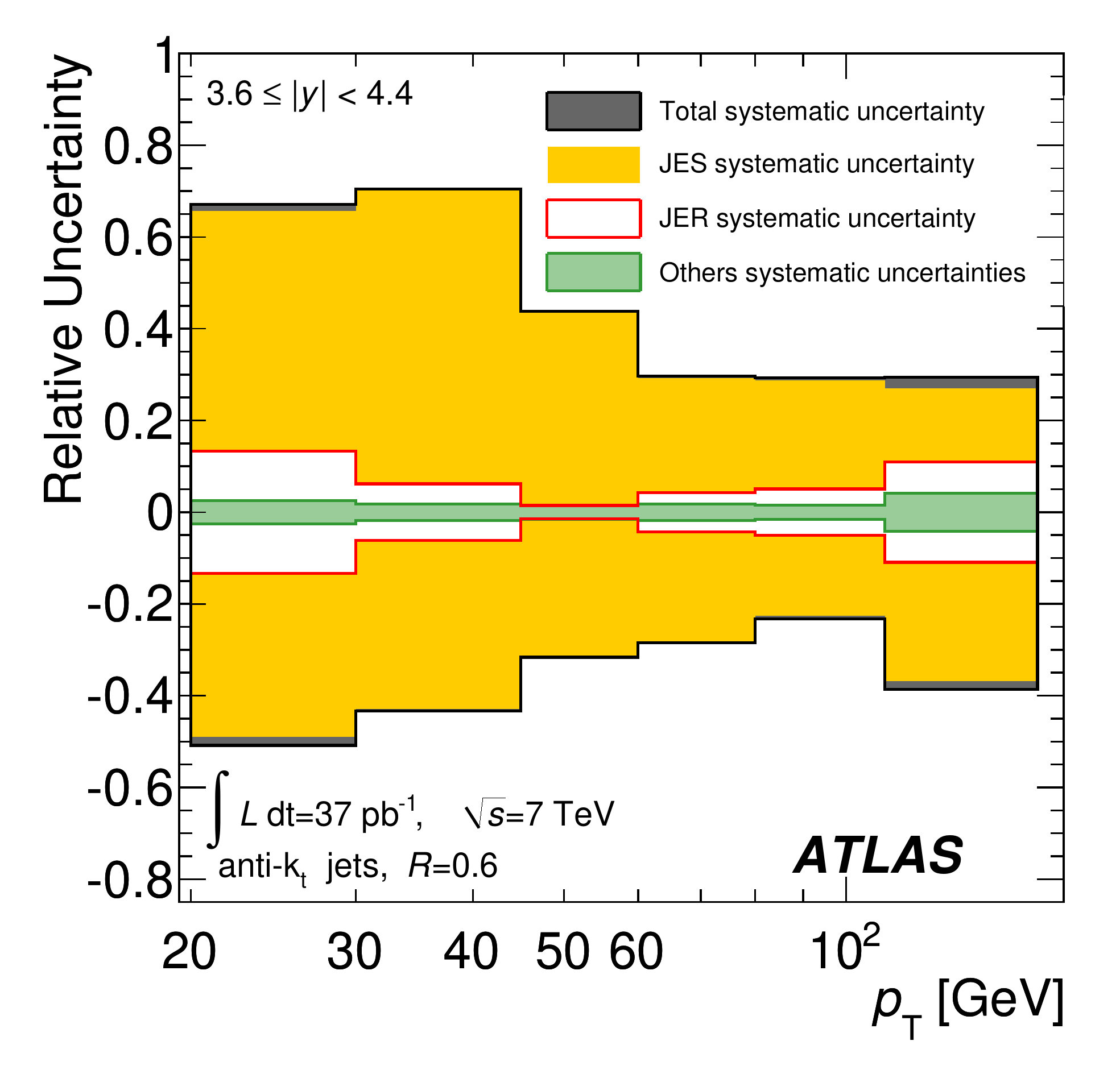}
  \includegraphics[width=.452\textwidth]{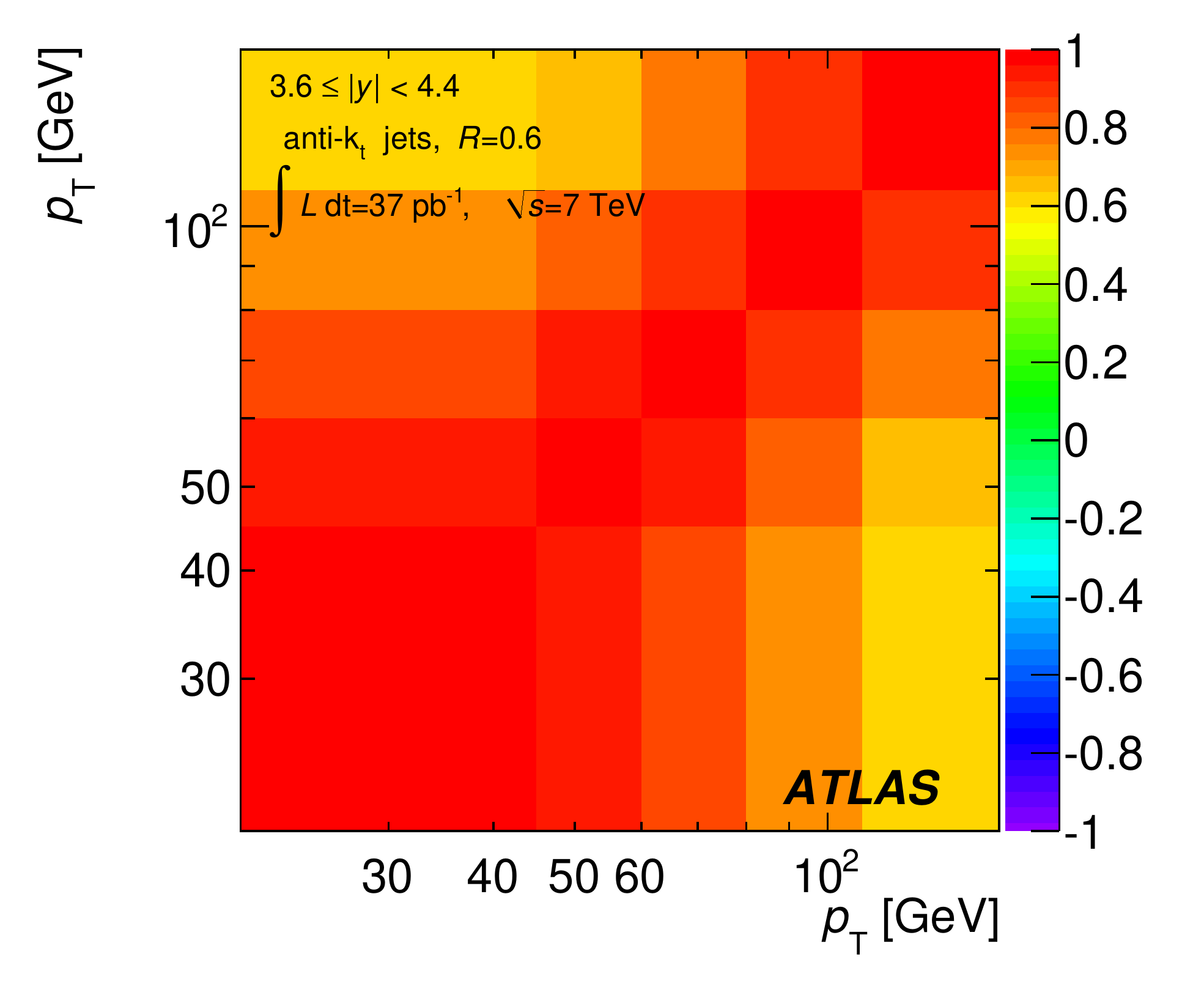}
  \caption{
    The magnitude (left) and correlation between \pt-bins (right) of
    the total systematic uncertainty on the inclusive jet cross
    section measurement for \AKT{} jets with $R=0.6$ in three
    representative $|y|$-bins.  The magnitudes of the uncertainties
    from the jet energy scale (JES), the jet energy resolution (JER),
    and other sources are shown separately.  The correlation matrix is
    calculated after symmetrising the uncertainties.  The statistical
    uncertainty and the 3.4\% uncertainty of the integrated luminosity
    are not shown here.
  }
  \label{fig:totSysAndCorr}
\end{figure*}

An example of the breakdown of the systematic uncertainties as a
function of the jet transverse momentum for the various rapidity bins
used in the inclusive jet measurement is shown in
Fig.~\ref{fig:totSysAndCorr}.

\begin{table}[t]
  \setlength{\extrarowheight}{3pt}
  \begin{center}
    \begin{tabular}{rrrrrrrr}\hline\hline
      $\pt$ [GeV] & $|y|$ & ~JES~ & JER & Trigger & Jet Rec. \\[2pt]\hline
      20--30 & 2.1--2.8 & $^{+35\%}_{-30\%}$ & 17\%  & $1\%$ & $2\%$ \\
      20--30 & 3.6--4.4 & $^{+65\%}_{-50\%}$ & 13\%  & $1\%$ & $2\%$ \\
      80--110 &  $<$ 0.3 &               10\% &  1\%  & $1\%$ & $1\%$ \\[2pt]
      \hline\hline
    \end{tabular}
  \end{center}
  \caption{
    The effect of the dominant systematic uncertainty sources on the
    inclusive jet cross section measurement, for representative $\pt$
    and $y$ regions for $\AKT$ jets with $R=0.6$.
}
  \label{tab:systsumm}
\end{table}

\begin{table}[t]
  \setlength{\extrarowheight}{3pt}
  \begin{center}
    \begin{tabular}{rrrrrrrr}\hline\hline
      $\twomass{1}{2}$ [TeV] & $y^{*}$ & ~JES~ & JER & Trigger & Jet Rec.
      \\[2pt]\hline
      0.37--0.44 & 2.0--2.5 & $^{+46\%}_{-27\%}$  & 7\% & $1\%$ & $2\%$
      \\
      2.55--3.04 & 4.0--4.4 & $^{+110\%}_{-50\%}$ & 8\% & $2\%$ & $2\%$
      \\
      0.21--0.26 &  $<$ 0.5 &              $10\%$ & 1\% & $1\%$ & $2\%$
      \\[2pt]\hline\hline
    \end{tabular}
  \end{center}
  \caption{
    The effect of the dominant systematic uncertainty sources on the
    dijet cross section measurement, for representative
    $\twomass{1}{2}$ and $y^{*}$ regions for $\AKT$ jets with $R=0.6$.
}
  \label{tab:systsummDijet}
\end{table}

\subsection{Correlations}
\label{sec:corr}

\begin{table*}[tb]
  \begin{center}
    \setlength{\extrarowheight}{1pt}
    \begin{tabular}{lrrrrrrr}
      \hline\hline
      & \multicolumn{7}{c}{$|y|$-bins} \\
      Uncertainty Source & 0-0.3 & 0.3-0.8 & 0.8-1.2 & 1.2-2.1 &
      2.1-2.8 & 2.8-3.6 & 3.6-4.4 \\ \hline
      JES 1: Noise threshold   &  1 &  1 &  2 &  3 &  4 &  5 &  6 \\ 
      JES 2: Theory UE         &  7 &  7 &  8 &  9 & 10 & 11 & 12 \\ 
      JES 3: Theory Showering  & 13 & 13 & 14 & 15 & 16 & 17 & 18 \\ 
      JES 4: Non-closure       & 19 & 19 & 20 & 21 & 22 & 23 & 24 \\ 
      JES 5: Dead material     & 25 & 25 & 26 & 27 & 28 & 29 & 30 \\ 
      JES 6: Forward JES       & 31 & 31 & 31 & 31 & 31 & 31 & 31 \\ 
      JES 7: $E/p$ response    & 32 & 32 & 33 & 34 & 35 & 36 & 37 \\ 
      JES 8: $E/p$ selection   & 38 & 38 & 39 & 40 & 41 & 42 & 43 \\ 
      JES 9: EM+neutrals       & 44 & 44 & 45 & 46 & 47 & 48 & 49 \\ 
      JES 10: HAD $E$-scale    & 50 & 50 & 51 & 52 & 53 & 54 & 55 \\ 
      JES 11: High $\pt$       & 56 & 56 & 57 & 58 & 59 & 60 & 61 \\ 
      JES 12: $E/p$ bias       & 62 & 62 & 63 & 64 & 65 & 66 & 67 \\ 
      JES 13: Test-beam bias   & 68 & 68 & 69 & 70 & 71 & 72 & 73 \\ 
      Unfolding                & 74 & 74 & 74 & 74 & 74 & 74 & 74 \\ 
      Jet matching             & 75 & 75 & 75 & 75 & 75 & 75 & 75 \\ 
      Jet energy resolution    & 76 & 76 & 77 & 78 & 79 & 80 & 81 \\ 
      $y$-resolution           & 82 & 82 & 82 & 82 & 82 & 82 & 82 \\ 
      Jet reconstruction eff.  & 83 & 83 & 83 & 83 & 84 & 85 & 86 \\ 
      Luminosity               & 87 & 87 & 87 & 87 & 87 & 87 & 87 \\ 
      JES 14: Pile-up $(u_1)$  &  u &  u &  u &  u &  u &  u &  u \\
      Trigger $(u_2)$          &  u &  u &  u &  u &  u &  u &  u \\ 
      Jet identification $(u_3)$ &  u &  u &  u &  u &  u &  u &  u \\
      \hline \hline
    \end{tabular}
  \end{center}
  \caption{
Description of bin-to-bin uncertainty correlation for the
inclusive jet measurement.  Each number corresponds to a nuisance
parameter for which the corresponding uncertainty is fully
correlated versus $\pt$.  Bins with the same nuisance parameter
are treated as fully correlated, while bins with different
nuisance parameters are uncorrelated.  The sources indicated by
the letter ``u'' are uncorrelated both between $\pt$- and
$|y|$-bins.  The one-standard-deviation amplitude of the
systematic effect associated with each nuisance parameter is
detailed in
Tables~\ref{tab:AntiKt4_EtaLow0_Results}--\ref{tab:AntiKt6_EtaLow3.6_Results}
in Appendix~\ref{sec:tablesIncJet}.  The JES uncertainties for
jets with $|y| \geq 0.8$ are determined relative to the JES of
jets with $|y|<0.8$.  As a consequence, several of the
uncertainties that are determined using jets with $|y|< 0.8$ are
also propagated to the more forward rapidities (such as the $E/p$
uncertainties).  Descriptions of the JES uncertainty sources can
be found in Refs.~\cite{Aad:2011he}
and~\cite{REL16-SINGLEPTCL-CONF}.  All tables are available on \hepdata~\cite{HEPDATA}.
  }
  \label{tab:correlations}
\end{table*}

\begin{table*}[tb]
  \begin{center}
    \setlength{\extrarowheight}{1pt}
    \begin{tabular}{lrrrrrrrrr}
      \hline\hline
      & \multicolumn{9}{c}{$y^{*}$-bins} \\
      Uncertainty Source & 0.0-0.5 & 0.5-1.0 & 1.0-1.5 & 1.5-2.0 &
      2.0-2.5 & 2.5-3.0 & 3.0-3.5 & 3.5-4.0 & 4.0-4.4 \\ \hline
      JES 1: Noise threshold   &  1 &  1 &  2 &  3 &  4 &  4 & 5  &  6 & 6 \\ 
      JES 2: Theory UE         &  7 &  7 &  8 &  9 & 10 & 10 & 11 & 12 & 12 \\ 
      JES 3: Theory Showering  & 13 & 13 & 14 & 15 & 16 & 16 & 17 & 18 & 18 \\ 
      JES 4: Non-closure       & 19 & 19 & 20 & 21 & 22 & 22 & 23 & 24 & 24 \\ 
      JES 5: Dead material     & 25 & 25 & 26 & 27 & 28 & 28 & 29 & 30 & 30 \\ 
      JES 6: Forward JES       & 31 & 31 & 31 & 31 & 31 & 31 & 31 & 31 & 31 \\ 
      JES 7: $E/p$ response    & 32 & 32 & 33 & 34 & 35 & 35 & 36 & 37 & 37 \\ 
      JES 8: $E/p$ selection   & 38 & 38 & 39 & 40 & 41 & 41 & 42 & 43 & 43 \\ 
      JES 9: EM+neutrals       & 44 & 44 & 45 & 46 & 47 & 47 & 48 & 49 & 49 \\ 
      JES 10: HAD $E$-scale    & 50 & 50 & 51 & 52 & 53 & 53 & 54 & 55 & 55 \\ 
      JES 11: High $\pt$       & 56 & 56 & 57 & 58 & 59 & 59 & 60 & 61 & 61 \\ 
      JES 12: $E/p$ bias       & 62 & 62 & 63 & 64 & 65 & 65 & 66 & 67 & 67 \\ 
      JES 13: Test-beam bias   & 68 & 68 & 69 & 70 & 71 & 71 & 72 & 73 & 73 \\ 
      Unfolding                & 74 & 74 & 74 & 74 & 74 & 74 & 74 & 74 & 74 \\ 
      Jet matching             & 75 & 75 & 75 & 75 & 75 & 75 & 75 & 75 & 75 \\ 
      Jet energy resolution    & 76 & 76 & 77 & 78 & 79 & 79 & 80 & 81 & 81 \\ 
      $y$-resolution           & 82 & 82 & 82 & 82 & 82 & 82 & 82 & 82 & 82 \\ 
      Jet reconstruction eff.  & 83 & 83 & 83 & 83 & 84 & 84 & 85 & 86 & 86 \\ 
      Luminosity               & 87 & 87 & 87 & 87 & 87 & 87 & 87 & 87 & 87 \\ 
      JES 14: Pile-up $(u_1)$  &  u &  u &  u &  u &  u &  u & u &  u & u \\
      Trigger $(u_2)$          &  u &  u &  u &  u &  u &  u & u &  u & u \\ 
      Jet identification $(u_3)$ &  u &  u &  u &  u &  u &  u & u &  u & u \\
      \hline \hline
    \end{tabular}
  \end{center}
  \caption{
    Description of bin-to-bin uncertainty correlation for the dijet
    measurement.  Each number corresponds to a nuisance parameter for
    which the corresponding uncertainty is fully correlated versus
    dijet mass, $\twomass{1}{2}$.  Bins with the same nuisance
    parameter are treated as fully correlated, while bins with
    different nuisance parameters are uncorrelated.  The sources
    indicated by the letter ``u'' are uncorrelated both between
    $\twomass{1}{2}$- and $y^{*}$-bins.  The one-standard-deviation
    amplitude of the systematic effect associated with each nuisance
    parameter is detailed in
    Tables~\ref{tab:DijetMassResults04_0}--\ref{tab:DijetMassResults06_8}
    in Appendix~\ref{sec:tablesDijet}.  Descriptions of the JES
    uncertainty sources can be found in Refs.~\cite{Aad:2011he}
    and~\cite{REL16-SINGLEPTCL-CONF}.  All tables are available on \hepdata~\cite{HEPDATA}.
  }
  \label{tab:correlationsDijet}
\end{table*}

The behaviour of various sources of systematic uncertainty in
different parts of the detector has been studied in detail in order to
understand their correlations across various $\pt$, $m_{12}$ and
rapidity bins.  As shown in Tables~\ref{tab:correlations}
and~\ref{tab:correlationsDijet}, 22 independent sources of systematic
uncertainty have been identified, including luminosity, jet energy
scale and resolution, and theory effects such as the uncertainty of
the modeling of the underlying event and the QCD showering.  For
example, the sources labeled ``JES 7--13'' in these tables correspond
to the calorimeter response to hadrons, which dominates the JES
uncertainty in the central region.  After examining the rapidity
dependence of all 22 sources, it was found that 87 independent
nuisance parameters are necessary to describe the correlations over
the whole phase space.  The systematic effect on the cross section
measurement associated with each nuisance parameter in its range of
use is completely correlated in $\pt$ and $y$ (dijet mass and
$y^{*}$).  These parameters represent correlations between the
uncertainties of the various bins.  Since many of the systematic
effects are not symmetric, it is not possible to provide a covariance
matrix containing the full information.  For symmetric uncertainties
corresponding to independent sources, the total covariance matrix is
given by:
\begin{equation}
  {\rm cov}({i,j})=\sum_{\lambda} \Gamma_{\lambda i}\Gamma_{\lambda
  j},
  \label{eq:cov}
\end{equation}
where $\lambda$ is an index running over the nuisance parameters, and
$\Gamma_{\lambda i}$ is the one-standard-deviation amplitude of the
systematic effect due to source $\lambda$ in bin $i$.  The full list
of relative uncertainties, $\gamma_{\lambda}$, where each uncertainty
may be asymmetric, is given for all sources $\lambda$ and bins of this
analysis in
Tables~\ref{tab:AntiKt4_EtaLow0_Results}--\ref{tab:AntiKt6_EtaLow3.6_Results}
and~\ref{tab:DijetMassResults04_0}--\ref{tab:DijetMassResults06_8}.
Fig.~\ref{fig:totSysAndCorr} shows the magnitude and approximate
bin-to-bin correlations of the total systematic uncertainty of the
inclusive jet cross section measurement.  The correlation matrix is
here converted from the covariance matrix, which is obtained using
Eq.~\ref{eq:cov}, after symmetrising the uncertainties:
$\Gamma_{\lambda i}=(\Gamma^+_{\lambda i} + \Gamma^-_{\lambda i})/2$.
The inclusive jet and dijet data should not be used simultaneously for
PDF fits due to significant correlations between the two measurements.

\section{Results and Discussion}
\label{sec:results}

\subsection{Inclusive Jet Cross Sections}
\label{sec:inclusive}
\begin{figure*}[p]
\begin{center}
\includegraphics[width=0.95\textwidth]{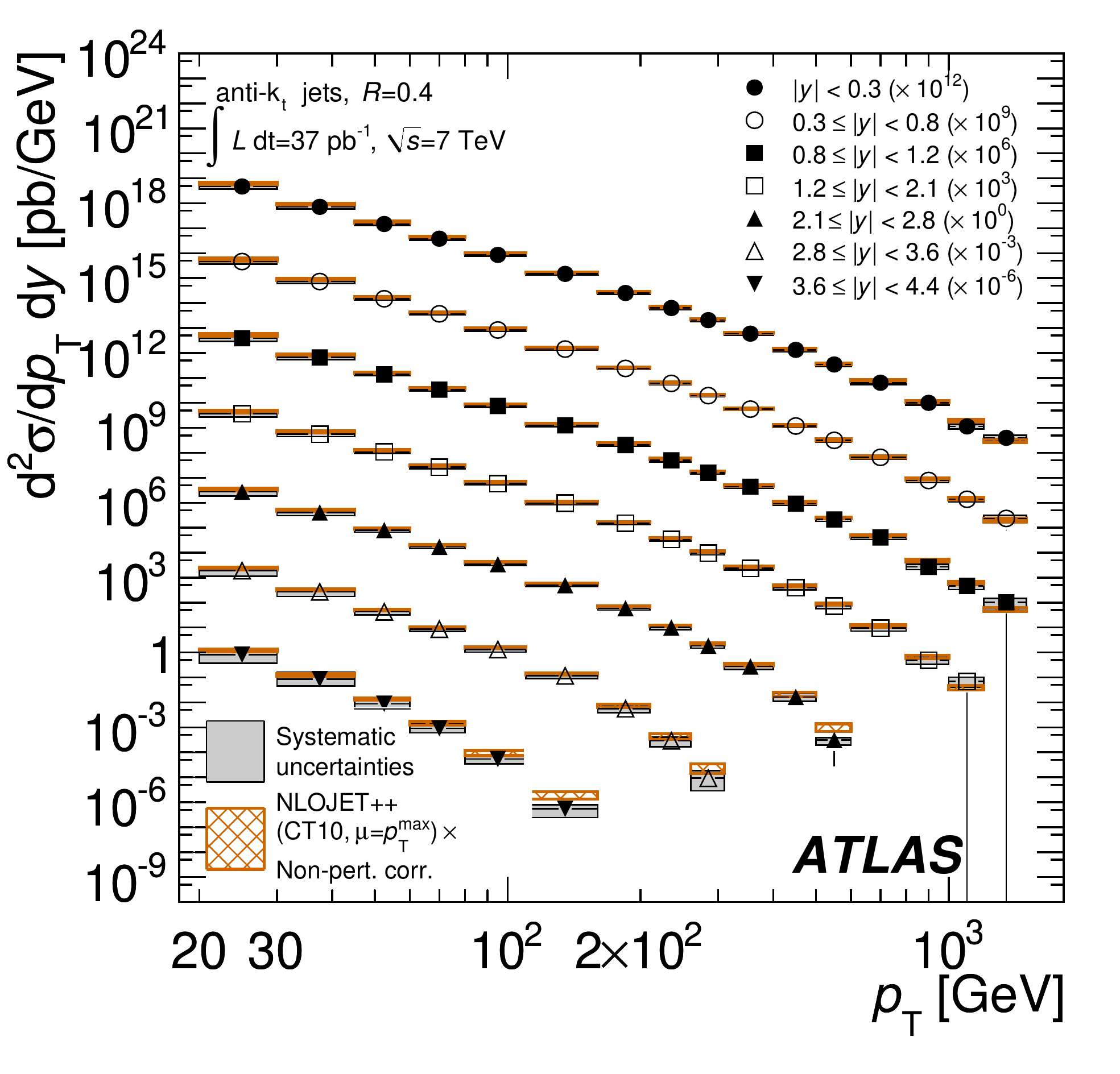}
\caption{
Inclusive jet double-differential cross section as a function of jet
$\pt$ in different regions of $|y|$ for jets identified using the \AKT
algorithm with $R=0.4$. For convenience, the cross sections are
multiplied by the factors indicated in the legend.  The data are
compared to NLO pQCD calculations using \nlojet to which
non-perturbative corrections have been applied.  The error bars, which
are usually smaller than the symbols, indicate the statistical
uncertainty on the measurement. The dark-shaded band indicates the
quadratic sum of the experimental systematic uncertainties, dominated
by the jet energy scale uncertainty. There is an additional overall
uncertainty of 3.4\% due to the luminosity measurement that is not
shown. The theory uncertainty, shown as the light, hatched band, is
the quadratic sum of uncertainties from the choice of the
renormalisation and factorisation scales, parton distribution
functions, $\alpha_s(M_Z)$, and the modeling of non-perturbative
effects, as described in the text.
}
\label{fig:incjetptsummary04}
\end{center}
\end{figure*}

The inclusive jet double-differential cross section is shown in
Figs.~\ref{fig:incjetptsummary04} and~\ref{fig:incjetptsummary06} and
Tables~\ref{tab:AntiKt4_EtaLow0_Results}--\ref{tab:AntiKt6_EtaLow3.6_Results}
in Appendix~\ref{sec:tablesIncJet} for jets reconstructed with the
\AKT algorithm with $R = 0.4$ and $R = 0.6$.  The measurement extends
from jet transverse momentum of 20~GeV to almost 1.5~TeV, spanning two
orders of magnitude in $\pt$ and ten orders of magnitude in the value
of the cross section.  The measured cross sections have been corrected
for all detector effects using the unfolding procedure described in
Sec.~\ref{sec:unfolding}.  The results are compared to \nlojet
predictions (using the CT10 PDF set) corrected for non-perturbative
effects, where the theoretical uncertainties from scale variations,
parton distribution functions, and non-perturbative corrections have
been accounted for.

\begin{figure}[tbh]
\begin{center}
\includegraphics[width=0.45\textwidth]{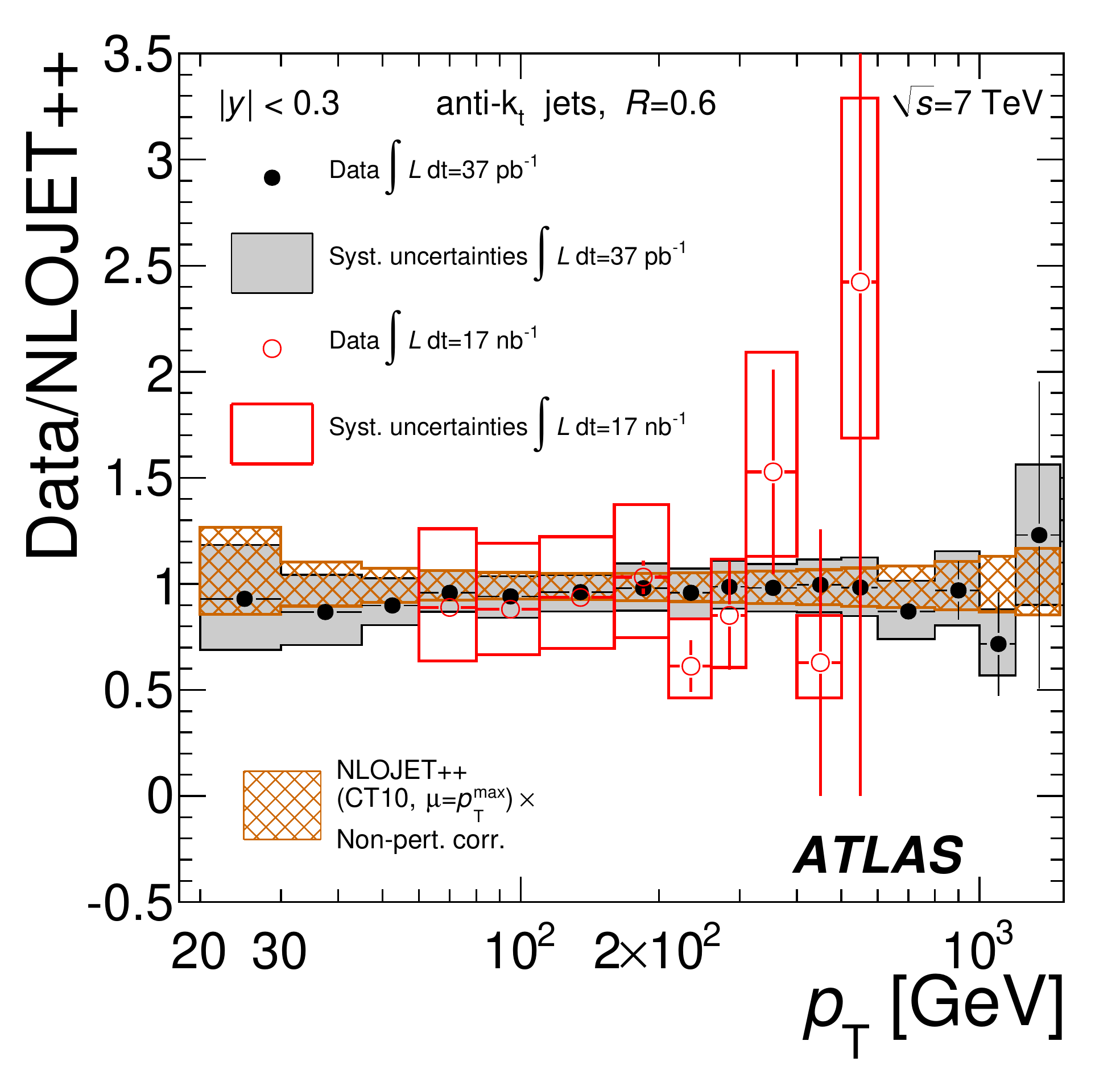}
\caption{
Ratio of inclusive jet cross section to the theoretical prediction
obtained using \nlojet with the CT10 PDF set.  The ratio is shown as a
function of jet $\pt$ in the rapidity region $|y|<0.3$, for jets
identified using the \AKT algorithm with $R=0.6$.  The current result
is compared to that published in Ref.~\cite{Collaboration:2010wv}.
}
\label{fig:comp19nb}
\end{center}
\end{figure}

\begin{figure*}[p]
\begin{center}
\includegraphics[width=0.54\textwidth]{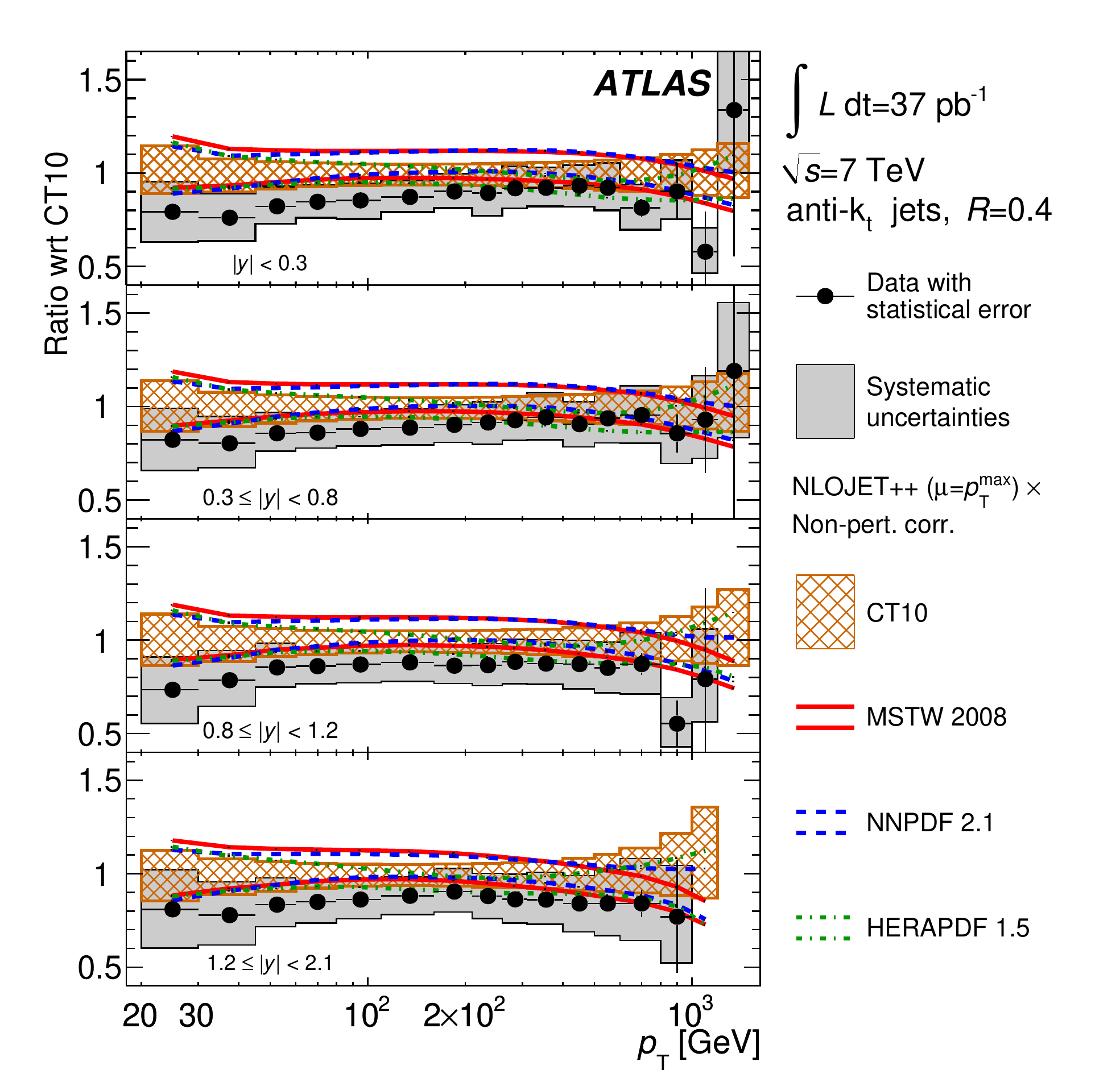}
\hspace*{-31.5mm}
\includegraphics[width=0.54\textwidth]{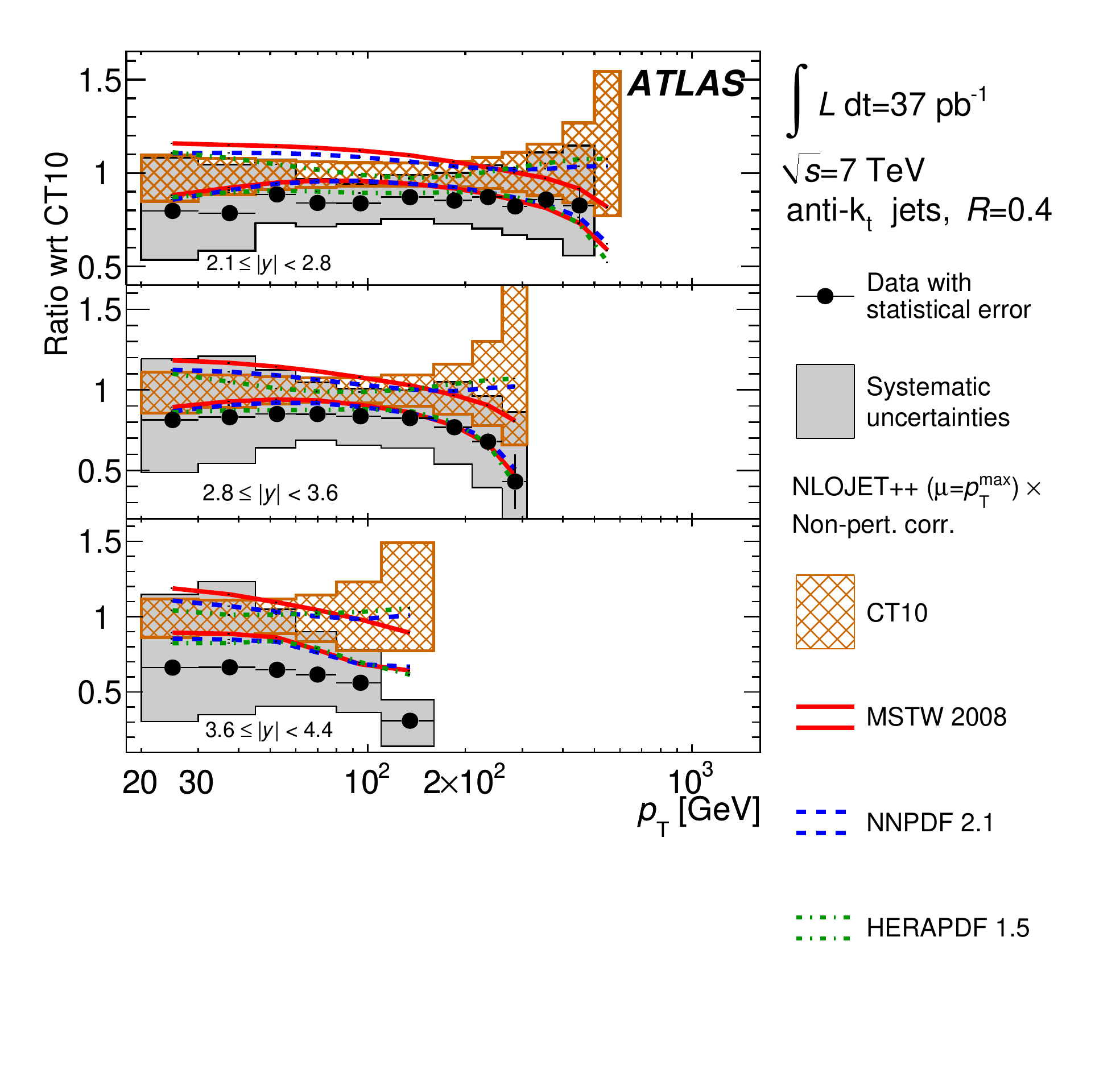}\\
\includegraphics[width=0.54\textwidth]{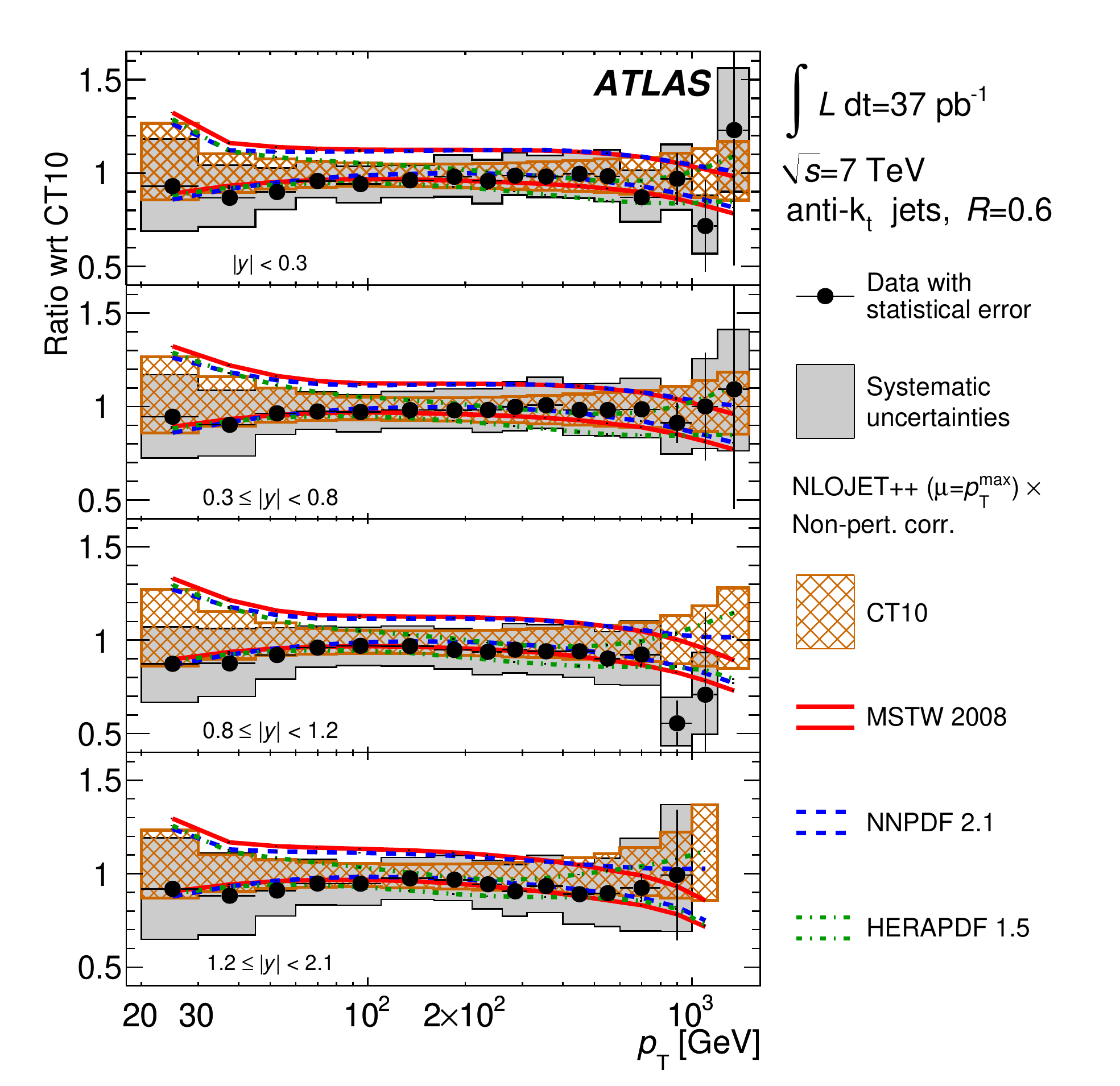}
\hspace*{-31.5mm}
\includegraphics[width=0.54\textwidth]{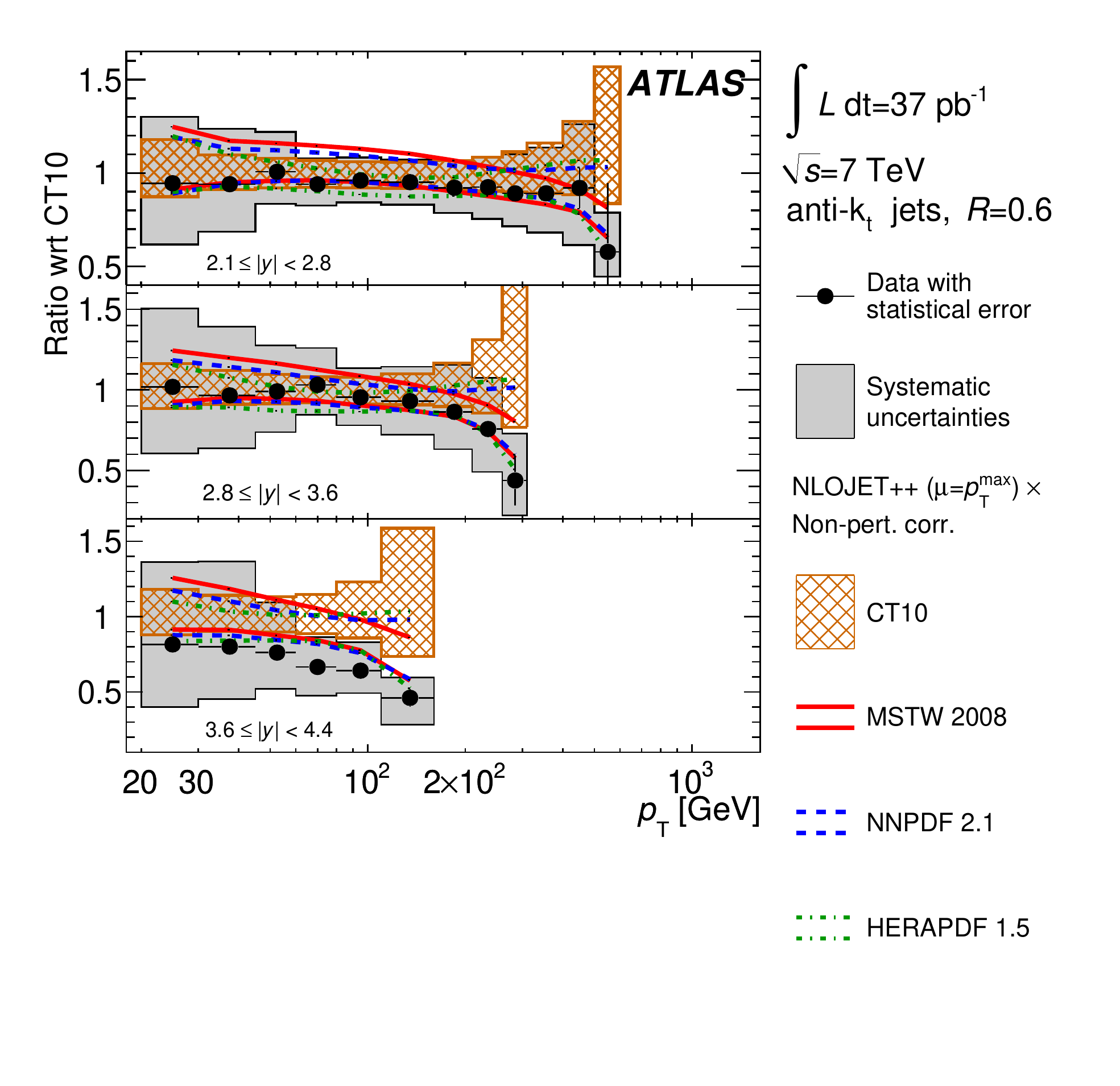}
\caption{
Ratios of inclusive jet double-differential cross section to the
theoretical prediction obtained using \nlojet with the CT10 PDF set.
The ratios are shown as a function of jet $\pt$ in different regions
of $|y|$ for jets identified using the \AKT algorithm with $R=0.4$
(upper plots) and $R=0.6$ (lower plots).  The theoretical error bands
obtained by using \nlojet with different PDF sets (CT10, MSTW 2008,
NNPDF 2.1, HERAPDF 1.5) are shown.  Statistically insignificant data
points at large $\pt$ are omitted in the ratio.
}
\label{fig:ratio_PDF}
\end{center}
\end{figure*}

\begin{figure*}[p]
\begin{center}
\includegraphics[width=0.54\textwidth]{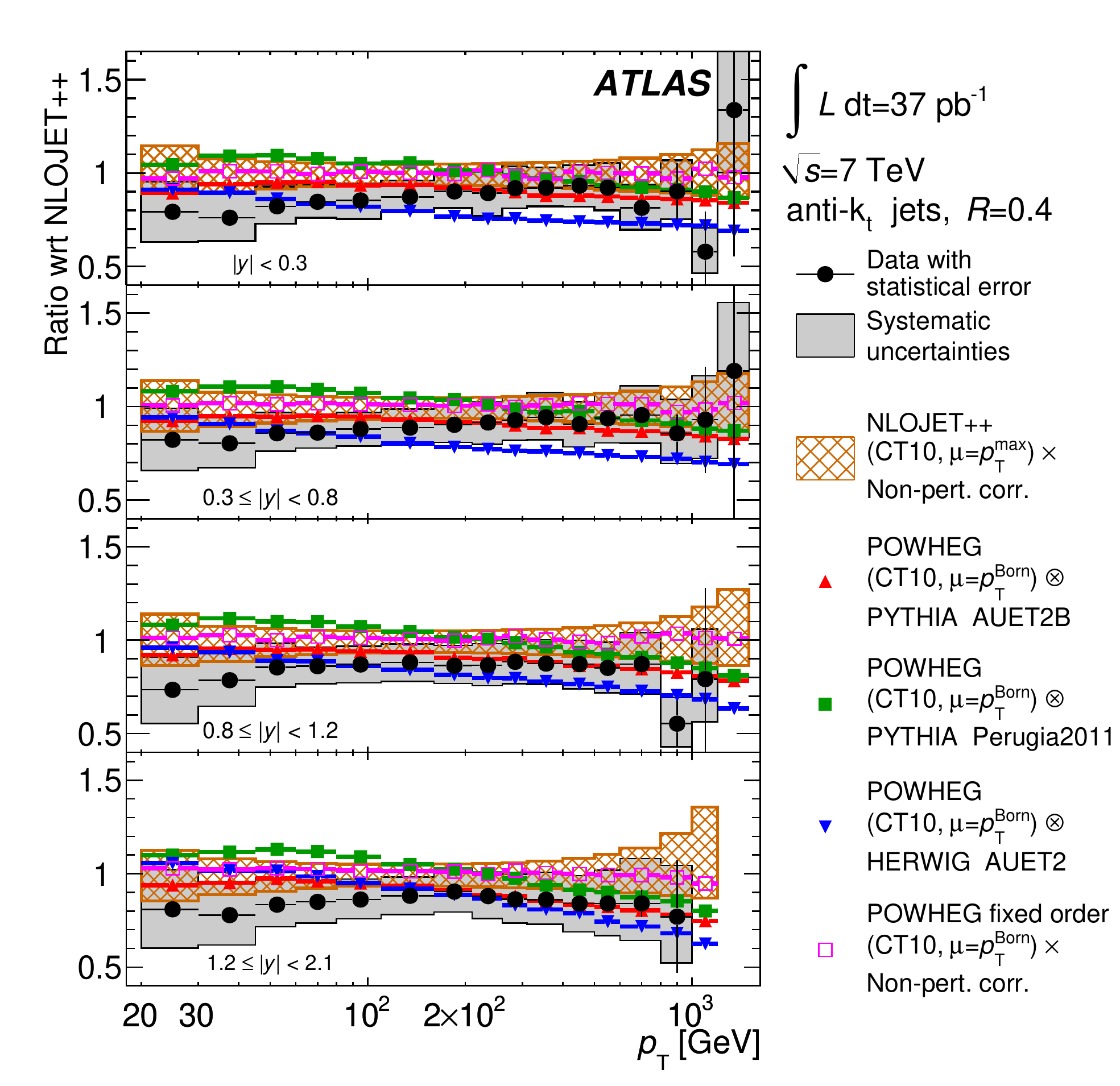}
\hspace*{-31.5mm}
\includegraphics[width=0.54\textwidth]{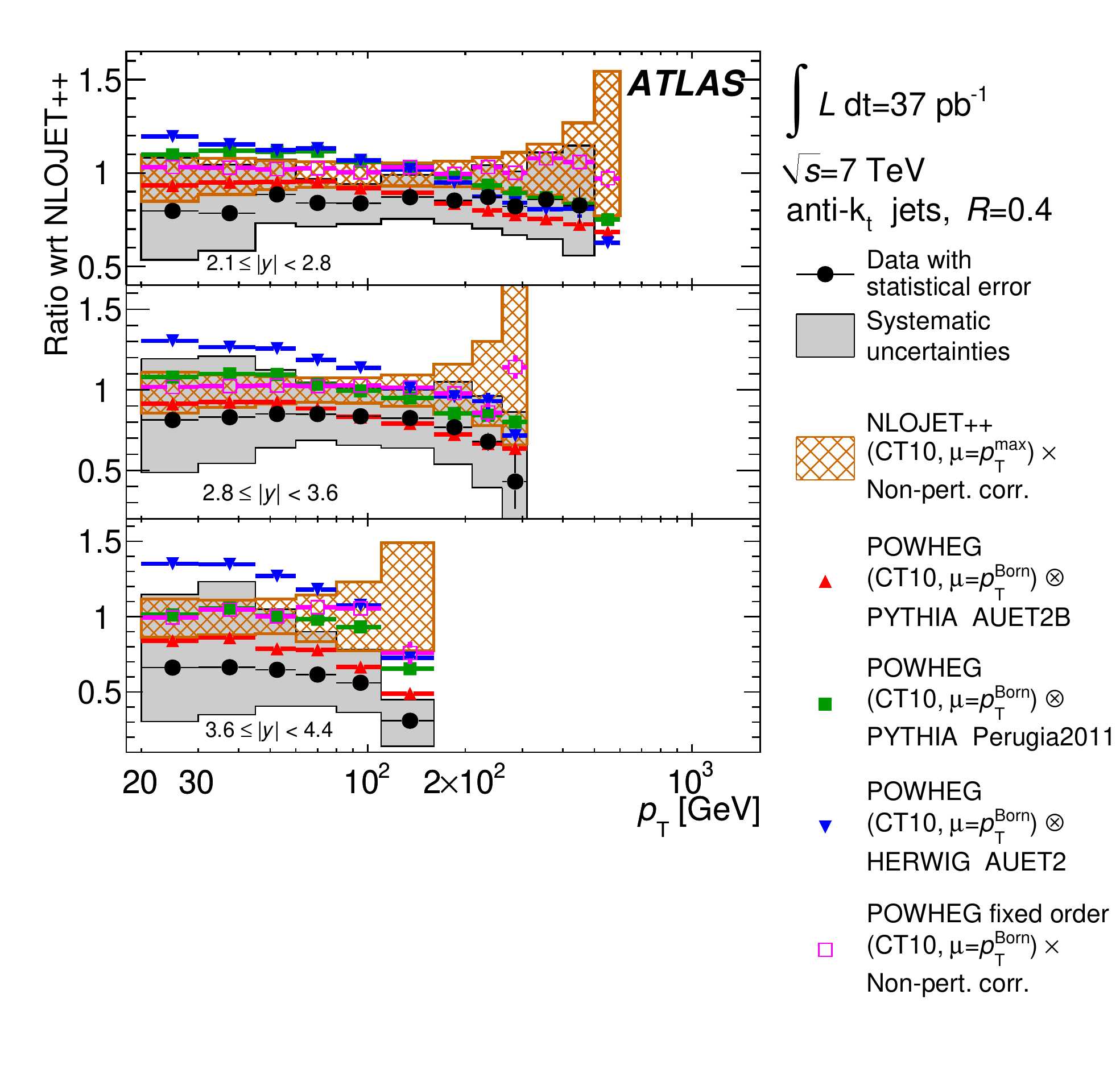}\\
\includegraphics[width=0.54\textwidth]{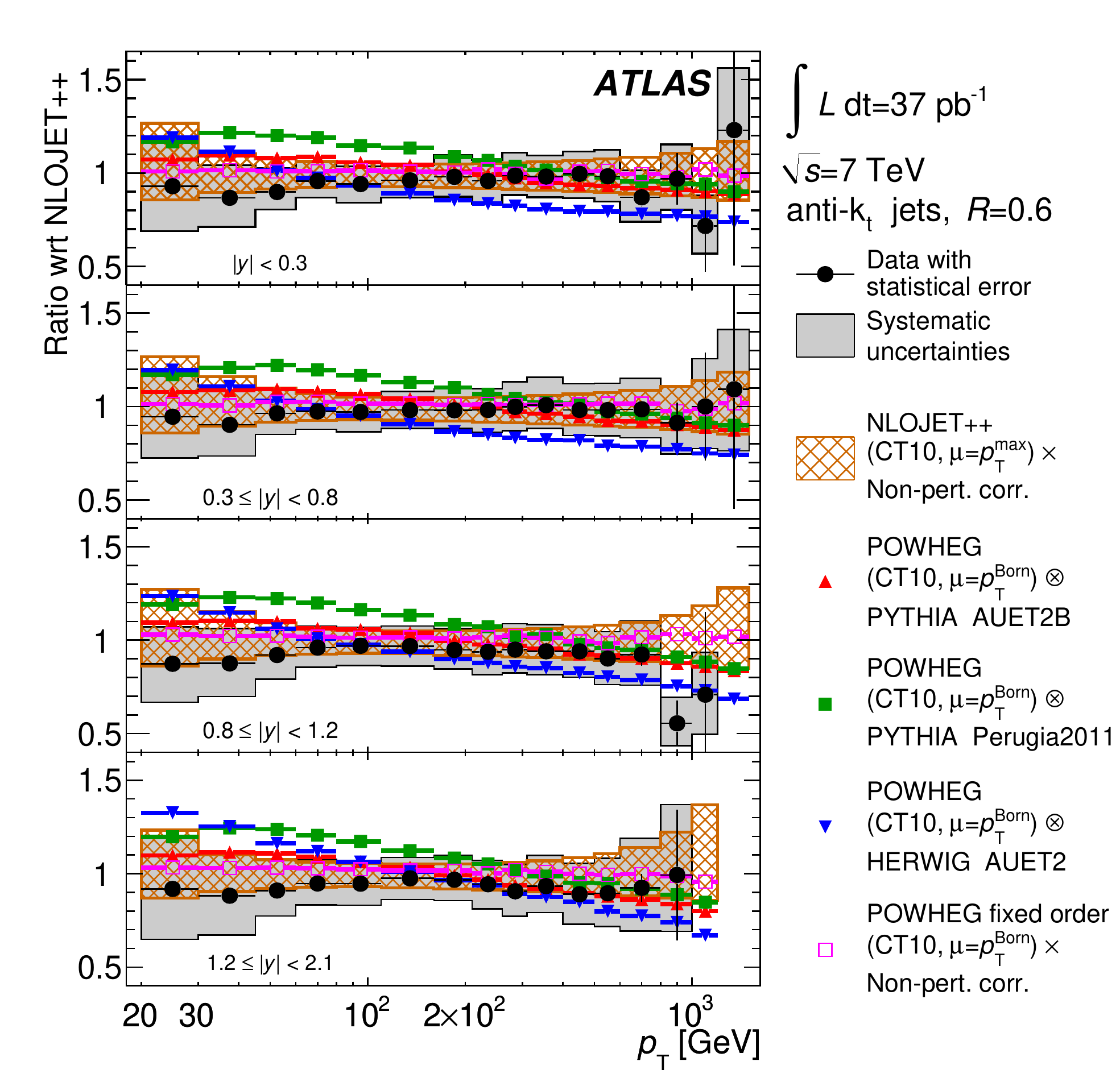}
\hspace*{-31.5mm}
\includegraphics[width=0.54\textwidth]{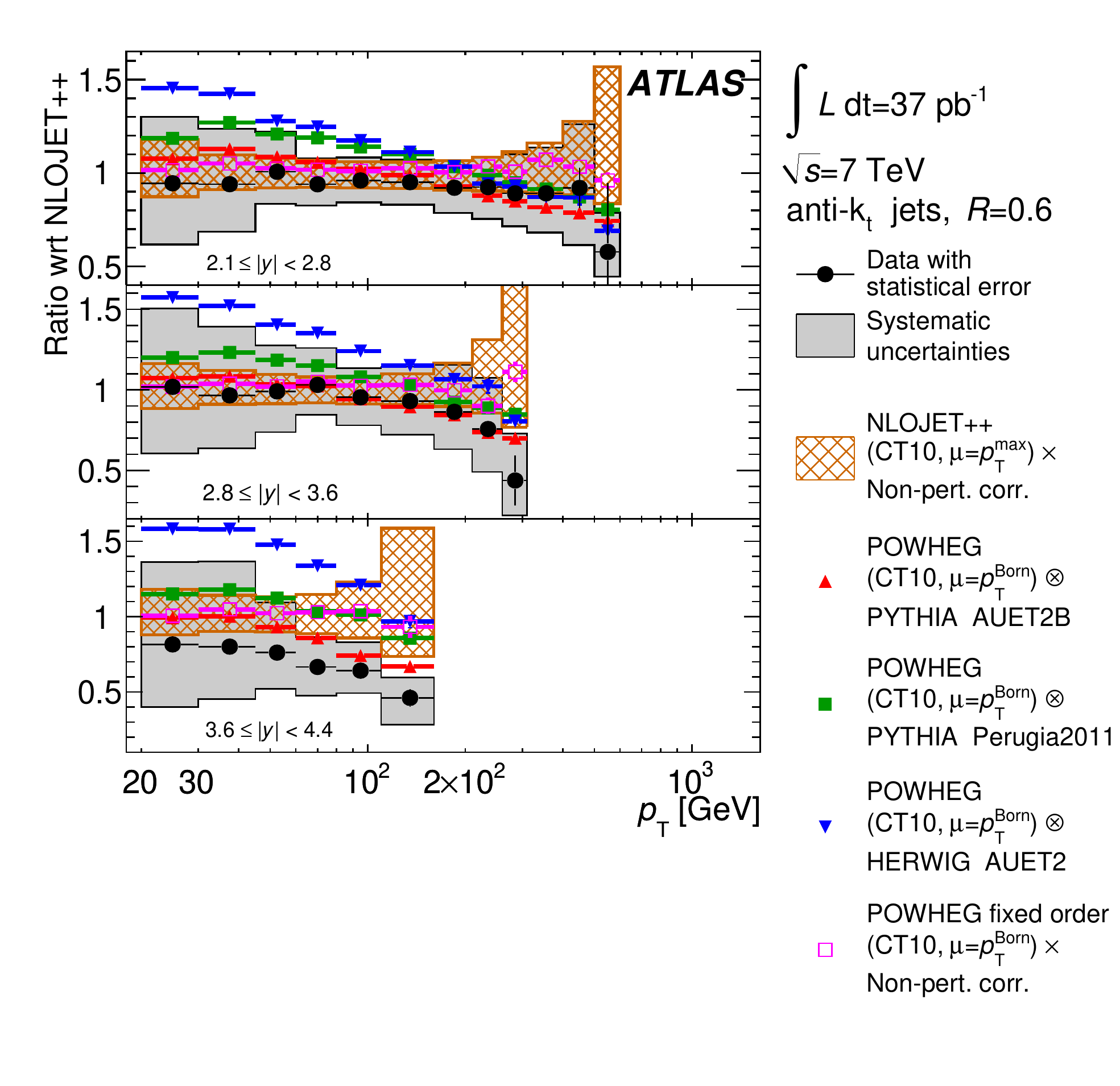}
\caption{
Ratios of inclusive jet double-differential cross section to the
theoretical prediction obtained using \nlojet with the CT10 PDF set.
The ratios are shown as a function of jet $\pt$ in different regions
of $|y|$ for jets identified using the \AKT algorithm with $R=0.4$
(upper plots) and $R=0.6$ (lower plots).  The ratios of \powheg
predictions showered using either \pythia or \herwig to the \nlojet
predictions corrected for non-perturbative effects are shown and can
be compared to the corresponding ratios for data.  Only the
statistical uncertainty on the \powheg predictions is shown.  The
total systematic uncertainties on the theory and the measurement are
indicated.  The \nlojet prediction and the \powheg ME calculations use
the CT10 PDF set. Statistically insignificant data points at large
$\pt$ are omitted in the ratio.
}
\label{fig:ratio_powhegct10}
\end{center}
\end{figure*}

In Figs.~\ref{fig:comp19nb}--\ref{fig:ratio_powhegct10}, the inclusive
jet results are presented in terms of the ratio with respect to the
\nlojet predictions using the CT10 PDF set.  Fig.~\ref{fig:comp19nb}
compares the current results to the previous measurements published by
ATLAS~\cite{Collaboration:2010wv}, for jets reconstructed with the
$\AKT$ algorithm with parameter $R=0.6$.  This figure is limited to
the central region, but similar conclusions can be drawn in all
rapidity bins.  In particular the two measurements are in good
agreement, although the new results cover a much larger kinematic
range with much reduced statistical and systematic uncertainties.

\begin{figure*}[p]
\begin{center}
\includegraphics[width=0.95\textwidth]{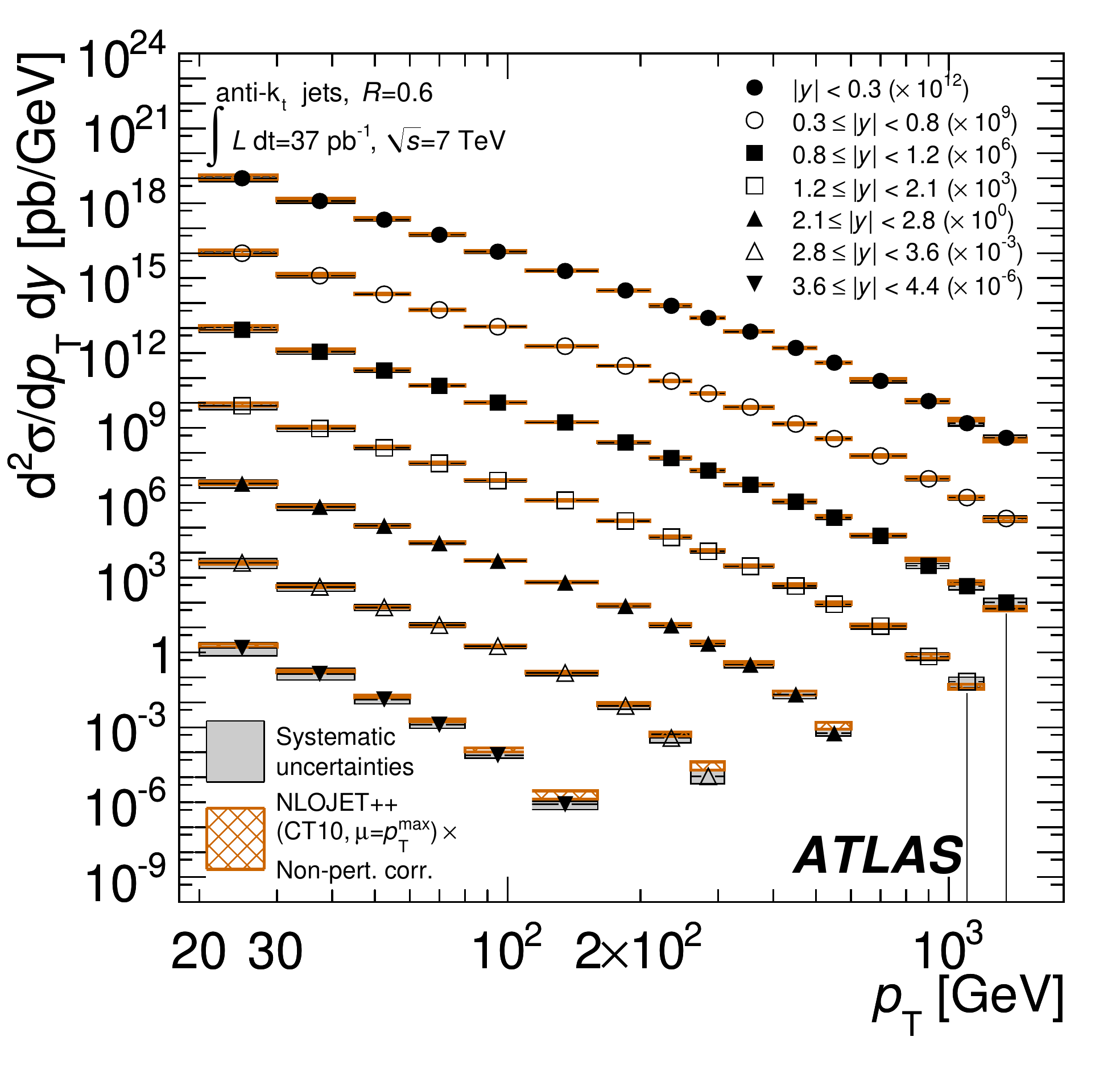}
\caption{
Inclusive jet double-differential cross section as a function of jet
$\pt$ in different regions of $|y|$ for jets identified using the \AKT
algorithm with $R=0.6$.  For convenience, the cross sections are
multiplied by the factors indicated in the legend.  The data are
compared to NLO pQCD calculations using \nlojet to which
non-perturbative corrections have been applied.  The theoretical and
experimental uncertainties indicated are calculated as described in
Fig.~\ref{fig:incjetptsummary04}.
}
\label{fig:incjetptsummary06}
\end{center}
\end{figure*}

Fig.~\ref{fig:ratio_PDF} shows the ratio of the measured cross
sections to the \nlojet theoretical predictions for various PDF
sets. Predictions obtained using CT10, MTSW~2008, NNPDF 2.1, and
HERAPDF~1.5, including uncertainty bands, are compared to the measured
cross sections, where data and theoretical predictions are normalised
to the prediction from the CT10 PDF set.  The data show a marginally
smaller cross section than the predictions from each of the PDF sets.
This trend is more pronounced for the measurements corresponding to
the \AKT algorithm with parameter $R = 0.4$, compared to $R = 0.6$.

The description becomes worse for large jet transverse momenta and
rapidities, where the MSTW 2008 PDF set follows the measured trend
better.  However, the differences between the measured cross section
and the prediction of each PDF set are of the same order as the total
systematic uncertainty on the measurement, including both experimental
and theoretical uncertainty sources.  A $\chi^2$ test of the
compatibility between data and the PDF curves, accounting for
correlations between bins, provides reasonable probabilities for all
sets, with non-significant differences between
them.\footnote{Comparisons to HERAPDF~1.0, CTEQ~6.6 and NNPDF~2.0 were
also performed, but they are not shown as they are very similar to
those for HERAPDF~1.5, CT10, and NNPDF~2.1, respectively.}

The comparison of the data with the \powheg prediction, using the
CT10~NLO PDF set, is shown for $\AKT$ jets with $R=0.4$ and $R=0.6$ in
different rapidity regions in Fig.~\ref{fig:ratio_powhegct10}.  The
data are compared with four theory curves, all of which are normalised
to the same common denominator of the \nlojet prediction corrected for
non-perturbative effects: \powheg showered with \pythia with the
default AUET2B tune; the same with the Perugia 2011 tune; \powheg
showered with \herwig{}; and \powheg run in ``pure NLO'' mode
(fixed-order calculation), without matching to parton shower, after
application of soft corrections calculated using \pythia and the
AUET2B tune.  Scale uncertainties are not shown for the \powheg
curves, but they have been found to be similar to those obtained with
\nlojet.

Good agreement at the level of a few percent is observed between NLO
fixed-order calculations based on \nlojet and \powheg, as described in
Sec.~\ref{sec:nlo_predictions}.  However, significant differences
reaching $O(30\%)$ are observed if \powheg is interfaced to different
showering and soft physics models, particularly at low $\pt$ and
forward rapidity, but also at high $\pt$.  These differences exceed
the uncertainties on the non-perturbative corrections, which are not
larger than 10\% for the inclusive jet measurements with $R=0.4$, thus
indicating a significant impact of the parton shower.  The Perugia
2011 tune tends to produce a consistently larger cross section than
the standard AUET2B tune over the full rapidity range.  The technique
of correcting fixed-order calculations for non-perturbative effects
remains the convention to define the baseline theory prediction until
NLO parton shower generators become sufficiently mature to describe
data well.  The corrected NLO result predicts a consistently larger
cross section than that seen in the data.  Good agreement in
normalisation is found between the data and the prediction from
\powheg showered with the default tune of \pythia.  These results are
confirmed by a $\chi^2$ test of the compatibility of the \powheg
results with the data, where the curve obtained using the \herwig
shower results in a much worse $\chi^2$ after all error correlations
have been accounted for.

\clearpage

\subsection{Dijet Cross Sections}
\label{sec:dijet}
\begin{figure*}[tb]
\begin{center}
\includegraphics[width=0.95\textwidth]{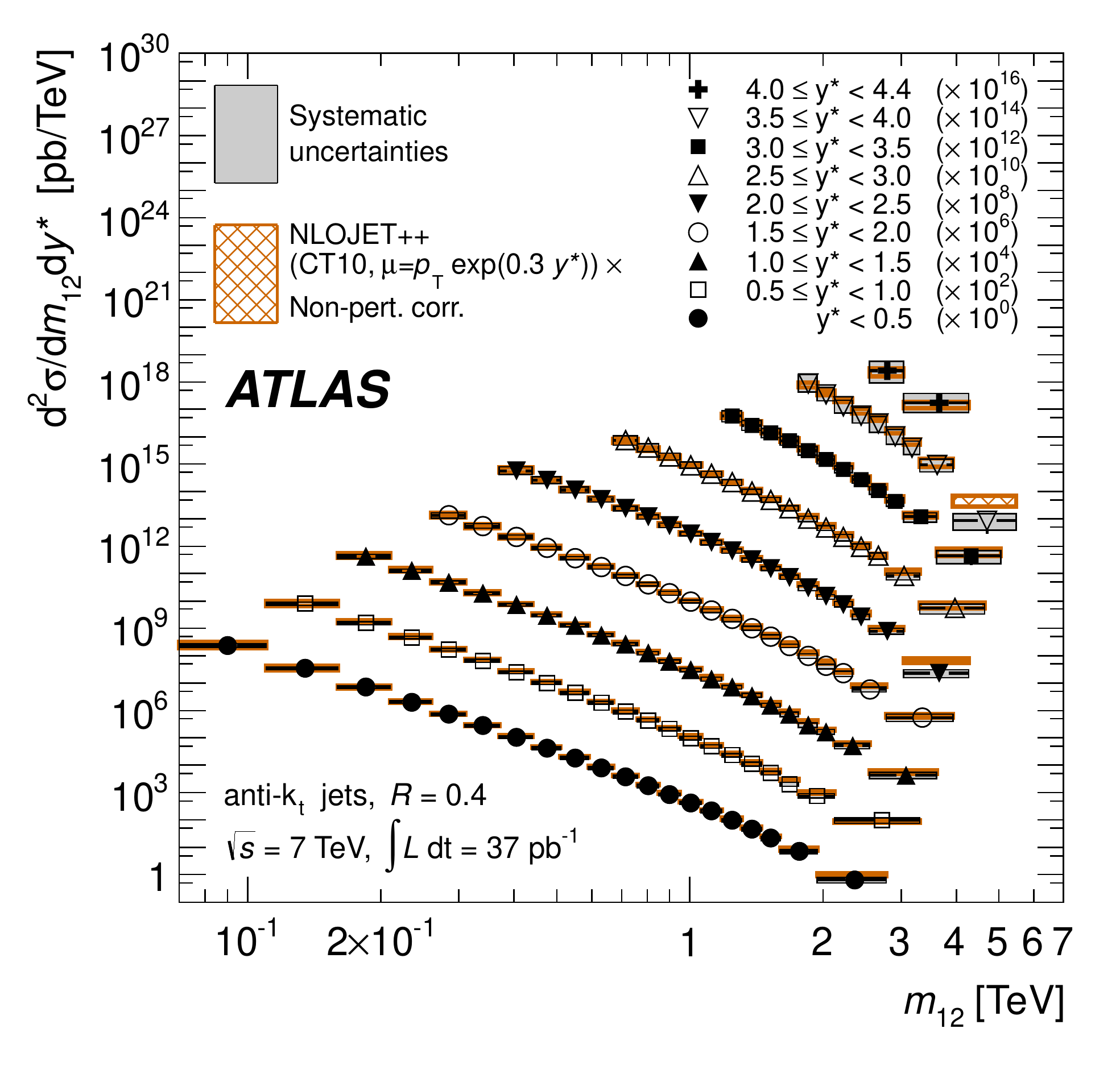}
\caption{
Dijet double-differential cross section as a function of dijet mass,
binned in half the rapidity separation between the two leading jets,
$y^* = |y_1 - y_2|/2$.  The results are shown for jets identified
using the \AKT algorithm with $R=0.4$.  For convenience, the cross
sections are multiplied by the factors indicated in the legend.  The
data are compared to NLO pQCD calculations using \nlojet to which
non-perturbative corrections have been applied.  The error bars, which
are usually smaller than the symbols, indicate the statistical
uncertainty on the measurement. The dark-shaded band indicates the
quadratic sum of the experimental systematic uncertainties, dominated
by the jet energy scale uncertainty. There is an additional overall
uncertainty of 3.4\% due to the luminosity measurement that is not
shown. The theory uncertainty, shown as the light, hatched band, is
the quadratic sum of uncertainties from the choice of the
renormalisation and factorisation scales, parton distribution
functions, $\alpha_s(M_Z)$, and the modeling of non-perturbative
effects, as described in the text.
}
\label{fig:MassSummary04}
\end{center}
\end{figure*}

The dijet double-differential cross section has been measured as a
function of the dijet invariant mass for various bins of the variable
$y^*$, which is the rapidity in the two-parton centre-of-mass frame.
The quantity $y^{*}$ is calculated as half the absolute value of the
rapidity difference of the two leading jets, ranging from 0 to 4.4.
The results are shown in Figs.~\ref{fig:MassSummary04}
and~\ref{fig:MassSummary06} and
Tables~\ref{tab:DijetMassResults04_0}--\ref{tab:DijetMassResults06_8}
in Appendix~\ref{sec:tablesDijet} for $\AKT$ jets with $R=0.4$ and
$R=0.6$.  The cross section measurements extend from dijet masses of
70~GeV to almost 5~TeV, covering two orders of magnitude in invariant
mass and nine orders of magnitude in the cross section.  The dijet
measurements are fully corrected for detector effects and are compared
to \nlojet predictions calculated using the scale defined in
Eq.~\ref{eq:dijetscale} (see Sec.~\ref{sec:nlo_predictions}) and the
CT10 PDF set, with non-perturbative corrections applied to the theory
prediction.  The theoretical uncertainties have been assessed as
described for the inclusive jet measurements in
Sec.~\ref{sec:inclusive}.

\begin{figure*}[tb]
\begin{minipage}{0.54\linewidth}
\includegraphics[width=1\textwidth]{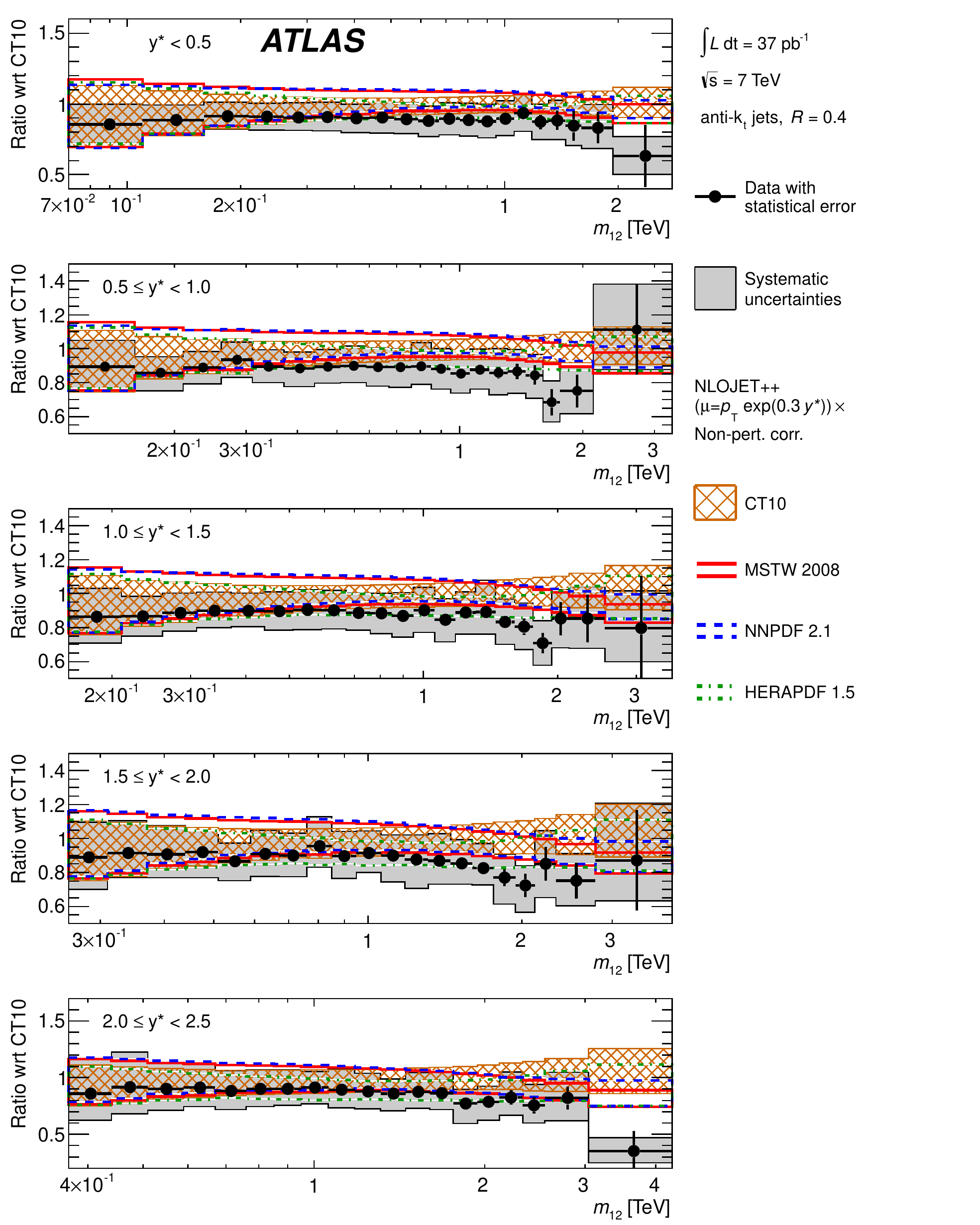}
\end{minipage}
\hspace*{-29.5mm}\begin{minipage}{0.54\linewidth}\vspace*{-24mm}
\includegraphics[width=1\textwidth]{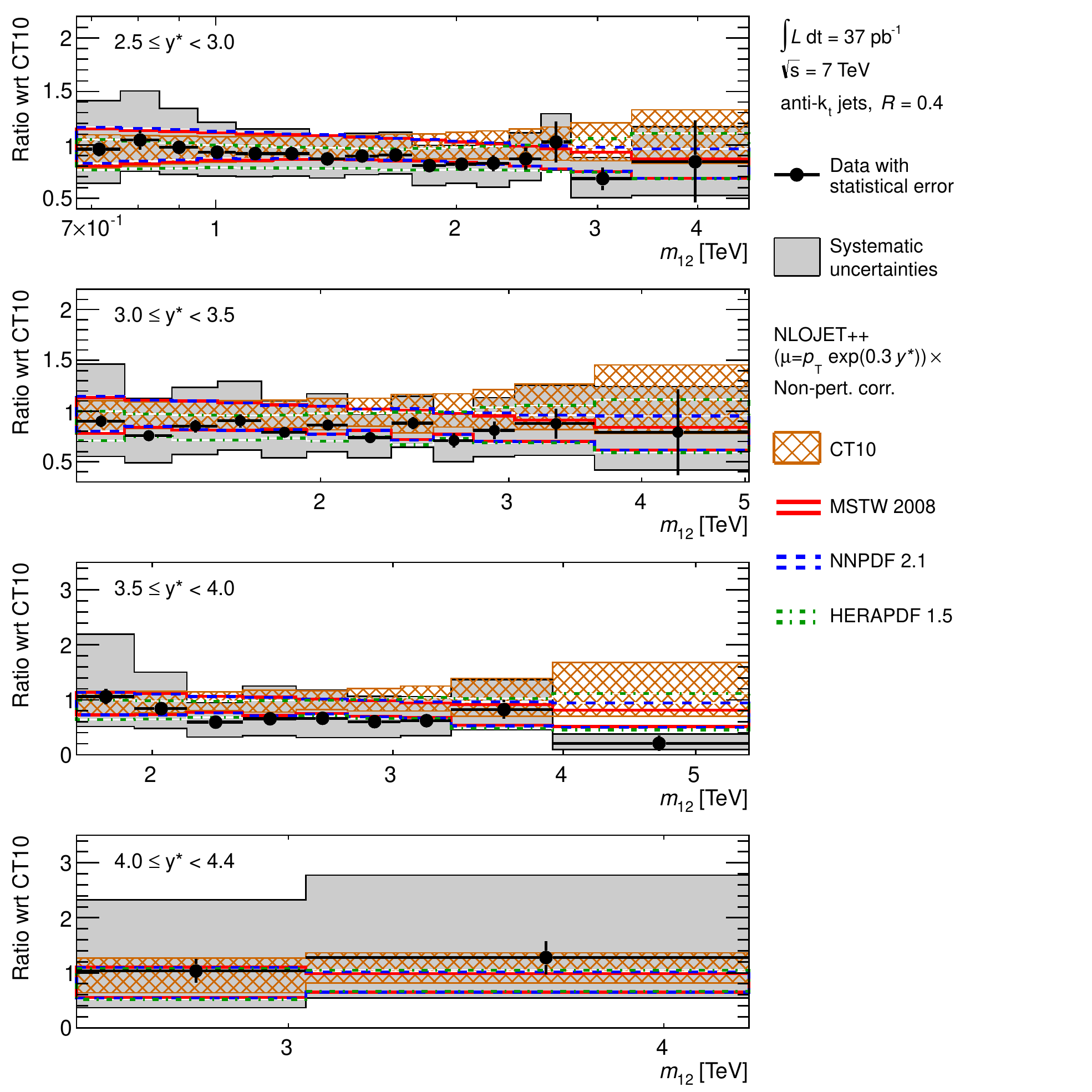}
\end{minipage}\\
\caption{
Ratios of dijet double-differential cross section to the theoretical
prediction obtained using \nlojet with the CT10 PDF set.  The ratios
are shown as a function of dijet mass, binned in half the rapidity
separation between the two leading jets, $y^* = |y_1 - y_2|/2$.  The
results are shown for jets identified using the \AKT algorithm with
$R=0.4$.  The theoretical error bands obtained by using \nlojet with
different PDF sets (CT10, MSTW 2008, NNPDF 2.1, HERAPDF 1.5) are
shown.  The systematic and theoretical uncertainties are calculated as
described in Fig.~\ref{fig:MassSummary04}.
}
\label{fig:dijetMassPDF_04}
\end{figure*}

\begin{figure*}[tb]
\begin{minipage}{0.54\linewidth}
\includegraphics[width=1\textwidth]{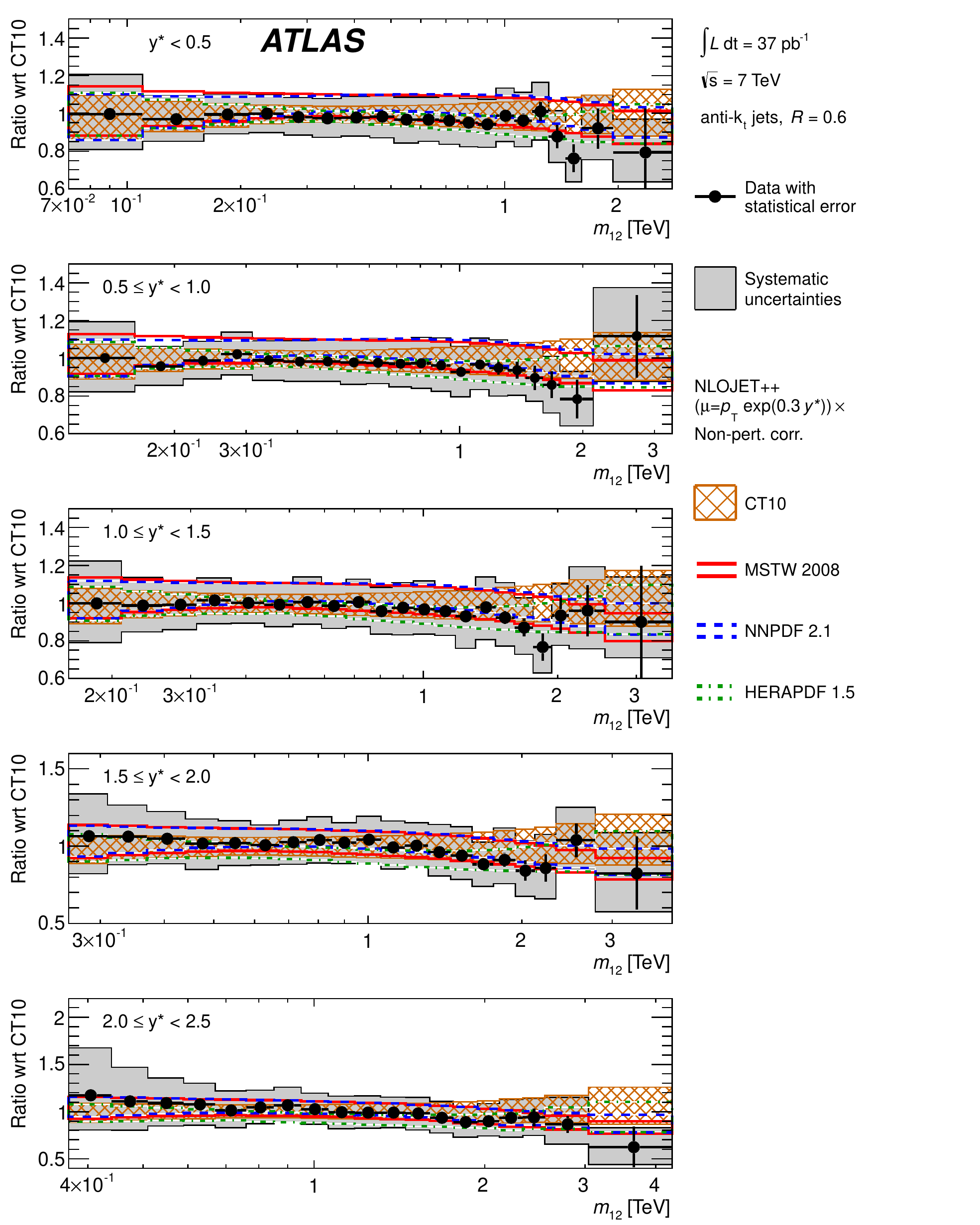}
\end{minipage}
\hspace*{-29.5mm}\begin{minipage}{0.54\linewidth}\vspace*{-24mm}
\includegraphics[width=1\textwidth]{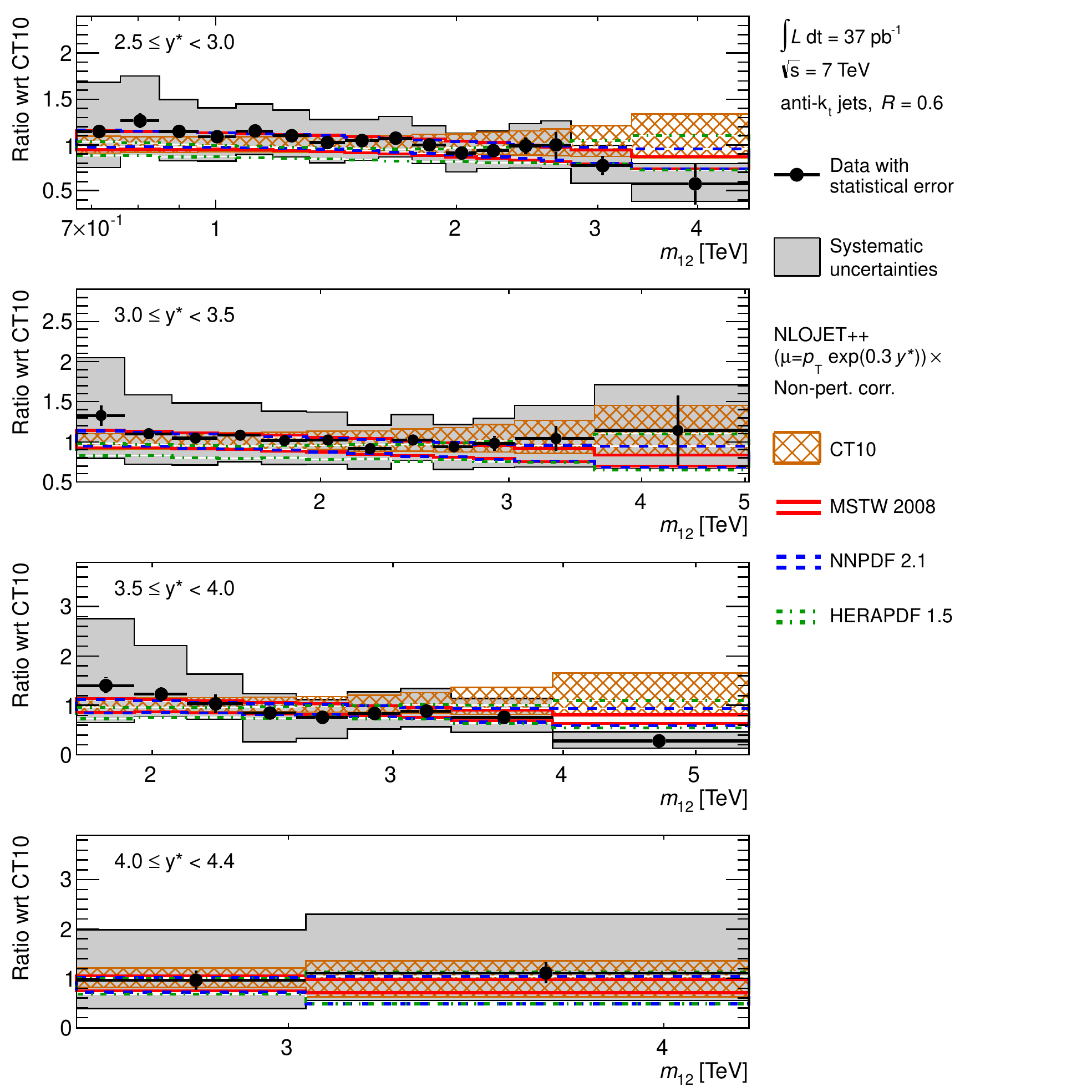}
\end{minipage}\\
\caption{
Ratios of dijet double-differential cross section to the theoretical
prediction obtained using \nlojet with the CT10 PDF set.  The ratios
are shown as a function of dijet mass, binned in half the rapidity
separation between the two leading jets, $y^* = |y_1 - y_2|/2$.  The
results are shown for jets identified using the \AKT algorithm with
$R=0.6$.  The theoretical error bands obtained by using \nlojet with
different PDF sets (CT10, MSTW 2008, NNPDF 2.1, HERAPDF 1.5) are
shown.  The systematic and theoretical uncertainties are calculated as
described in Fig.~\ref{fig:MassSummary04}.
}
\label{fig:dijetMassPDF_06}
\end{figure*}

The dijet data are also compared with \nlojet predictions obtained
using the MSTW~2008, NNPDF~2.1, and HERAPDF~1.5 PDF sets.
Figs.~\ref{fig:dijetMassPDF_04} and \ref{fig:dijetMassPDF_06} show the
dijet mass spectra for $\AKT$ jets with $R=0.4$ and $R=0.6$
respectively, where both the data and the predictions from the
above-mentioned PDF sets have been normalised to the CT10 prediction.
The data for $R=0.6$ exhibit a slight falling slope with respect to
the CT10 prediction and appear to be described better by other PDF
sets, a similar behaviour to that observed in the inclusive jet data.
However, in all cases, the differences between the data and each PDF
set lie well within the systematic and theory uncertainties,
indicating a reasonable agreement with the dijet data, particularly in
the kinematic region at low $y^{*}$.

\begin{figure*}[tb]
\begin{minipage}{0.54\linewidth}
\includegraphics[width=1\textwidth]{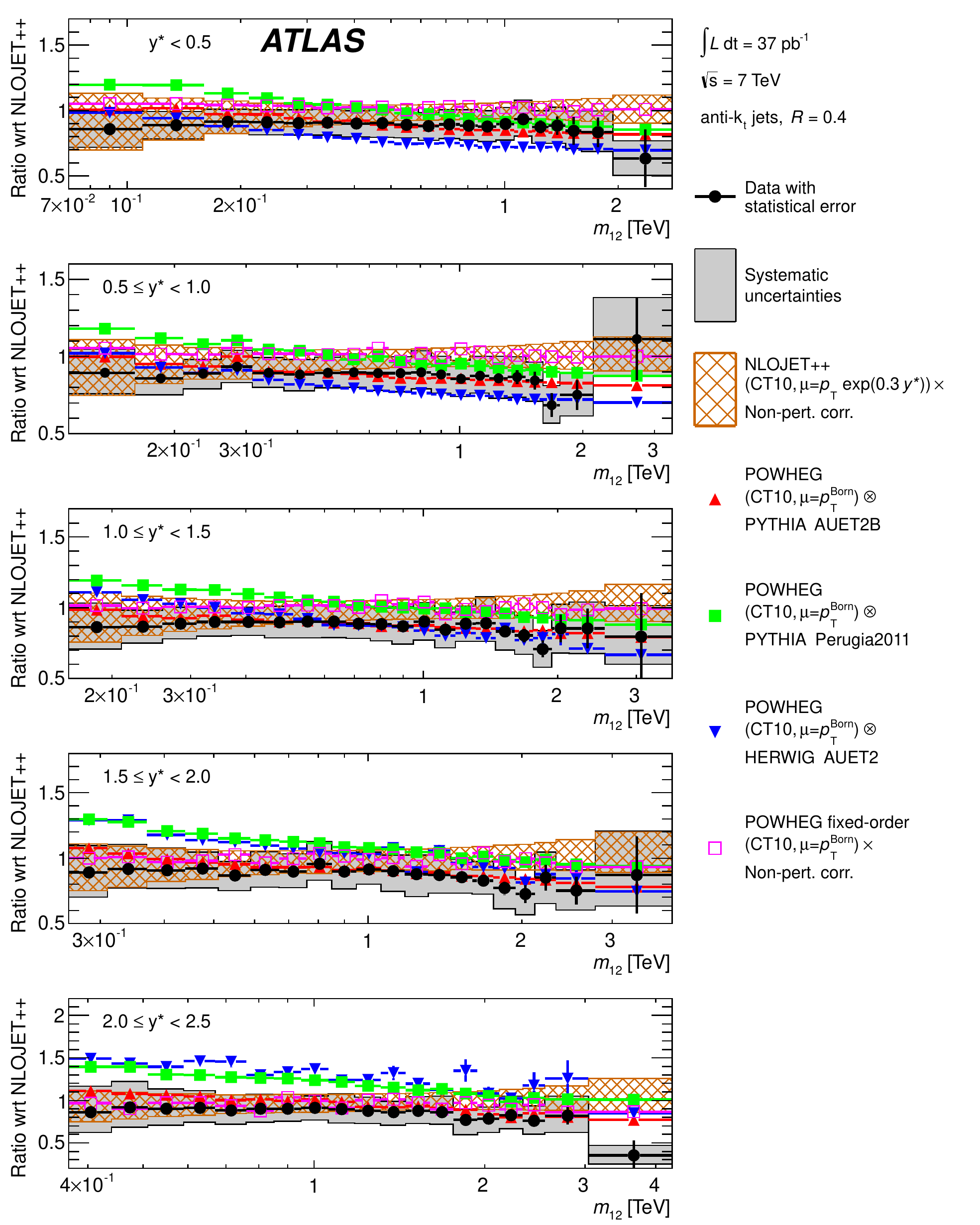}
\end{minipage}
\hspace*{-29.5mm}\begin{minipage}{0.54\linewidth}\vspace*{-24mm}
\includegraphics[width=1\textwidth]{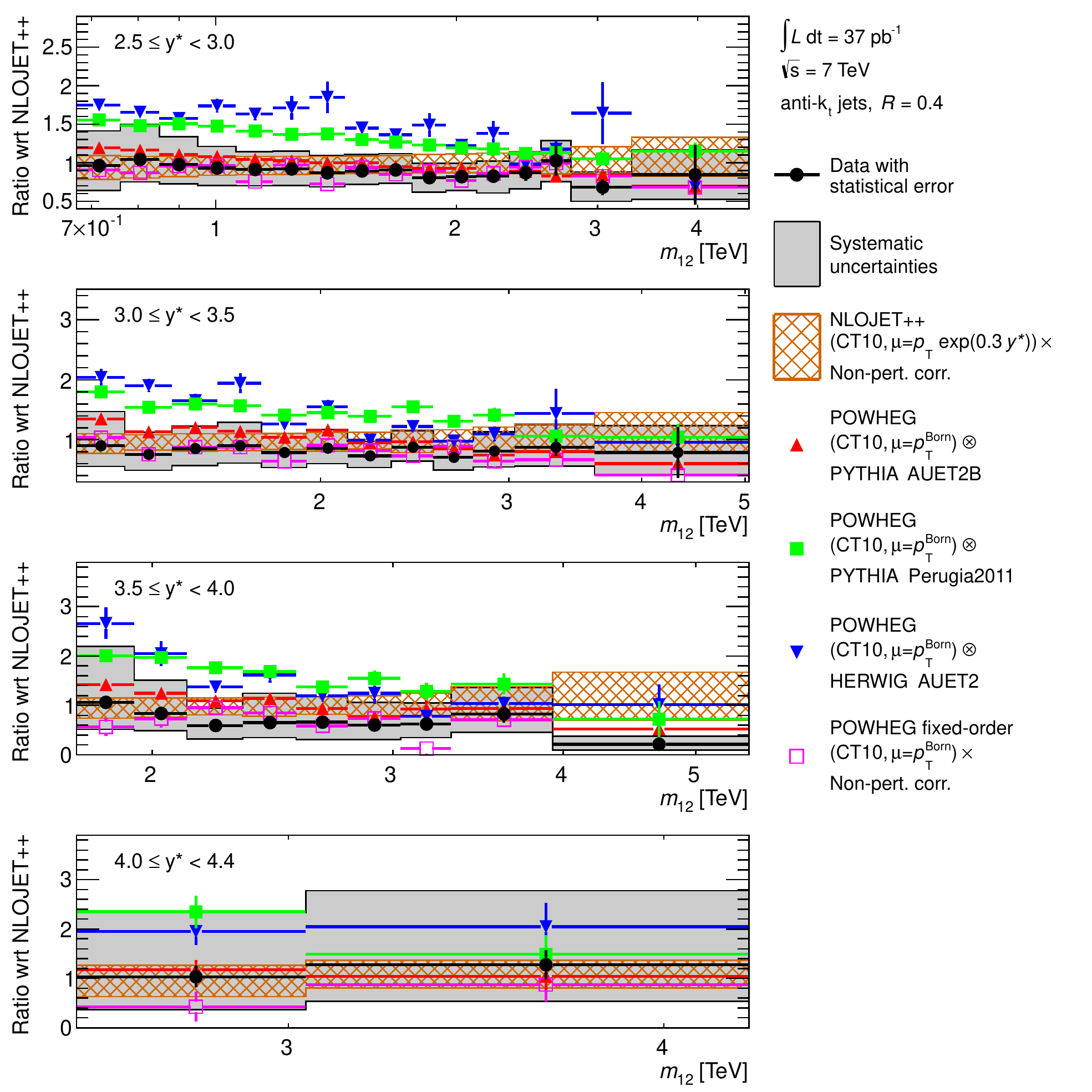}\\
\end{minipage}
\caption{
Ratios of dijet double-differential cross section to the theoretical
prediction obtained using \nlojet with the CT10 PDF set.  The ratios
are shown as a function of dijet mass, binned in half the rapidity
separation between the two leading jets, $y^* = |y_1 - y_2|/2$.  The
results are shown for jets identified using the \AKT algorithm with
$R=0.4$.  The ratios of \powheg predictions showered using either
\pythia or \herwig to the \nlojet predictions corrected for
non-perturbative effects are shown and can be compared to the
corresponding ratios for data.  Only the statistical uncertainty on
the \powheg predictions is shown.  The total systematic uncertainties
on the theory and the measurement are indicated.  The \nlojet
prediction and the \powheg ME calculations use the CT10 PDF set.
}
\label{fig:dijetMassPowheg_04}
\end{figure*}

\begin{figure*}[tb]
\begin{minipage}{0.54\linewidth}
\includegraphics[width=1\textwidth]{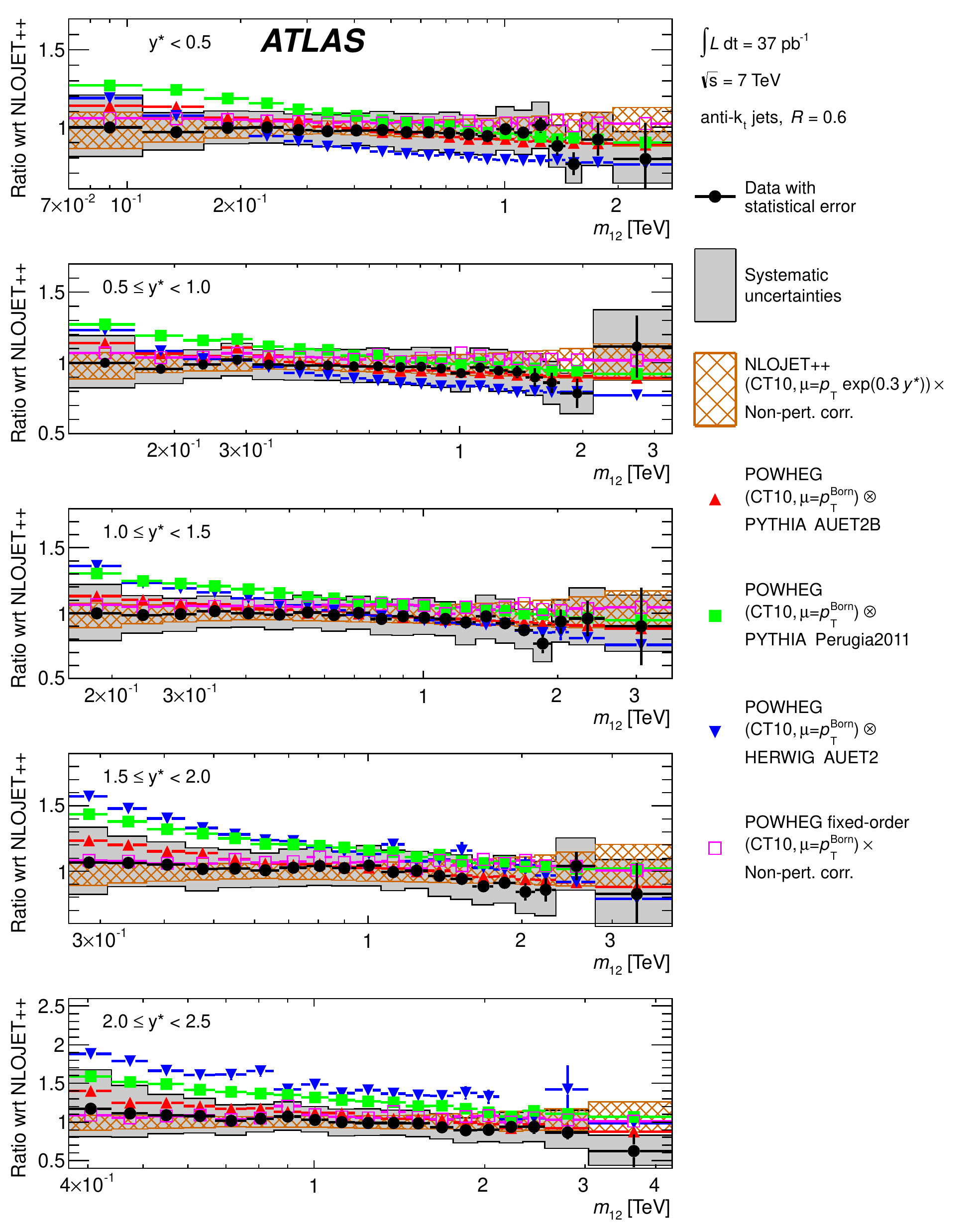}
\end{minipage}
\hspace*{-29.5mm}\begin{minipage}{0.54\linewidth}\vspace*{-24mm}
\includegraphics[width=1\textwidth]{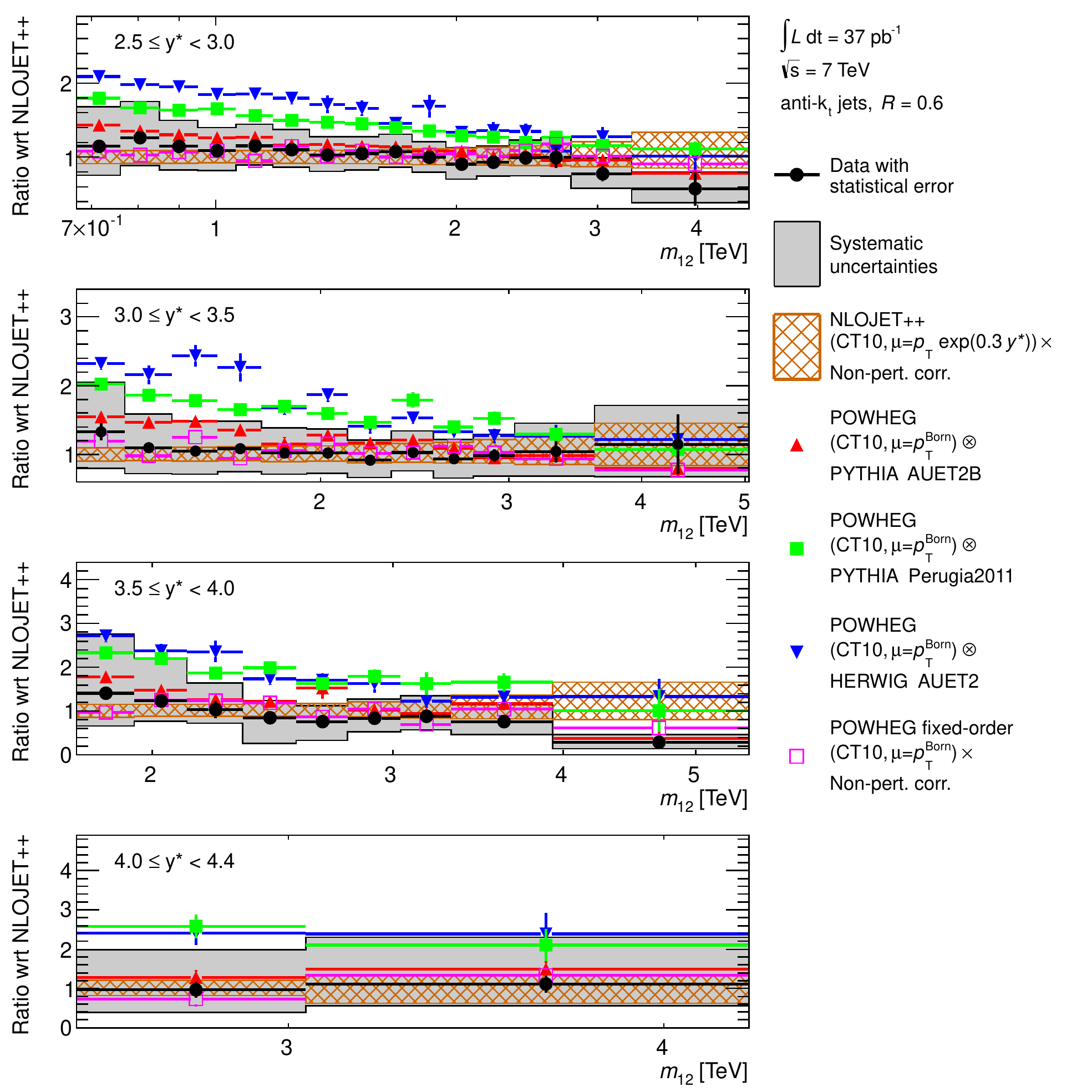}
\end{minipage}
\caption{
Ratios of dijet double-differential cross section to the theoretical
prediction obtained using \nlojet with the CT10 PDF set.  The ratios
are shown as a function of dijet mass, binned in half the rapidity
separation between the two leading jets, $y^* = |y_1 - y_2|/2$.  The
results are shown for jets identified using the \AKT algorithm with
$R=0.6$.  The ratios of \powheg predictions showered using either
\pythia or \herwig to the \nlojet predictions corrected for
non-perturbative effects are shown and can be compared to the
corresponding ratios for data.  Only the statistical uncertainty on
the \powheg predictions is shown.  The total systematic uncertainties
on the theory and the measurement are indicated.  The \nlojet
prediction and the \powheg ME calculations use the CT10 PDF set.
}
\label{fig:dijetMassPowheg_06}
\end{figure*}

The data are also compared with \powheg predictions produced using the
CT10 PDF set and showered with different tunes of the \pythia or
\herwig generator.  These comparisons are shown for $R=0.4$ and
$R=0.6$ respectively in Figs.~\ref{fig:dijetMassPowheg_04}
and~\ref{fig:dijetMassPowheg_06}, where the data and all theory
predictions have been normalised to the \nlojet prediction with CT10.
The \nlojet prediction has been corrected for non-perturbative effects
calculated using the \pythia MC with the AUET2B tune.  The \powheg
predictions shown are interfaced to the \pythia parton shower with the
AUET2B or Perugia2011 tune, and to the \herwig parton shower using the
AUET2 tune.  The data are also compared to the \powheg fixed-order NLO
prediction (corrected for non-perturbative effects), where the POWHEG
prediction has been calculated using a scale choice of $\mu_{\rm R} =
\mu_{\rm F} = \pt^{\rm Born}$.

The data are in best agreement with the \powheg prediction showered
with \pythia using the AUET2B tune.  The other \powheg showered
predictions exhibit discrepancies at low dijet mass in all $y^{*}$
slices, where they predict larger cross sections than are observed in
the data.

\section{Conclusions}
Cross section measurements have been presented for inclusive jets and
dijets reconstructed with the $\AKT$ algorithm using two values of the
clustering parameter ($R=0.4$ and $R=0.6$).  Inclusive jet production
has been measured as a function of jet transverse momentum, in bins of
jet rapidity.  Dijet production has been measured as a function of the
invariant mass of the two leading jets, in bins of half their rapidity
difference.  These results are based on the data sample collected with
the ATLAS detector during 2010, which corresponds to ($37.3 \pm
1.2$)~pb$^{-1}$ of integrated luminosity.

Two different sizes of the jet clustering parameter have been used in
order to probe the relative effects of the parton shower,
hadronisation, and the underlying event.  The measurements have been
corrected for all detector effects to the particle level so that they
can be compared to any theoretical calculation.  In this paper, they
have been compared to fixed-order NLO pQCD calculations corrected for
non-perturbative effects, as well as to parton shower Monte Carlo
simulations with NLO matrix elements.  The latter predictions have
only recently become available for inclusive jet and dijet production.

The current results reflect a number of significant experimental
accomplishments:
\begin{itemize}
\item The cross section measurements extend to 1.5~TeV in jet
transverse momentum and 5~TeV in dijet invariant mass, the highest
ever measured.  These results probe NLO pQCD in a large, new kinematic
regime.
\item Using data taken with minimum bias and forward jet triggers,
these measurements extend to both the low-$\pt$ region (down to jet
transverse momentum of 20~GeV and dijet invariant mass of 70~GeV) and
to the forward region (out to rapidities of $|y|=4.4$).  The forward
region, in particular, has never been explored before with such
precision at a hadron-hadron collider.
\item High-precision measurements of the data collected during LHC
beam position scans have determined the uncertainty on the collected
luminosity to 3.4\%.
\item Detailed understanding of the detector performance has precisely
determined systematic uncertainties, in particular those arising from
the jet energy scale.  In the central region ($|\eta| < 0.8$) the JES
uncertainty is lower than 4.6\% for all jets with $\pt{} > 20$~GeV,
while for jet transverse momenta between 60 and 800 GeV the JES
uncertainty is below 2.5\%.
\item The correlations of the cross section measurement across various 
$\pt$, $m_{12}$, and rapidity bins have been studied for 22
independent sources of systematic uncertainty.  These have been
provided in the form of 87 nuisance parameters, each of which is
fully correlated in $\pt$ and $y$ (dijet mass and $y^{*}$), for use
in PDF fits.
\end{itemize}
The experimental uncertainties achieved are similar in size to the
theoretical uncertainties in some regions of phase space, thereby
providing some sensitivity to different theoretical predictions.

The measurements are compared to fixed-order NLO pQCD calculations, as
well as to new calculations in which NLO pQCD matrix elements are
matched to leading-logarithmic parton showers. Overall, both sets of
calculations agree with the data over many orders of magnitude,
although the cross sections predicted by the theory tend to be larger
than the measured values at large jet transverse momentum and dijet
invariant mass. The matched NLO parton shower calculations predict
significant effects of the parton shower in some regions of phase
space, in some cases improving and in others degrading the agreement
with data with respect to the fixed-order calculations.

These measurements probe and may constrain the largely unexplored area
of parton distribution functions at large $x$ and high momentum
transfer.  The results reported here constitute a comprehensive test
of QCD across a large kinematic regime.

\section{Acknowledgements}

We thank S. Ellis, M. Mangano, D. Soper, and R. Thorne for useful
discussions regarding the scales used for the dijet theory
predictions; the \powheg authors for assistance with the \powheg
predictions; and P. Skands for advice regarding the non-perturbative
corrections.

We thank CERN for the very successful operation of the LHC, as well as
the support staff from our institutions without whom ATLAS could not
be operated efficiently.

We acknowledge the support of ANPCyT, Argentina; YerPhI, Armenia; ARC,
Australia; BMWF, Austria; ANAS, Azerbaijan; SSTC, Belarus; CNPq and FAPESP,
Brazil; NSERC, NRC and CFI, Canada; CERN; CONICYT, Chile; CAS, MOST and
NSFC, China; COLCIENCIAS, Colombia; MSMT CR, MPO CR and VSC CR, Czech
Republic; DNRF, DNSRC and Lundbeck Foundation, Denmark; ARTEMIS, European
Union; IN2P3-CNRS, CEA-DSM/IRFU, France; GNAS, Georgia; BMBF, DFG, HGF, MPG
and AvH Foundation, Germany; GSRT, Greece; ISF, MINERVA, GIF, DIP and
Benoziyo Center, Israel; INFN, Italy; MEXT and JSPS, Japan; CNRST, Morocco;
FOM and NWO, Netherlands; RCN, Norway; MNiSW, Poland; GRICES and FCT,
Portugal; MERYS (MECTS), Romania; MES of Russia and ROSATOM, Russian
Federation; JINR; MSTD, Serbia; MSSR, Slovakia; ARRS and MVZT, Slovenia;
DST/NRF, South Africa; MICINN, Spain; SRC and Wallenberg Foundation,
Sweden; SER, SNSF and Cantons of Bern and Geneva, Switzerland; NSC, Taiwan;
TAEK, Turkey; STFC, the Royal Society and Leverhulme Trust, United Kingdom;
DOE and NSF, United States of America.

The crucial computing support from all WLCG partners is acknowledged
gratefully, in particular from CERN and the ATLAS Tier-1 facilities at
TRIUMF (Canada), NDGF (Denmark, Norway, Sweden), CC-IN2P3 (France),
KIT/GridKA (Germany), INFN-CNAF (Italy), NL-T1 (Netherlands), PIC (Spain),
ASGC (Taiwan), RAL (UK) and BNL (USA) and in the Tier-2 facilities
worldwide.

\begin{figure*}[tb]
\begin{center}
\includegraphics[width=0.95\textwidth]{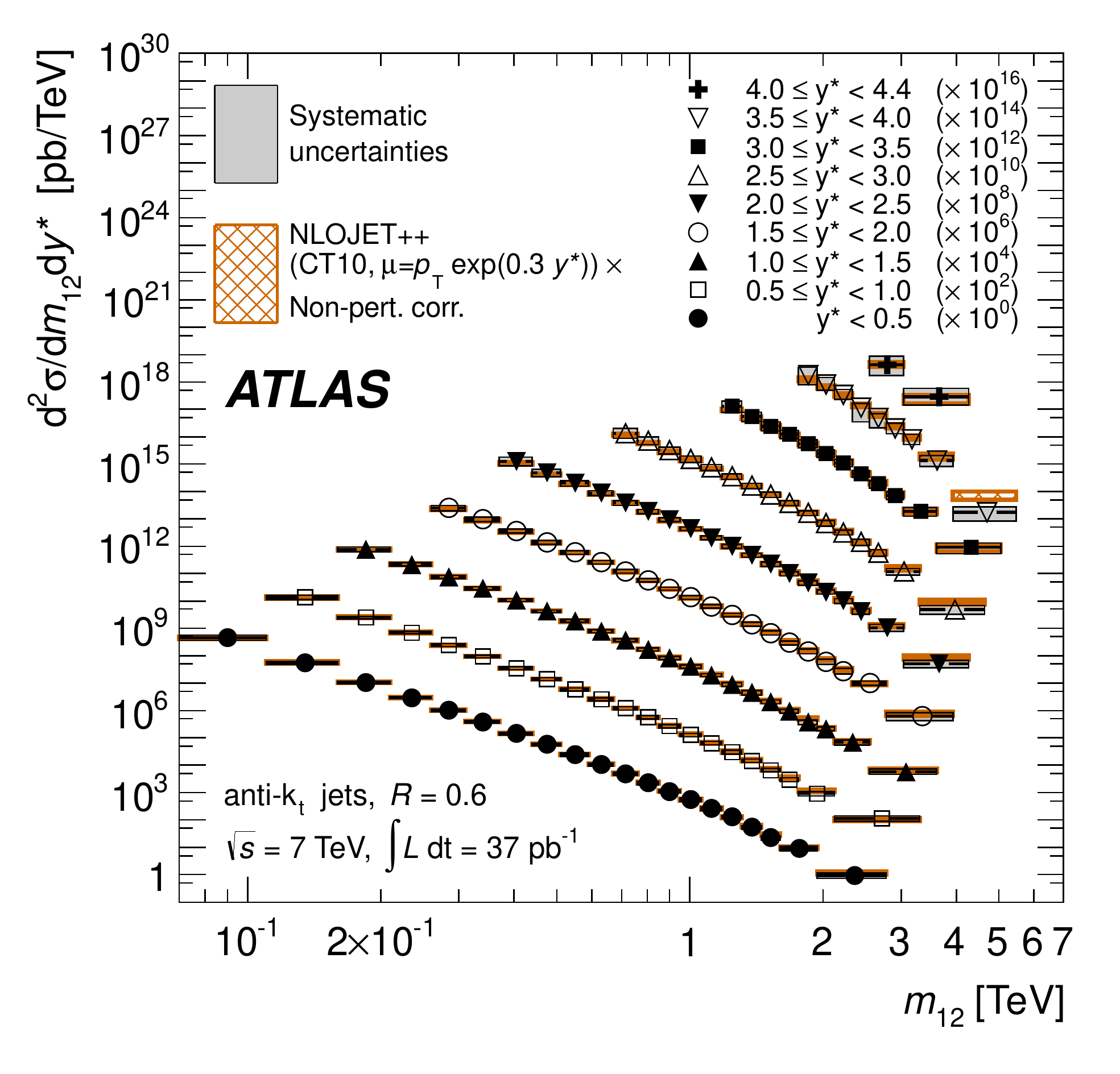}
\caption{
Dijet double-differential cross section as a function of dijet mass,
binned in half the rapidity separation between the two leading jets,
$y^* = |y_1 - y_2|/2$.  The results are shown for jets identified
using the \AKT algorithm with $R=0.6$.  For convenience, the cross
sections are multiplied by the factors indicated in the legend.  The
data are compared to NLO pQCD calculations using \nlojet to which
non-perturbative corrections have been applied.  The theoretical and
experimental uncertainties indicated are calculated as described in
Fig.~\ref{fig:MassSummary04}.
}
\label{fig:MassSummary06}
\end{center}
\end{figure*}

\bibliography{JetPaper}

\appendix

\vspace{-10cm}
\begin{figure*}[!ht]
\section{Non-perturbative corrections}
\label{sec:nonPert}
  \centering
  \includegraphics[width=.4\textwidth]{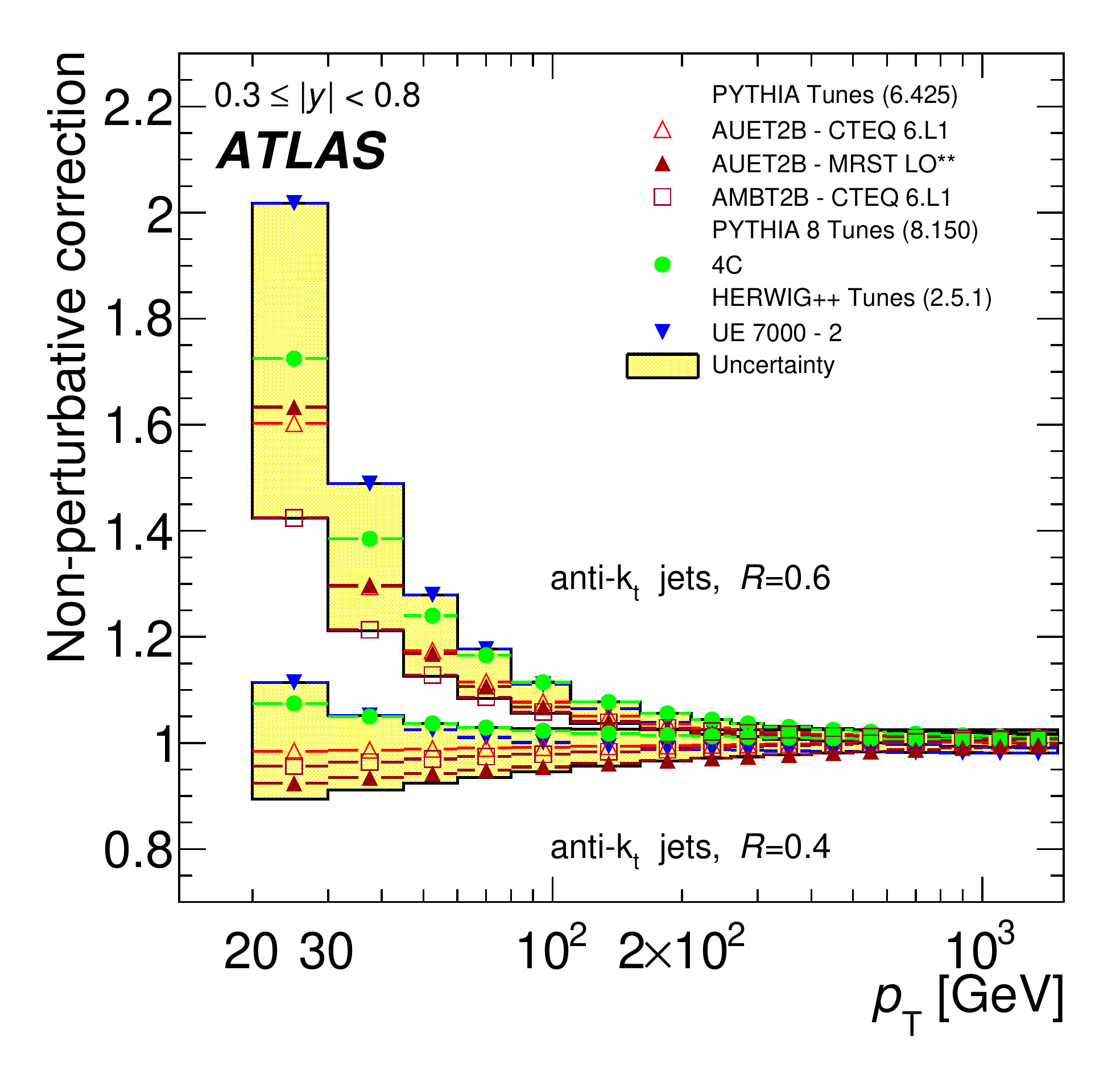}
  \includegraphics[width=.4\textwidth]{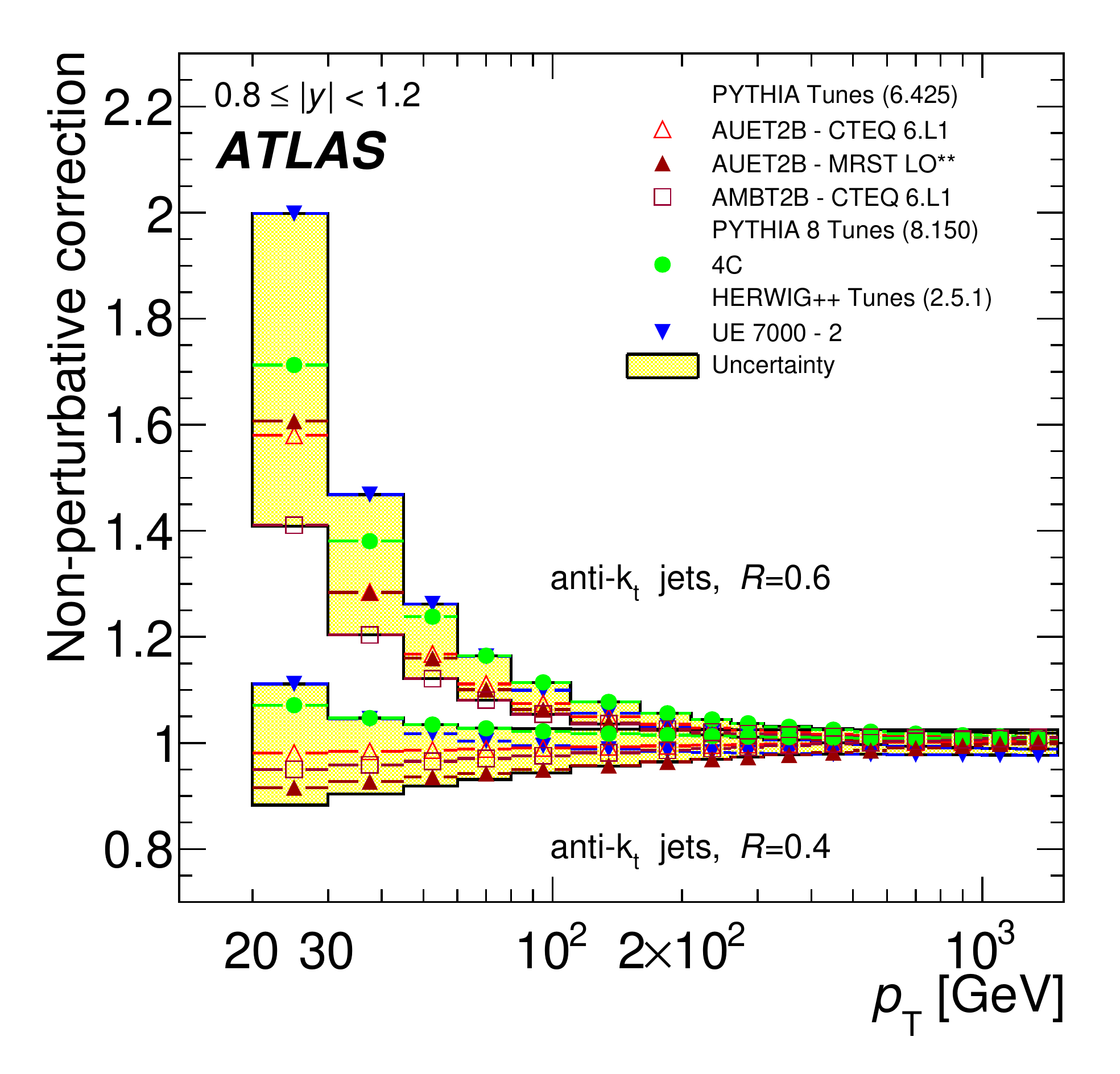}\\
  \includegraphics[width=.4\textwidth]{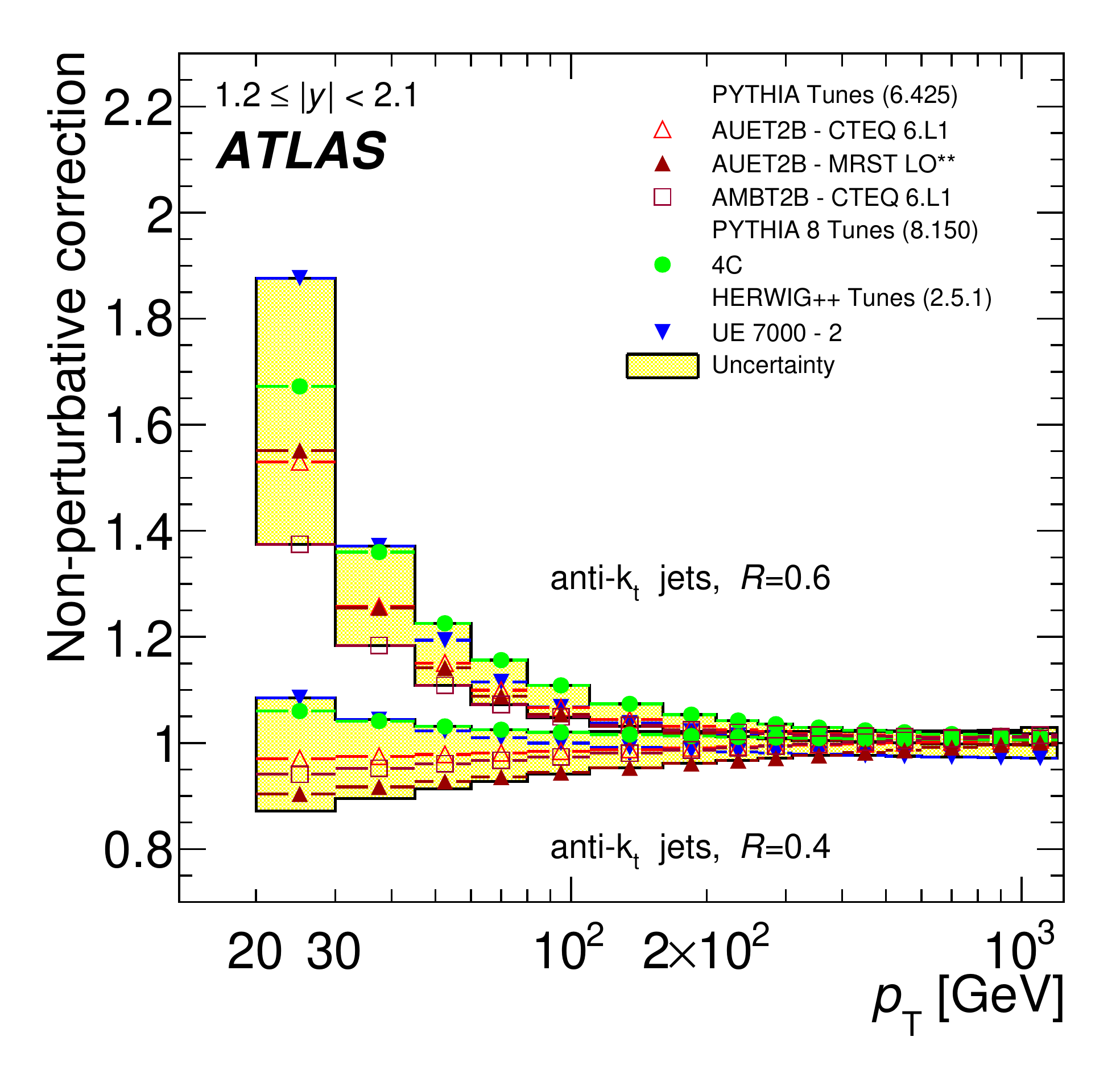}
  \includegraphics[width=.4\textwidth]{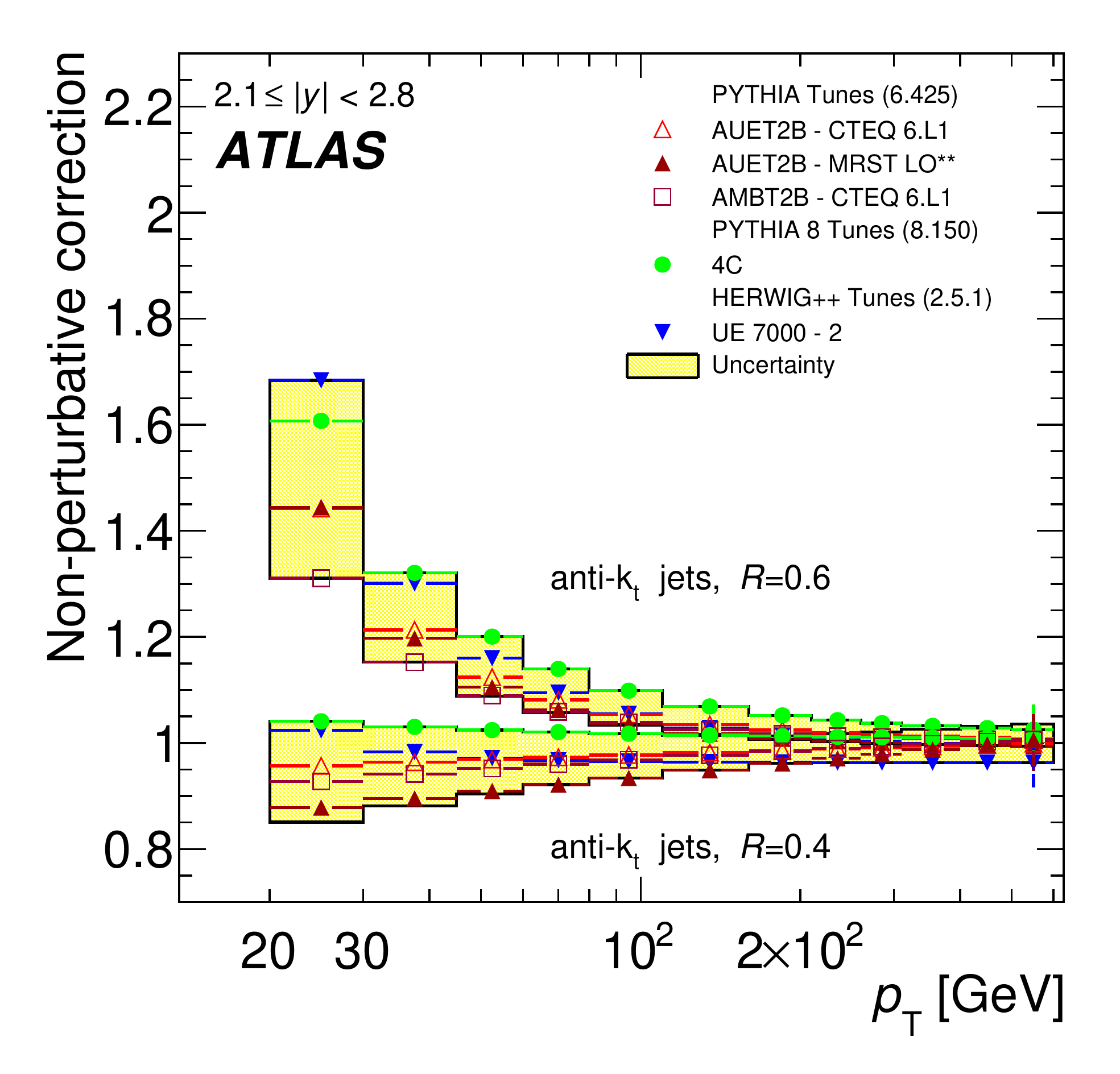}\\
  \includegraphics[width=.4\textwidth]{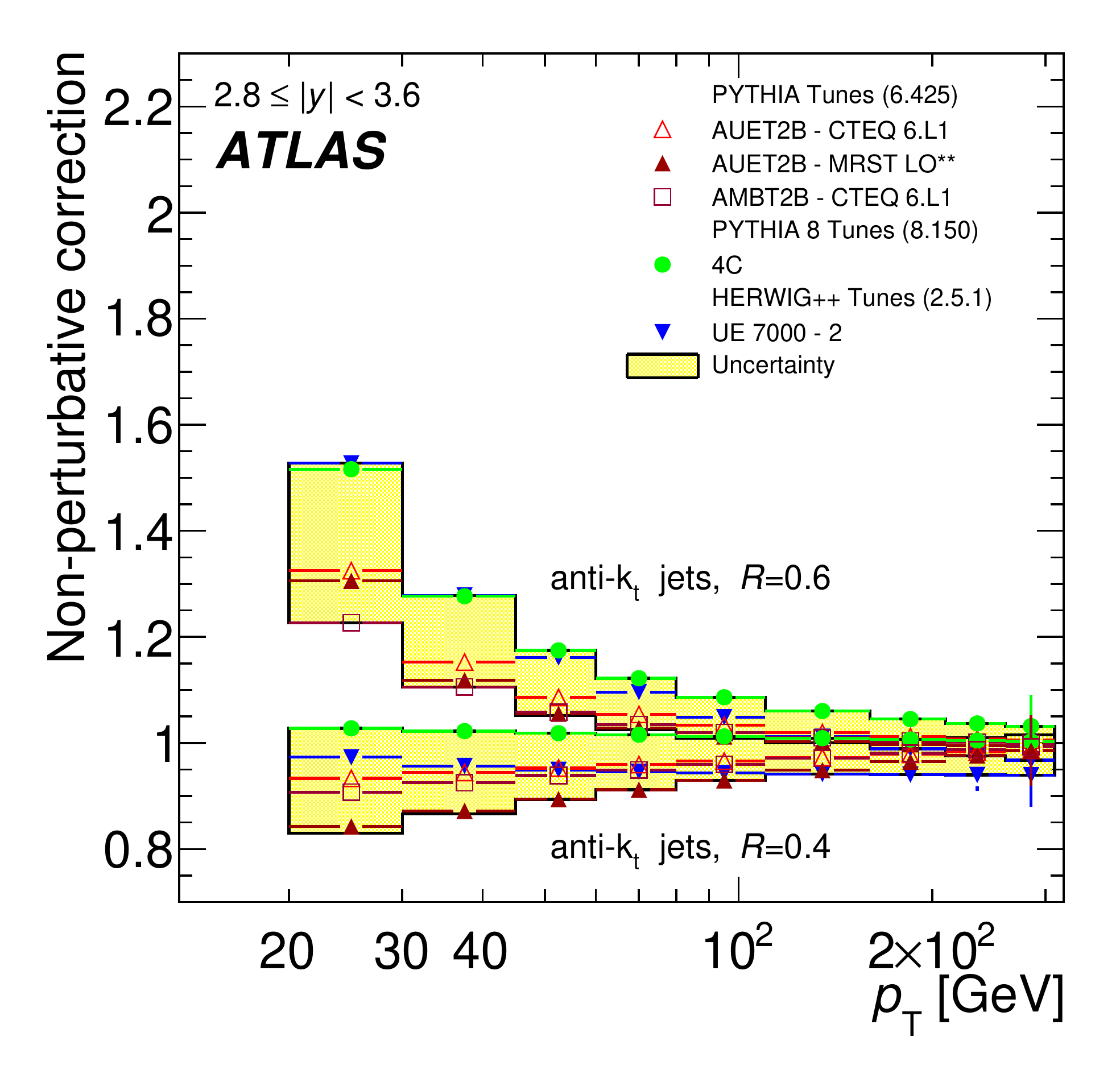}
  \includegraphics[width=.4\textwidth]{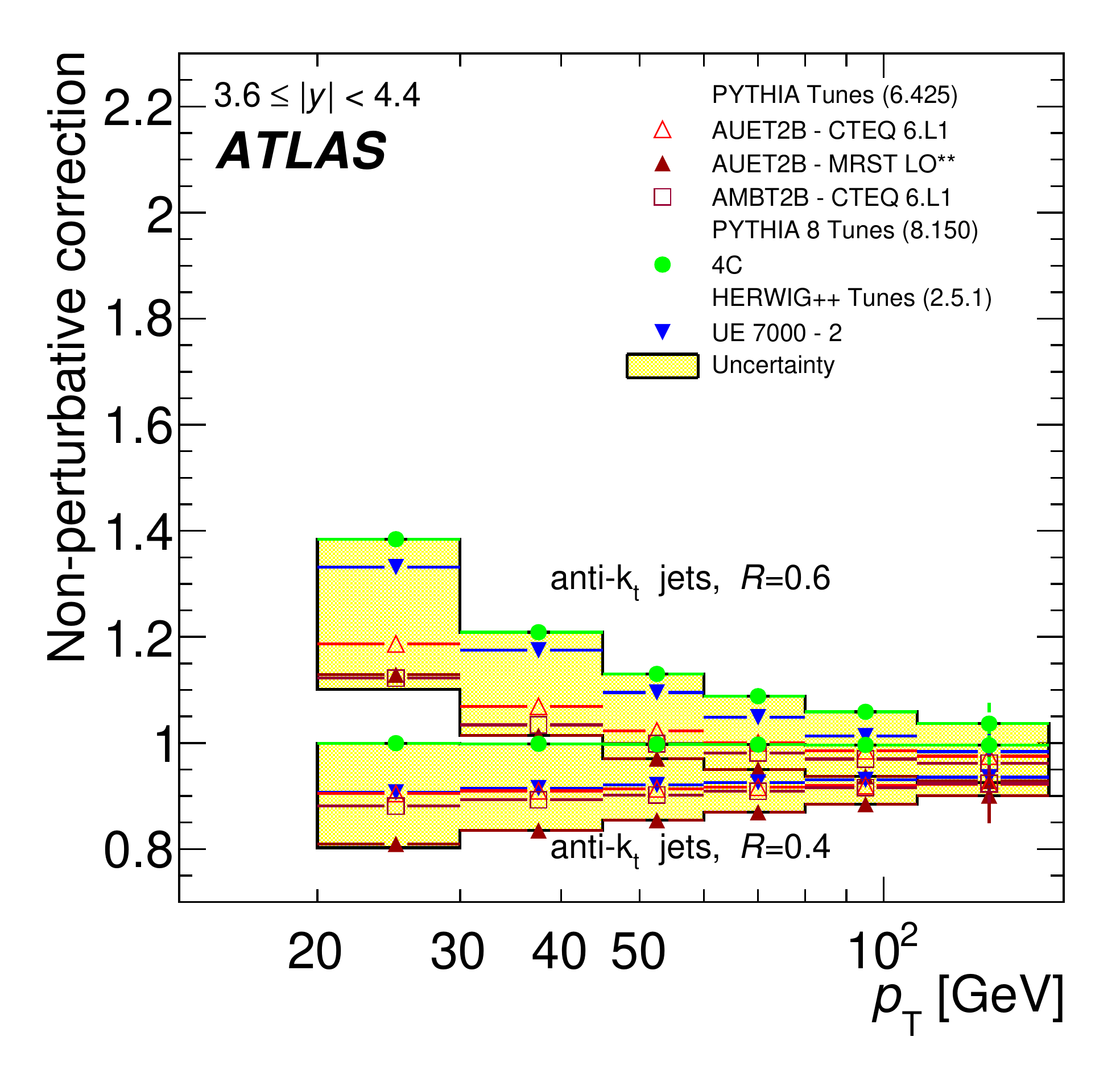}
  \caption{
Non-perturbative correction factors for inclusive jets identified
using the \AKT algorithm with distance parameters $R=0.4$ and $R=0.6$
in various rapidity regions, derived using various Monte Carlo
generators. The correction derived using \pythia 6.425 with the
AUET2B~CTEQ6L1 tune is used for the fixed-order NLO calculations
presented in this analysis.
  }
  \label{fig:NP-other-bins}
\end{figure*}

\begin{table*}
\section{Inclusive Jet Tables}
\label{sec:tablesIncJet}
  \begin{tiny}
  \setlength{\extrarowheight}{4pt}
  \begin{tabular}{rrrrrrrrrrrrrrrrrrrrrrrrr}\hline\hline
  $\pt$-bin & NPC & \multicolumn{1}{c}{$\sigma$} & $\delta_{\rm stat}$
& $\gamma_{1}$ & $\gamma_{7}$ & $\gamma_{13}$ & $\gamma_{19}$ &
$\gamma_{25}$ & $\gamma_{31}$ & $\gamma_{32}$ & $\gamma_{38}$ &
$\gamma_{44}$ & $\gamma_{50}$ & $\gamma_{56}$ & $\gamma_{62}$ &
$\gamma_{68}$ & $\gamma_{74}$ & $\gamma_{75}$ & $\gamma_{76}$ &
$\gamma_{82}$ & $\gamma_{83}$ & $u_{1}$ & $u_{2}$ & $u_{3}$ \\
$[$GeV$]$ & & $[$pb/GeV$]$ & \% &\\ \hline $ 20$-$30$ & 0.99(11) &
$4.70\cdot 10^{ 6}$ & 0.86 & $^{+ 10}_{-9.6}$ & $^{+ 10}_{-9.9}$ &
$^{+ 7.0}_{-7.1}$ & $^{+ 8.3}_{-8.6}$ & $^{+ 3.7}_{-4.0}$ & 0.0 & $^{+
2.0}_{-2.2}$ & $^{+ 4.9}_{-5.0}$ & $^{+ 4.1}_{-4.7}$ & 0.0 & 0.0 &
$^{+ 1.8}_{-2.0}$ & 0.0 & 0.0 & $\pm 1.5$ & $\pm 4.2$ & $\pm 0.2$ &
$\pm 2.0$ & 0.70 & 1.00 & 0.33\\ $ 30$-$45$ & 0.99(7) & $7.17\cdot
10^{ 5}$ & 1.33 & $^{+ 4.7}_{-4.8}$ & $^{+ 8.9}_{-8.3}$ & $^{+
5.9}_{-6.0}$ & $^{+ 10}_{-9.0}$ & $^{+ 1.4}_{-1.7}$ & 0.0 & $^{+
2.3}_{-2.4}$ & $^{+ 4.0}_{-3.8}$ & $^{+ 4.1}_{-4.4}$ & 0.0 & 0.0 &
$^{+ 3.3}_{-3.5}$ & 0.0 & 0.0 & $\pm 1.1$ & $\pm 0.9$ & 0.0 & $\pm
1.0$ & 0.41 & 1.00 & 0.22\\ $ 45$-$60$ & 0.99(5) & $1.48\cdot 10^{ 5}$
& 3.03 & $^{+ 1.8}_{-1.7}$ & $^{+ 7.8}_{-6.8}$ & $^{+ 4.1}_{-3.5}$ &
$^{+ 5.3}_{-4.7}$ & $^{+ 0.4}_{-0.1}$ & 0.0 & $^{+ 2.9}_{-2.8}$ & $^{+
2.4}_{-2.3}$ & $^{+ 4.0}_{-3.7}$ & $^{+ 0.2}_{+ 0.2}$ & $^{+ 0.3}_{+
0.0}$ & $^{+ 4.6}_{-4.4}$ & 0.0 & 0.0 & $\pm 0.7$ & $\pm 0.7$ & 0.0 &
$\pm 1.0$ & 0.27 & 1.00 & 0.19\\ $ 60$-$80$ & 0.99(5) & $3.81\cdot
10^{ 4}$ & 1.10 & $\mp 0.6$ & $^{+ 6.2}_{-6.0}$ & $^{+ 2.5}_{-2.4}$ &
$^{+ 3.3}_{-3.4}$ & $^{+ 0.9}_{-1.0}$ & 0.0 & $\pm 3.2$ & $^{+
1.5}_{-1.6}$ & $^{+ 4.1}_{-3.9}$ & $^{+ 0.4}_{-0.3}$ & 0.0 & $^{+
4.6}_{-4.2}$ & $\mp 0.1$ & 0.0 & $\pm 0.4$ & 0.0 & 0.0 & $\pm 1.0$ &
0.27 & 1.00 & 0.15\\ $ 80$-$110$ & 0.99(4) & $8.52\cdot 10^{ 3}$ &
0.68 & $\mp 0.4$ & $^{+ 6.4}_{-6.8}$ & $^{+ 3.6}_{-4.1}$ & $^{+
3.2}_{-3.3}$ & $^{+ 1.7}_{-2.2}$ & 0.0 & $^{+ 2.7}_{-3.1}$ & $^{+
1.1}_{-1.6}$ & $^{+ 4.2}_{-4.9}$ & $^{+ 0.5}_{-1.0}$ & 0.0 & $\pm 2.4$
& $^{-0.6}_{+ 0.2}$ & 0.0 & $\pm 2.0$ & $\pm 2.3$ & 0.0 & $\pm 1.0$ &
0.40 & 1.00 & 0.10\\ $ 110$-$160$ & 0.99(3) & $1.48\cdot 10^{ 3}$ &
0.62 & 0.0 & $^{+ 5.4}_{-4.9}$ & $^{+ 3.5}_{-3.3}$ & $^{+ 3.3}_{-3.4}$
& $\pm 0.7$ & 0.0 & $^{+ 3.2}_{-2.9}$ & $^{+ 1.2}_{-1.0}$ & $^{+
5.3}_{-4.9}$ & $^{+ 1.7}_{-1.6}$ & 0.0 & 0.0 & $^{+ 1.3}_{-1.1}$ & 0.0
& 0.0 & $\pm 0.8$ & 0.0 & $\pm 1.0$ & 0.35 & 1.00 & 0.07\\ $
160$-$210$ & 1.00(3) & $2.54\cdot 10^{ 2}$ & 0.69 & 0.0 & $^{+
3.8}_{-3.5}$ & $^{+ 3.8}_{-3.9}$ & $^{+ 3.0}_{-2.6}$ & $^{+
1.4}_{-0.9}$ & 0.0 & $^{+ 2.5}_{-2.6}$ & $^{+ 0.8}_{-0.4}$ & $^{+
6.1}_{-5.5}$ & $^{+ 2.5}_{-1.9}$ & $^{+ 0.3}_{+ 0.2}$ & $\mp 0.3$ &
$\pm 4.5$ & $\pm 0.1$ & $\pm 0.1$ & $\pm 0.4$ & 0.0 & $\pm 1.0$ & 0.34
& 1.00 & 0.06\\ $ 210$-$260$ & 1.00(3) & $6.34\cdot 10^{ 1}$ & 0.91 &
0.0 & $^{+ 5.4}_{-5.6}$ & $^{+ 5.7}_{-5.9}$ & $^{+ 2.1}_{-2.7}$ & $^{+
1.1}_{-1.7}$ & 0.0 & $^{+ 2.8}_{-2.9}$ & $^{+ 0.5}_{-0.8}$ & $^{+
6.1}_{-6.5}$ & $^{+ 3.2}_{-3.8}$ & $^{+ 0.2}_{-0.5}$ & 0.0 & $^{+
6.2}_{-6.6}$ & $\pm 0.2$ & 0.0 & $\pm 0.4$ & $\pm 0.1$ & $\pm 1.0$ &
0.37 & 1.00 & 0.05\\ $ 260$-$310$ & 1.00(3) & $2.07\cdot 10^{ 1}$ &
0.86 & 0.0 & $^{+ 5.4}_{-4.9}$ & $^{+ 4.0}_{-3.7}$ & $^{+ 2.1}_{-1.9}$
& $^{+ 1.8}_{-1.5}$ & 0.0 & $^{+ 2.4}_{-2.0}$ & $^{+ 0.6}_{-0.3}$ &
$^{+ 6.1}_{-5.7}$ & $^{+ 4.1}_{-3.8}$ & $^{+ 0.6}_{-0.4}$ & 0.0 & $^{+
6.9}_{-6.4}$ & $\pm 0.2$ & 0.0 & $\pm 0.7$ & 0.0 & $\pm 1.0$ & 0.28 &
1.00 & 0.05\\ $ 310$-$400$ & 1.00(2) & $5.96\cdot 10^{ 0}$ & 1.03 &
0.0 & $^{+ 3.7}_{-3.2}$ & $^{+ 2.3}_{-1.9}$ & $^{+ 1.1}_{-0.5}$ & $^{+
1.9}_{-1.8}$ & 0.0 & $^{+ 2.0}_{-1.9}$ & $^{+ 0.7}_{-0.1}$ & $^{+
6.3}_{-6.1}$ & $^{+ 4.5}_{-3.8}$ & $^{+ 0.7}_{-0.2}$ & 0.0 & $^{+
7.0}_{-6.5}$ & $\pm 0.1$ & 0.0 & $\pm 0.8$ & 0.0 & $\pm 1.0$ & 0.26 &
1.00 & 0.05\\ $ 400$-$500$ & 1.00(2) & $1.33\cdot 10^{ 0}$ & 2.02 &
0.0 & $^{+ 0.4}_{-0.8}$ & 0.0 & $^{+ 1.6}_{-2.0}$ & $^{+ 1.8}_{-2.2}$
& 0.0 & $^{+ 1.1}_{-1.5}$ & $^{+ 0.3}_{-0.7}$ & $^{+ 6.7}_{-6.6}$ &
$^{+ 5.0}_{-5.3}$ & $^{+ 0.1}_{-0.6}$ & 0.0 & $^{+ 7.7}_{-7.8}$ & $\pm
0.2$ & 0.0 & $\pm 0.4$ & 0.0 & $\pm 1.0$ & 0.21 & 1.00 & 0.05\\ $
500$-$600$ & 1.00(2) & $3.47\cdot 10^{-1}$ & 3.22 & 0.0 & $\pm 2.3$ &
$^{+ 1.0}_{-0.8}$ & $^{+ 1.9}_{-1.8}$ & $^{+ 3.0}_{-2.7}$ & 0.0 & $^{+
1.6}_{-1.3}$ & $^{+ 0.3}_{-0.4}$ & $^{+ 7.2}_{-6.7}$ & $^{+
6.2}_{-5.5}$ & $^{+ 0.5}_{-0.6}$ & 0.0 & $^{+ 9.6}_{-8.8}$ & $\pm 0.3$
& 0.0 & $\pm 0.7$ & 0.0 & $\pm 1.0$ & 0.19 & 1.00 & 0.05\\ $
600$-$800$ & 1.00(2) & $6.44\cdot 10^{-2}$ & 5.73 & 0.0 & $\pm 4.2$ &
$^{+ 3.3}_{-3.4}$ & $^{+ 0.8}_{-0.6}$ & $^{+ 1.8}_{-2.1}$ & 0.0 & $^{+
1.2}_{-1.5}$ & $^{+ 0.5}_{-0.4}$ & $^{+ 7.7}_{-6.8}$ & $^{+
6.0}_{-5.8}$ & $^{+ 1.0}_{-0.9}$ & 0.0 & $^{+ 10}_{-9.4}$ & $\pm 0.3$
& 0.0 & $\pm 0.7$ & 0.0 & $\pm 1.0$ & 0.17 & 1.00 & 0.21\\ $
800$-$1000$ & 1.00(2) & $1.01\cdot 10^{-2}$ & 16.7 & 0.0 & $^{+
3.0}_{-2.6}$ & $^{+ 5.4}_{-4.9}$ & $^{+ 1.1}_{-0.9}$ & $^{+
1.7}_{-1.5}$ & 0.0 & $^{+ 1.1}_{-1.0}$ & $^{+ 0.3}_{+ 0.1}$ & $^{+
8.4}_{-7.8}$ & $^{+ 7.9}_{-6.8}$ & $^{+ 1.8}_{-2.1}$ & 0.0 & $^{+
12}_{ -11}$ & $\pm 0.5$ & 0.0 & $\pm 0.9$ & 0.0 & $\pm 1.0$ & 0.11 &
1.00 & 0.21\\ $1000$-$1200$ & 1.00(2) & $1.14\cdot 10^{-3}$ & 37.3 &
0.0 & $^{+ 2.4}_{-2.5}$ & $^{+ 5.6}_{-5.3}$ & $^{+ 0.9}_{-0.7}$ & $\pm
3.1$ & 0.0 & $^{+ 1.0}_{-0.6}$ & $^{+ 0.5}_{-0.3}$ & $^{+ 8.8}_{-7.9}$
& $^{+ 8.8}_{-8.4}$ & $\pm 4.9$ & 0.0 & $^{+ 16}_{ -14}$ & $\pm 0.7$ &
0.0 & $\pm 0.4$ & 0.0 & $\pm 1.0$ & 0.12 & 1.00 & 0.21\\ $1200$-$1500$
& 1.00(2) & $4.00\cdot 10^{-4}$ & 58.6 & 0.0 & $^{+ 2.3}_{-2.5}$ &
$^{+ 6.3}_{-6.4}$ & $^{+ 1.0}_{-0.8}$ & $^{+ 2.4}_{-2.7}$ & 0.0 & $^{+
0.5}_{-0.8}$ & $^{+ 0.6}_{-0.5}$ & $^{+ 9.7}_{-8.9}$ & $^{+
10}_{-9.5}$ & $^{+ 9.3}_{-8.6}$ & 0.0 & $^{+ 20}_{ -18}$ & $\pm 0.5$ &
0.0 & $\pm 3.0$ & $\pm 0.1$ & $\pm 1.0$ & 0.08 & 1.00 & 0.21\\
\hline\hline
  \end{tabular}
  \end{tiny}\vspace*{-1.5mm}
  \caption{ Measured jet cross section for $R=0.4$, $|y|<0.3$.  NPC
    stands for multiplicative non-perturbative corrections with ${\rm
    error} \times 100$ in brackets, {\it i.e.} 1.25(10) means $1.25
    \pm 0.10$.  $\sigma$ is the measured cross section. $\delta_{\rm
    stat}$ is the statistical uncertainty. $\gamma_{i}$ and $u_{i}$
    are the correlated and uncorrelated systematic uncertainties, as
    described in Sec.~\ref{sec:corr} and
    Table~\ref{tab:correlations}. All uncertainties are given in $\%$.
    An overall luminosity uncertainty of $3.4\%$, which is applicable
    to all ATLAS data samples based on 2010 data, is not shown.
    All tables are available on \hepdata~\cite{HEPDATA}.}
  \label{tab:AntiKt4_EtaLow0_Results}
\vspace*{-2mm}
\end{table*}

\begin{table*}
  \begin{tiny}
  \setlength{\extrarowheight}{4pt}
\begin{tabular}{rrrrrrrrrrrrrrrrrrrrrrrrr}\hline\hline
  $\pt$-bin & NPC & \multicolumn{1}{c}{$\sigma$} & $\delta_{\rm stat}$
& $\gamma_{1}$ & $\gamma_{7}$ & $\gamma_{13}$ & $\gamma_{19}$ &
$\gamma_{25}$ & $\gamma_{31}$ & $\gamma_{32}$ & $\gamma_{38}$ &
$\gamma_{44}$ & $\gamma_{50}$ & $\gamma_{56}$ & $\gamma_{62}$ &
$\gamma_{68}$ & $\gamma_{74}$ & $\gamma_{75}$ & $\gamma_{76}$ &
$\gamma_{82}$ & $\gamma_{83}$ & $u_{1}$ & $u_{2}$ & $u_{3}$ \\
$[$GeV$]$ & & $[$pb/GeV$]$ & \% &\\ \hline $ 20$-$30$ & 0.98(11) &
$4.61\cdot 10^{ 6}$ & 0.61 & $^{+ 10}_{-9.5}$ & $^{+ 8.1}_{-8.3}$ &
$^{+ 10}_{-9.6}$ & $\pm 7.3$ & $\pm 4.2$ & 0.0 & $\pm 2.1$ & $\pm 4.3$
& $^{+ 4.6}_{-4.5}$ & 0.0 & 0.0 & $\pm 2.9$ & 0.0 & 0.0 & $\pm 1.3$ &
$\pm 3.9$ & $\pm 0.2$ & $\pm 2.0$ & 0.70 & 1.00 & 0.30\\ $ 30$-$45$ &
0.99(7) & $7.30\cdot 10^{ 5}$ & 1.19 & $^{+ 4.9}_{-4.7}$ & $^{+
11}_{-9.5}$ & $^{+ 6.8}_{-6.3}$ & $^{+ 8.8}_{-7.5}$ & $^{+
2.4}_{-2.1}$ & 0.0 & $\pm 2.4$ & $\pm 3.2$ & $^{+ 4.6}_{-4.4}$ & 0.0 &
0.0 & $\pm 3.2$ & $^{-0.2}_{+ 0.1}$ & 0.0 & $\pm 1.0$ & $\pm 1.7$ &
0.0 & $\pm 1.0$ & 0.43 & 1.00 & 0.20\\ $ 45$-$60$ & 0.99(6) &
$1.49\cdot 10^{ 5}$ & 2.15 & $^{+ 2.0}_{-1.5}$ & $^{+ 7.5}_{-6.7}$ &
$^{+ 6.0}_{-5.3}$ & $^{+ 4.9}_{-4.2}$ & $^{+ 1.7}_{-1.1}$ & 0.0 & $^{+
3.2}_{-2.6}$ & $^{+ 1.9}_{-1.3}$ & $^{+ 4.4}_{-3.8}$ & $^{+ 0.1}_{+
0.2}$ & $^{+ 0.3}_{+ 0.0}$ & $^{+ 2.7}_{-2.3}$ & $^{+ 0.3}_{+ 0.1}$ &
$\pm 0.1$ & $\pm 0.5$ & $\pm 1.2$ & 0.0 & $\pm 1.0$ & 0.28 & 1.00 &
0.19\\ $ 60$-$80$ & 0.99(5) & $3.71\cdot 10^{ 4}$ & 0.83 & $\mp 0.6$ &
$^{+ 6.2}_{-5.3}$ & $^{+ 5.1}_{-4.4}$ & $^{+ 3.9}_{-3.4}$ & $^{+
2.4}_{-2.1}$ & 0.0 & $^{+ 3.2}_{-2.7}$ & $^{+ 1.5}_{-1.2}$ & $^{+
4.6}_{-3.9}$ & $^{+ 0.5}_{-0.2}$ & 0.0 & $^{+ 0.9}_{-0.7}$ & $^{+
1.6}_{-1.3}$ & 0.0 & 0.0 & $\pm 0.6$ & 0.0 & $\pm 1.0$ & 0.30 & 1.00 &
0.16\\
\hline\hline
  \end{tabular}
  \end{tiny}\vspace*{-1.5mm}
  \caption{ Measured jet cross section for $R=0.4$, $0.3\leq|y|<0.8$.
    See Table~\ref{tab:AntiKt4_EtaLow0_Results} for description of the
    columns.  All tables are available on \hepdata~\cite{HEPDATA}.  }
  \label{tab:AntiKt4_EtaLow0.3_Results}
\vspace*{-2mm}
\end{table*}

\begin{table*}
  \begin{tiny}
  \setlength{\extrarowheight}{4pt}
  \begin{tabular}{rrrrrrrrrrrrrrrrrrrrrrrrr}\hline\hline
  $\pt$-bin & NPC & \multicolumn{1}{c}{$\sigma$} & $\delta_{\rm stat}$
& $\gamma_{2}$ & $\gamma_{8}$ & $\gamma_{14}$ & $\gamma_{20}$ &
$\gamma_{26}$ & $\gamma_{31}$ & $\gamma_{33}$ & $\gamma_{39}$ &
$\gamma_{45}$ & $\gamma_{51}$ & $\gamma_{57}$ & $\gamma_{63}$ &
$\gamma_{69}$ & $\gamma_{74}$ & $\gamma_{75}$ & $\gamma_{77}$ &
$\gamma_{82}$ & $\gamma_{83}$ & $u_{1}$ & $u_{2}$ & $u_{3}$ \\
$[$GeV$]$ & & $[$pb/GeV$]$ & \% &\\ \hline $ 20$-$30$ & 0.98(11) &
$3.92\cdot 10^{ 6}$ & 0.68 & $\pm 9.1$ & $^{+ 7.0}_{-8.0}$ & $^{+
8.9}_{-9.3}$ & $\pm 13$ & $^{+ 3.7}_{-3.9}$ & $^{+ 7.0}_{-7.5}$ & $^{+
1.8}_{-1.9}$ & $^{+ 3.9}_{-4.1}$ & $^{+ 4.1}_{-4.4}$ & 0.0 & 0.0 &
$^{+ 2.5}_{-2.8}$ & 0.0 & 0.0 & $\pm 1.0$ & $\pm 9.4$ & $\pm 1.2$ &
$\pm 2.0$ & 1.02 & 1.00 & 0.25\\ $ 30$-$45$ & 0.98(7) & $6.66\cdot
10^{ 5}$ & 1.51 & $^{+ 5.0}_{-4.5}$ & $^{+ 11}_{-9.4}$ & $^{+
7.2}_{-6.2}$ & $^{+ 5.7}_{-4.8}$ & $^{+ 2.5}_{-1.9}$ & $^{+
8.7}_{-7.4}$ & $^{+ 2.6}_{-2.3}$ & $^{+ 3.4}_{-3.0}$ & $^{+
4.9}_{-4.3}$ & 0.0 & 0.0 & $^{+ 3.4}_{-3.0}$ & $^{-0.2}_{+ 0.1}$ & 0.0
& $\pm 0.7$ & $\pm 6.5$ & $\pm 0.2$ & $\pm 1.0$ & 0.72 & 1.00 & 0.20\\
$ 45$-$60$ & 0.99(6) & $1.39\cdot 10^{ 5}$ & 2.59 & $^{+ 1.9}_{-1.4}$
& $^{+ 8.2}_{-6.7}$ & $^{+ 6.3}_{-5.3}$ & $^{+ 2.0}_{-1.2}$ & $^{+
1.8}_{-1.2}$ & $^{+ 7.2}_{-6.0}$ & $^{+ 3.2}_{-2.6}$ & $^{+
1.9}_{-1.4}$ & $^{+ 4.7}_{-3.8}$ & $^{+ 0.2}_{+ 0.2}$ & $^{+ 0.3}_{+
0.0}$ & $^{+ 2.7}_{-2.2}$ & $^{+ 0.2}_{+ 0.1}$ & 0.0 & $\pm 0.5$ &
$\pm 3.1$ & $\pm 0.3$ & $\pm 1.0$ & 0.46 & 1.00 & 0.21\\ $ 60$-$80$ &
0.99(5) & $3.39\cdot 10^{ 4}$ & 0.99 & $\mp 0.7$ & $^{+ 6.0}_{-5.3}$ &
$^{+ 4.8}_{-4.5}$ & $\pm 2.4$ & $^{+ 2.3}_{-2.2}$ & $^{+ 5.3}_{-5.0}$
& $\pm 2.8$ & $^{+ 1.5}_{-1.3}$ & $^{+ 4.4}_{-4.1}$ & $^{+
0.3}_{-0.2}$ & 0.0 & $\pm 0.8$ & $^{+ 1.6}_{-1.5}$ & 0.0 & $\pm 0.1$ &
$\pm 1.5$ & $\pm 0.2$ & $\pm 1.0$ & 0.47 & 1.00 & 0.18\\
\hline\hline
  \end{tabular}
  \end{tiny}\vspace*{-1.5mm}
  \caption{ Measured jet cross section for $R=0.4$, $0.8\leq|y|<1.2$.
    See Table~\ref{tab:AntiKt4_EtaLow0_Results} for description of the
    columns.  All tables are available on \hepdata~\cite{HEPDATA}.  }
  \label{tab:AntiKt4_EtaLow0.8_Results}
\vspace*{-2mm}
\end{table*}

\begin{table*}
  \begin{tiny}
  \setlength{\extrarowheight}{4pt}
  \begin{tabular}{rrrrrrrrrrrrrrrrrrrrrrrrr}\hline\hline
  $\pt$-bin & NPC & \multicolumn{1}{c}{$\sigma$} & $\delta_{\rm stat}$
& $\gamma_{3}$ & $\gamma_{9}$ & $\gamma_{15}$ & $\gamma_{21}$ &
$\gamma_{27}$ & $\gamma_{31}$ & $\gamma_{34}$ & $\gamma_{40}$ &
$\gamma_{46}$ & $\gamma_{52}$ & $\gamma_{58}$ & $\gamma_{64}$ &
$\gamma_{70}$ & $\gamma_{74}$ & $\gamma_{75}$ & $\gamma_{78}$ &
$\gamma_{82}$ & $\gamma_{83}$ & $u_{1}$ & $u_{2}$ & $u_{3}$ \\
$[$GeV$]$ & & $[$pb/GeV$]$ & \% &\\ \hline $ 20$-$30$ & 0.97(11) &
$3.70\cdot 10^{ 6}$ & 0.55 & $^{+ 9.5}_{-9.0}$ & $^{+ 7.0}_{-7.6}$ &
$^{+ 9.3}_{-9.1}$ & $^{+ 9.5}_{-8.7}$ & $\pm 3.7$ & $\pm 14$ & $^{+
1.7}_{-1.8}$ & $\pm 3.8$ & $^{+ 4.0}_{-4.1}$ & 0.0 & 0.0 & $^{+
2.5}_{-2.6}$ & 0.0 & $\mp 0.1$ & $\pm 0.9$ & $\pm 10$ & $\pm 0.6$ &
$\pm 2.0$ & 1.10 & 1.00 & 0.30\\ $ 30$-$45$ & 0.97(7) & $5.62\cdot
10^{ 5}$ & 1.17 & $^{+ 4.7}_{-4.6}$ & $^{+ 11}_{-9.5}$ & $^{+
6.8}_{-6.3}$ & $^{+ 2.4}_{-1.9}$ & $^{+ 2.3}_{-1.9}$ & $^{+ 16}_{
-14}$ & $^{+ 2.3}_{-2.2}$ & $^{+ 3.1}_{-3.0}$ & $^{+ 4.4}_{-4.3}$ &
0.0 & 0.0 & $^{+ 3.1}_{-3.0}$ & $\mp 0.2$ & 0.0 & $\pm 0.9$ & $\pm
4.7$ & 0.0 & $\pm 1.0$ & 0.77 & 1.00 & 0.15\\ $ 45$-$60$ & 0.98(6) &
$1.09\cdot 10^{ 5}$ & 2.01 & $^{+ 2.0}_{-1.8}$ & $^{+ 8.1}_{-6.7}$ &
$^{+ 6.3}_{-5.4}$ & $^{+ 0.2}_{ 0.0}$ & $^{+ 1.7}_{-1.4}$ & $^{+
12}_{-9.2}$ & $^{+ 3.1}_{-2.8}$ & $^{+ 2.0}_{-1.7}$ & $^{+
4.4}_{-3.8}$ & $^{+ 0.2}_{+ 0.2}$ & $^{+ 0.3}_{+ 0.0}$ & $^{+
2.8}_{-2.5}$ & $^{+ 0.3}_{+ 0.0}$ & 0.0 & $\pm 0.6$ & $\pm 1.8$ & 0.0
& $\pm 1.0$ & 0.51 & 1.00 & 0.13\\ $ 60$-$80$ & 0.98(5) & $2.67\cdot
10^{ 4}$ & 0.73 & $\mp 0.5$ & $^{+ 6.7}_{-6.6}$ & $\pm 5.4$ & $\pm
0.4$ & $^{+ 2.5}_{-2.6}$ & $^{+ 8.0}_{-7.6}$ & $\pm 3.1$ & $^{+
1.5}_{-1.7}$ & $\pm 4.9$ & $^{+ 0.3}_{-0.2}$ & 0.0 & $\pm 0.9$ & $^{+
1.7}_{-1.8}$ & $\pm 0.1$ & $\pm 0.4$ & $\pm 0.6$ & 0.0 & $\pm 1.0$ &
0.48 & 1.00 & 0.10\\
\hline\hline
  \end{tabular}
  \end{tiny}\vspace*{-1.5mm}
  \caption{ Measured jet cross section for $R=0.4$, $1.2\leq|y|<2.1$.
    See Table~\ref{tab:AntiKt4_EtaLow0_Results} for description of the
    columns.  All tables are available on \hepdata~\cite{HEPDATA}.  }
  \label{tab:AntiKt4_EtaLow1.2_Results}
\vspace*{-2mm}
\end{table*}

\begin{table*}
  \begin{tiny}
  \setlength{\extrarowheight}{4pt}
  \begin{tabular}{rrrrrrrrrrrrrrrrrrrrrrrrr}\hline\hline
  $\pt$-bin & NPC & \multicolumn{1}{c}{$\sigma$} & $\delta_{\rm stat}$
& $\gamma_{4}$ & $\gamma_{10}$ & $\gamma_{16}$ & $\gamma_{22}$ &
$\gamma_{28}$ & $\gamma_{31}$ & $\gamma_{35}$ & $\gamma_{41}$ &
$\gamma_{47}$ & $\gamma_{53}$ & $\gamma_{59}$ & $\gamma_{65}$ &
$\gamma_{71}$ & $\gamma_{74}$ & $\gamma_{75}$ & $\gamma_{79}$ &
$\gamma_{82}$ & $\gamma_{84}$ & $u_{1}$ & $u_{2}$ & $u_{3}$ \\
$[$GeV$]$ & & $[$pb/GeV$]$ & \% &\\ \hline $ 20$-$30$ & 0.96(10) &
$2.76\cdot 10^{ 6}$ & 0.69 & $^{+ 10}_{-9.1}$ & $\pm 7.4$ & $^{+
10}_{-9.1}$ & $^{+ 6.5}_{-6.1}$ & $\pm 4.0$ & $^{+ 26}_{ -23}$ & $^{+
2.0}_{-1.9}$ & $^{+ 4.2}_{-4.1}$ & $^{+ 4.4}_{-4.3}$ & 0.0 & 0.0 &
$\pm 2.7$ & 0.0 & 0.0 & $\pm 1.2$ & $\pm 15$ & $\pm 0.1$ & $\pm 2.0$ &
1.27 & 1.00 & 0.02\\ $ 30$-$45$ & 0.96(7) & $4.07\cdot 10^{ 5}$ & 1.37
& $^{+ 4.8}_{-4.3}$ & $^{+ 11}_{-9.1}$ & $^{+ 6.7}_{-5.8}$ & $^{+
4.7}_{-4.0}$ & $^{+ 2.3}_{-1.8}$ & $^{+ 28}_{ -21}$ & $^{+
2.4}_{-2.2}$ & $^{+ 3.3}_{-2.9}$ & $^{+ 4.5}_{-4.1}$ & 0.0 & 0.0 &
$^{+ 3.2}_{-2.9}$ & $^{-0.2}_{+ 0.1}$ & 0.0 & $\pm 0.7$ & $\pm 5.1$ &
0.0 & $\pm 1.0$ & 0.85 & 1.00 & 0.02\\ $ 45$-$60$ & 0.97(6) &
$8.00\cdot 10^{ 4}$ & 2.84 & $^{+ 2.5}_{-2.1}$ & $^{+ 8.1}_{-7.1}$ &
$^{+ 6.7}_{-5.8}$ & $^{+ 2.5}_{-2.0}$ & $^{+ 2.1}_{-1.4}$ & $^{+ 16}_{
-13}$ & $^{+ 3.5}_{-2.6}$ & $^{+ 2.2}_{-1.7}$ & $^{+ 4.7}_{-3.9}$ &
$^{+ 0.1}_{+ 0.2}$ & $^{+ 0.3}_{+ 0.0}$ & $^{+ 3.1}_{-2.6}$ & $^{+
0.3}_{+ 0.2}$ & $\pm 0.2$ & $\pm 0.1$ & $\pm 1.6$ & 0.0 & $\pm 1.0$ &
0.55 & 1.00 & 0.02\\ $ 60$-$80$ & 0.97(5) & $1.71\cdot 10^{ 4}$ & 1.05
& $^{-0.5}_{+ 0.4}$ & $\pm 7.0$ & $^{+ 5.5}_{-5.6}$ & $^{+
1.3}_{-1.5}$ & $^{+ 2.5}_{-2.8}$ & $^{+ 10}_{-9.4}$ & $^{+
3.1}_{-3.5}$ & $^{+ 1.7}_{-1.9}$ & $^{+ 5.0}_{-5.2}$ & $\pm 0.3$ & 0.0
& $^{+ 0.8}_{-1.0}$ & $^{+ 1.7}_{-1.9}$ & $\pm 0.2$ & $\pm 0.3$ & $\pm
0.8$ & 0.0 & $\pm 1.0$ & 0.57 & 1.00 & 0.02\\
\hline\hline
  \end{tabular}
  \end{tiny}\vspace*{-1.5mm}
  \caption{ Measured jet cross section for $R=0.4$, $2.1\leq|y|<2.8$.
    See Table~\ref{tab:AntiKt4_EtaLow0_Results} for description of the
    columns.  All tables are available on \hepdata~\cite{HEPDATA}.  }
  \label{tab:AntiKt4_EtaLow2.1_Results}
\vspace*{-2mm}
\end{table*}

\begin{table*}
  \begin{tiny}
  \setlength{\extrarowheight}{4pt}
  \begin{tabular}{rrrrrrrrrrrrrrrrrrrrrrrrr}\hline\hline
  $\pt$-bin & NPC & \multicolumn{1}{c}{$\sigma$} & $\delta_{\rm stat}$
& $\gamma_{5}$ & $\gamma_{11}$ & $\gamma_{17}$ & $\gamma_{23}$ &
$\gamma_{29}$ & $\gamma_{31}$ & $\gamma_{36}$ & $\gamma_{42}$ &
$\gamma_{48}$ & $\gamma_{54}$ & $\gamma_{60}$ & $\gamma_{66}$ &
$\gamma_{72}$ & $\gamma_{74}$ & $\gamma_{75}$ & $\gamma_{80}$ &
$\gamma_{82}$ & $\gamma_{85}$ & $u_{1}$ & $u_{2}$ & $u_{3}$ \\
$[$GeV$]$ & & $[$pb/GeV$]$ & \% &\\ \hline $ 20$-$30$ & 0.93(10) &
$1.89\cdot 10^{ 6}$ & 0.78 & $^{+ 11}_{-9.0}$ & $^{+ 7.8}_{-7.6}$ &
$^{+ 10}_{-9.2}$ & $^{+ 17}_{ -14}$ & $^{+ 4.1}_{-3.9}$ & $^{+ 37}_{
-30}$ & $^{+ 1.8}_{-1.9}$ & $^{+ 4.3}_{-4.0}$ & $^{+ 4.5}_{-4.2}$ &
0.0 & 0.0 & $^{+ 2.7}_{-2.6}$ & 0.0 & $\mp 0.2$ & $\pm 0.6$ & $\pm 14$
& $\pm 0.3$ & $\pm 2.0$ & 3.83 & 1.00 & 0.00\\ $ 30$-$45$ & 0.94(8) &
$2.67\cdot 10^{ 5}$ & 1.67 & $^{+ 4.9}_{-5.0}$ & $^{+ 12}_{ -11}$ &
$^{+ 6.8}_{-7.0}$ & $^{+ 10}_{-9.3}$ & $^{+ 2.5}_{-2.4}$ & $^{+ 41}_{
-28}$ & $^{+ 2.6}_{-2.7}$ & $^{+ 3.4}_{-3.6}$ & $^{+ 4.8}_{-5.1}$ &
0.0 & 0.0 & $^{+ 3.3}_{-3.6}$ & $\mp 0.1$ & 0.0 & $\pm 0.9$ & $\pm
6.5$ & 0.0 & $\pm 1.0$ & 2.67 & 1.00 & 0.00\\ $ 45$-$60$ & 0.95(6) &
$4.25\cdot 10^{ 4}$ & 3.69 & $^{+ 2.2}_{-2.0}$ & $^{+ 9.1}_{-8.0}$ &
$^{+ 7.3}_{-6.5}$ & $^{+ 7.5}_{-6.8}$ & $^{+ 2.2}_{-1.7}$ & $^{+ 28}_{
-20}$ & $^{+ 3.8}_{-3.3}$ & $^{+ 2.4}_{-1.9}$ & $^{+ 5.4}_{-4.7}$ &
$^{+ 0.3}_{+ 0.2}$ & $^{+ 0.4}_{+ 0.0}$ & $^{+ 3.1}_{-2.7}$ & $^{+
0.5}_{-0.1}$ & $\pm 0.3$ & $\pm 0.2$ & $\pm 2.1$ & 0.0 & $\pm 1.0$ &
1.78 & 1.00 & 0.00\\ $ 60$-$80$ & 0.96(5) & $8.57\cdot 10^{ 3}$ & 1.69
& $\mp 0.7$ & $^{+ 7.4}_{-6.4}$ & $^{+ 5.6}_{-5.1}$ & $^{+
7.9}_{-6.7}$ & $\pm 2.5$ & $^{+ 18}_{ -14}$ & $^{+ 3.5}_{-3.2}$ & $^{+
1.7}_{-1.6}$ & $^{+ 5.3}_{-4.6}$ & $\pm 0.4$ & 0.0 & $^{+ 1.0}_{-0.7}$
& $^{+ 2.1}_{-2.0}$ & $\pm 0.3$ & $\pm 0.3$ & $\pm 1.7$ & 0.0 & $\pm
1.0$ & 3.35 & 1.00 & 0.00\\
\hline\hline
  \end{tabular}
  \end{tiny}\vspace*{-1.5mm}
  \caption{ Measured jet cross section for $R=0.4$, $2.8\leq|y|<3.6$.
    See Table~\ref{tab:AntiKt4_EtaLow0_Results} for description of the
    columns.  All tables are available on \hepdata~\cite{HEPDATA}.  }
  \label{tab:AntiKt4_EtaLow2.8_Results}
\vspace*{-2mm}
\end{table*}

\begin{table*}
  \begin{tiny}
  \setlength{\extrarowheight}{4pt}
  \begin{tabular}{rrrrrrrrrrrrrrrrrrrrrrrrr}\hline\hline
  $\pt$-bin & NPC & \multicolumn{1}{c}{$\sigma$} & $\delta_{\rm stat}$
& $\gamma_{6}$ & $\gamma_{12}$ & $\gamma_{18}$ & $\gamma_{24}$ &
$\gamma_{30}$ & $\gamma_{31}$ & $\gamma_{37}$ & $\gamma_{43}$ &
$\gamma_{49}$ & $\gamma_{55}$ & $\gamma_{61}$ & $\gamma_{67}$ &
$\gamma_{73}$ & $\gamma_{74}$ & $\gamma_{75}$ & $\gamma_{81}$ &
$\gamma_{82}$ & $\gamma_{86}$ & $u_{1}$ & $u_{2}$ & $u_{3}$ \\
$[$GeV$]$ & & $[$pb/GeV$]$ & \% &\\ \hline $ 20$-$30$ & 0.90(10) &
$8.19\cdot 10^{ 5}$ & 1.04 & $^{+ 12}_{ -10}$ & $^{+ 8.8}_{-8.7}$ &
$^{+ 11}_{ -10}$ & $^{+ 3.3}_{-3.4}$ & $^{+ 4.7}_{-4.5}$ & $^{+ 68}_{
-48}$ & $^{+ 2.3}_{-2.2}$ & $^{+ 4.9}_{-4.5}$ & $^{+ 5.2}_{-4.9}$ &
0.0 & 0.0 & $^{+ 3.1}_{-3.2}$ & 0.0 & 0.0 & $\pm 0.3$ & $\pm 15$ &
$\pm 1.0$ & $\pm 2.0$ & 4.69 & 1.00 & 0.00\\ $ 30$-$45$ & 0.91(8) &
$8.64\cdot 10^{ 4}$ & 1.13 & $\pm 5.2$ & $^{+ 13}_{ -11}$ & $^{+
7.9}_{-7.2}$ & $^{+ 2.2}_{-2.0}$ & $^{+ 2.6}_{-2.2}$ & $^{+ 83}_{
-44}$ & $^{+ 2.8}_{-2.9}$ & $^{+ 3.4}_{-3.5}$ & $^{+ 5.4}_{-5.3}$ &
0.0 & 0.0 & $^{+ 3.5}_{-3.6}$ & 0.0 & 0.0 & $\pm 0.7$ & $\pm 4.4$ &
$\pm 0.5$ & $\pm 1.0$ & 7.98 & 1.00 & 0.00\\ $ 45$-$60$ & 0.91(7) &
$8.83\cdot 10^{ 3}$ & 1.95 & $^{+ 2.3}_{-2.6}$ & $^{+ 10}_{-9.1}$ &
$^{+ 8.4}_{-7.8}$ & $^{+ 1.7}_{-1.9}$ & $^{+ 2.3}_{-2.6}$ & $^{+ 60}_{
-34}$ & $^{+ 3.9}_{-4.2}$ & $^{+ 2.4}_{-2.7}$ & $^{+ 6.3}_{-5.5}$ &
$^{+ 0.3}_{+ 0.1}$ & $^{+ 0.4}_{+ 0.0}$ & $^{+ 3.3}_{-3.4}$ & $^{+
0.7}_{-0.5}$ & $\pm 0.2$ & $\pm 0.7$ & $\pm 1.0$ & $\pm 0.4$ & $\pm
1.0$ & 5.37 & 1.00 & 0.00\\ $ 60$-$80$ & 0.92(6) & $9.20\cdot 10^{ 2}$
& 2.78 & $\mp 0.9$ & $^{+ 11}_{ -10}$ & $^{+ 8.9}_{-8.4}$ & $^{+
3.0}_{-1.5}$ & $^{+ 5.0}_{-3.6}$ & $^{+ 41}_{ -29}$ & $^{+
6.5}_{-5.7}$ & $^{+ 3.8}_{-2.4}$ & $^{+ 8.7}_{-8.1}$ & $^{+
1.1}_{-0.7}$ & 0.0 & $^{+ 1.2}_{-1.1}$ & $^{+ 5.0}_{-3.5}$ & 0.0 &
$\pm 1.8$ & $\pm 2.1$ & $\pm 0.4$ & $\pm 1.0$ & 5.40 & 1.00 & 0.02\\
\hline\hline
  \end{tabular}
  \end{tiny}\vspace*{-1.5mm}
  \caption{ Measured jet cross section for $R=0.4$, $3.6\leq|y|<4.4$.
    See Table~\ref{tab:AntiKt4_EtaLow0_Results} for description of the
    columns.  All tables are available on \hepdata~\cite{HEPDATA}.  }
  \label{tab:AntiKt4_EtaLow3.6_Results}
\vspace*{-2mm}
\end{table*}

\begin{table*}
  \begin{tiny}
  \setlength{\extrarowheight}{4pt}
  \begin{tabular}{rrrrrrrrrrrrrrrrrrrrrrrrr}\hline\hline
  $\pt$-bin & NPC & \multicolumn{1}{c}{$\sigma$} & $\delta_{\rm stat}$
& $\gamma_{1}$ & $\gamma_{7}$ & $\gamma_{13}$ & $\gamma_{19}$ &
$\gamma_{25}$ & $\gamma_{31}$ & $\gamma_{32}$ & $\gamma_{38}$ &
$\gamma_{44}$ & $\gamma_{50}$ & $\gamma_{56}$ & $\gamma_{62}$ &
$\gamma_{68}$ & $\gamma_{74}$ & $\gamma_{75}$ & $\gamma_{76}$ &
$\gamma_{82}$ & $\gamma_{83}$ & $u_{1}$ & $u_{2}$ & $u_{3}$ \\
$[$GeV$]$ & & $[$pb/GeV$]$ & \% &\\ \hline $ 20$-$30$ & 1.61(30) &
$1.00\cdot 10^{ 7}$ & 0.57 & $^{+ 13}_{ -12}$ & $\pm 11$ & $^{+ 16}_{
-15}$ & $^{+ 9.7}_{-9.1}$ & $^{+ 5.7}_{-5.6}$ & 0.0 & $^{+
2.2}_{-2.3}$ & $^{+ 5.6}_{-5.4}$ & $^{+ 4.7}_{-5.1}$ & 0.0 & 0.0 &
$^{+ 2.1}_{-2.2}$ & 0.0 & 0.0 & $\pm 1.3$ & $\pm 3.3$ & 0.0 & $\pm
2.0$ & 0.69 & 1.00 & 0.33\\ $ 30$-$45$ & 1.30(10) & $1.23\cdot 10^{
6}$ & 1.21 & $^{+ 8.4}_{-7.3}$ & $^{+ 10}_{-8.9}$ & $^{+ 10}_{-8.7}$ &
$^{+ 8.0}_{-6.9}$ & $^{+ 2.4}_{-2.1}$ & 0.0 & $\pm 2.5$ & $^{+
4.1}_{-3.8}$ & $^{+ 4.4}_{-4.3}$ & 0.0 & 0.0 & $^{+ 3.5}_{-3.6}$ & 0.0
& 0.0 & $\pm 0.8$ & $\pm 3.0$ & 0.0 & $\pm 1.0$ & 0.47 & 1.00 & 0.22\\
$ 45$-$60$ & 1.18(6) & $2.19\cdot 10^{ 5}$ & 2.63 & $^{+ 2.1}_{-1.2}$
& $^{+ 8.4}_{-6.4}$ & $^{+ 5.1}_{-3.5}$ & $^{+ 4.6}_{-3.1}$ & $^{+
1.2}_{-0.5}$ & 0.0 & $^{+ 3.5}_{-2.4}$ & $^{+ 2.8}_{-1.8}$ & $^{+
4.9}_{-3.4}$ & $^{+ 0.2}_{+ 0.2}$ & $^{+ 0.4}_{+ 0.0}$ & $^{+
5.6}_{-4.1}$ & 0.0 & 0.0 & $\pm 1.0$ & $\pm 1.4$ & $\pm 0.1$ & $\pm
1.0$ & 0.30 & 1.00 & 0.19\\ $ 60$-$80$ & 1.12(4) & $5.45\cdot 10^{ 4}$
& 0.96 & $\mp 1.1$ & $^{+ 6.9}_{-5.9}$ & $^{+ 1.1}_{-0.8}$ & $^{+
3.7}_{-3.1}$ & $^{+ 0.8}_{-0.6}$ & 0.0 & $^{+ 3.2}_{-2.7}$ & $^{+
1.7}_{-1.3}$ & $^{+ 4.3}_{-3.6}$ & $^{+ 0.4}_{-0.3}$ & $^{+ 0.0}_{+
0.1}$ & $^{+ 4.6}_{-3.7}$ & $^{-0.1}_{+ 0.2}$ & 0.0 & $\pm 0.4$ & $\pm
1.3$ & 0.0 & $\pm 1.0$ & 0.28 & 1.00 & 0.15\\
\hline\hline
  \end{tabular}
  \end{tiny}\vspace*{-1.5mm}
  \caption{ Measured jet cross section for $R=0.6$, $|y|<0.3$.  NPC
    stands for multiplicative non-perturbative corrections with ${\rm
    error} \times 100$ in brackets, {\it i.e.} 1.25(10) means $1.25
    \pm 0.10$.  $\sigma$ is the measured cross section. $\delta_{\rm
    stat}$ is the statistical uncertainty. $\gamma_{i}$ and $u_{i}$
    are the correlated and uncorrelated systematic uncertainties, as
    described in Sec.~\ref{sec:corr} and
    Table~\ref{tab:correlations}. All uncertainties are given in $\%$.
    An overall luminosity uncertainty of $3.4\%$, which is applicable
    to all ATLAS data samples based on 2010 data, is not shown.
    All tables are available on \hepdata~\cite{HEPDATA}.  }
  \label{tab:AntiKt6_EtaLow0_Results}
\vspace*{-2mm}
\end{table*}

\begin{table*}
  \begin{tiny}
  \setlength{\extrarowheight}{4pt}
\begin{tabular}{rrrrrrrrrrrrrrrrrrrrrrrrr}\hline\hline
  $\pt$-bin & NPC & \multicolumn{1}{c}{$\sigma$} & $\delta_{\rm stat}$
& $\gamma_{1}$ & $\gamma_{7}$ & $\gamma_{13}$ & $\gamma_{19}$ &
$\gamma_{25}$ & $\gamma_{31}$ & $\gamma_{32}$ & $\gamma_{38}$ &
$\gamma_{44}$ & $\gamma_{50}$ & $\gamma_{56}$ & $\gamma_{62}$ &
$\gamma_{68}$ & $\gamma_{74}$ & $\gamma_{75}$ & $\gamma_{76}$ &
$\gamma_{82}$ & $\gamma_{83}$ & $u_{1}$ & $u_{2}$ & $u_{3}$ \\
$[$GeV$]$ & & $[$pb/GeV$]$ & \% &\\ \hline $ 20$-$30$ & 1.60(30) &
$9.95\cdot 10^{ 6}$ & 0.48 & $^{+ 12}_{ -11}$ & $^{+ 9.4}_{-9.5}$ &
$^{+ 14}_{ -13}$ & $^{+ 8.0}_{-8.3}$ & $^{+ 4.7}_{-4.6}$ & 0.0 & $^{+
2.2}_{-2.1}$ & $^{+ 4.4}_{-4.3}$ & $^{+ 4.7}_{-4.6}$ & 0.0 & 0.0 &
$^{+ 3.0}_{-2.9}$ & 0.0 & 0.0 & $\pm 0.9$ & $\pm 2.7$ & 0.0 & $\pm
2.0$ & 0.71 & 1.00 & 0.30\\ $ 30$-$45$ & 1.30(14) & $1.25\cdot 10^{
6}$ & 1.03 & $^{+ 8.6}_{-7.6}$ & $^{+ 9.9}_{-9.2}$ & $^{+ 10}_{-9.3}$
& $^{+ 7.9}_{-7.1}$ & $^{+ 3.0}_{-2.5}$ & 0.0 & $^{+ 2.6}_{-2.5}$ &
$^{+ 3.5}_{-3.2}$ & $^{+ 4.9}_{-4.6}$ & 0.0 & 0.0 & $^{+ 3.5}_{-3.2}$
& $^{-0.2}_{+ 0.1}$ & 0.0 & $\pm 0.7$ & $\pm 2.8$ & $\pm 0.1$ & $\pm
1.0$ & 0.47 & 1.00 & 0.20\\ $ 45$-$60$ & 1.17(8) & $2.25\cdot 10^{ 5}$
& 2.04 & $^{+ 2.2}_{-1.6}$ & $^{+ 8.9}_{-7.2}$ & $^{+ 7.4}_{-5.8}$ &
$^{+ 4.2}_{-3.4}$ & $^{+ 1.9}_{-1.3}$ & 0.0 & $^{+ 3.4}_{-2.7}$ & $^{+
2.1}_{-1.6}$ & $^{+ 4.7}_{-3.9}$ & $^{+ 0.2}_{+ 0.2}$ & $^{+ 0.4}_{+
0.0}$ & $^{+ 2.9}_{-2.4}$ & $^{+ 0.4}_{+ 0.1}$ & 0.0 & $\pm 0.5$ &
$\pm 1.6$ & $\pm 0.1$ & $\pm 1.0$ & 0.31 & 1.00 & 0.19\\ $ 60$-$80$ &
1.12(5) & $5.32\cdot 10^{ 4}$ & 0.81 & $^{-1.1}_{+ 1.0}$ & $^{+
6.4}_{-5.5}$ & $^{+ 4.0}_{-3.5}$ & $^{+ 2.7}_{-2.6}$ & $^{+
3.5}_{-3.4}$ & 0.0 & $^{+ 3.0}_{-2.9}$ & $^{+ 1.5}_{-1.3}$ & $^{+
4.7}_{-4.2}$ & $^{+ 0.4}_{-0.2}$ & 0.0 & $^{+ 0.9}_{-0.8}$ & $^{+
1.7}_{-1.5}$ & 0.0 & $\pm 0.3$ & $\pm 1.2$ & 0.0 & $\pm 1.0$ & 0.30 &
1.00 & 0.16\\
\hline\hline
  \end{tabular}
  \end{tiny}\vspace*{-1.5mm}
  \caption{ Measured jet cross section for $R=0.6$, $0.3\leq|y|<0.8$.
    See Table~\ref{tab:AntiKt6_EtaLow0_Results} for description of the
    columns.  All tables are available on \hepdata~\cite{HEPDATA}.
  }
  \label{tab:AntiKt6_EtaLow0.3_Results}
\vspace*{-2mm}
\end{table*}

\begin{table*}
  \begin{tiny}
  \setlength{\extrarowheight}{4pt}
  \begin{tabular}{rrrrrrrrrrrrrrrrrrrrrrrrr}\hline\hline
  $\pt$-bin & NPC & \multicolumn{1}{c}{$\sigma$} & $\delta_{\rm stat}$
& $\gamma_{2}$ & $\gamma_{8}$ & $\gamma_{14}$ & $\gamma_{20}$ &
$\gamma_{26}$ & $\gamma_{31}$ & $\gamma_{33}$ & $\gamma_{39}$ &
$\gamma_{45}$ & $\gamma_{51}$ & $\gamma_{57}$ & $\gamma_{63}$ &
$\gamma_{69}$ & $\gamma_{74}$ & $\gamma_{75}$ & $\gamma_{77}$ &
$\gamma_{82}$ & $\gamma_{83}$ & $u_{1}$ & $u_{2}$ & $u_{3}$ \\
$[$GeV$]$ & & $[$pb/GeV$]$ & \% &\\ \hline $ 20$-$30$ & 1.58(29) &
$8.54\cdot 10^{ 6}$ & 0.55 & $^{+ 10}_{ -11}$ & $^{+ 7.9}_{-8.6}$ &
$\pm 12$ & $^{+ 8.0}_{-8.4}$ & $^{+ 3.9}_{-4.1}$ & $\pm 4.2$ & $\pm
1.8$ & $^{+ 3.7}_{-3.8}$ & $^{+ 3.9}_{-4.0}$ & 0.0 & 0.0 & $^{+
2.5}_{-2.6}$ & 0.0 & 0.0 & $\pm 1.6$ & $\pm 8.1$ & $\pm 1.4$ & $\pm
2.0$ & 0.97 & 1.00 & 0.25\\ $ 30$-$45$ & 1.28(13) & $1.12\cdot 10^{
6}$ & 1.15 & $^{+ 8.3}_{-7.6}$ & $^{+ 9.6}_{-9.4}$ & $^{+ 10}_{-9.5}$
& $^{+ 5.3}_{-4.8}$ & $^{+ 3.0}_{-2.6}$ & $^{+ 5.3}_{-4.9}$ & $^{+
2.6}_{-2.5}$ & $^{+ 3.5}_{-3.2}$ & $^{+ 4.9}_{-4.6}$ & 0.0 & 0.0 &
$^{+ 3.5}_{-3.3}$ & $\mp 0.1$ & 0.0 & $\pm 0.4$ & $\pm 8.3$ & 0.0 &
$\pm 1.0$ & 0.84 & 1.00 & 0.20\\ $ 45$-$60$ & 1.17(7) & $1.99\cdot
10^{ 5}$ & 2.37 & $^{+ 2.2}_{-1.8}$ & $^{+ 9.2}_{-8.1}$ & $^{+
7.7}_{-6.5}$ & $^{+ 3.6}_{-2.7}$ & $^{+ 1.8}_{-1.5}$ & $^{+
4.9}_{-4.5}$ & $^{+ 3.3}_{-3.0}$ & $^{+ 2.0}_{-1.7}$ & $^{+
4.9}_{-4.5}$ & $^{+ 0.2}_{+ 0.2}$ & $^{+ 0.4}_{+ 0.0}$ & $^{+
2.8}_{-2.6}$ & $^{+ 0.3}_{-0.1}$ & $\pm 0.1$ & $\pm 0.3$ & $\pm 3.8$ &
$\pm 0.4$ & $\pm 1.0$ & 0.54 & 1.00 & 0.21\\ $ 60$-$80$ & 1.11(4) &
$4.84\cdot 10^{ 4}$ & 0.91 & $\mp 1.1$ & $^{+ 6.8}_{-6.1}$ & $^{+
4.2}_{-3.7}$ & $^{+ 2.0}_{-1.9}$ & $^{+ 3.7}_{-3.5}$ & $^{+
4.0}_{-3.6}$ & $^{+ 3.2}_{-3.1}$ & $\pm 1.5$ & $^{+ 5.0}_{-4.5}$ &
$^{+ 0.4}_{-0.3}$ & 0.0 & $\pm 0.9$ & $\pm 1.8$ & 0.0 & $\pm 0.2$ &
$\pm 1.5$ & $\pm 0.2$ & $\pm 1.0$ & 0.51 & 1.00 & 0.18\\
\hline\hline
  \end{tabular}
  \end{tiny}\vspace*{-1.5mm}
  \caption{ Measured jet cross section for $R=0.6$, $0.8\leq|y|<1.2$.
    See Table~\ref{tab:AntiKt6_EtaLow0_Results} for description of the
    columns.  All tables are available on \hepdata~\cite{HEPDATA}.
  }
  \label{tab:AntiKt6_EtaLow0.8_Results}
\vspace*{-2mm}
\end{table*}

\begin{table*}
  \begin{tiny}
  \setlength{\extrarowheight}{4pt}
  \begin{tabular}{rrrrrrrrrrrrrrrrrrrrrrrrr}\hline\hline
  $\pt$-bin & NPC & \multicolumn{1}{c}{$\sigma$} & $\delta_{\rm stat}$
& $\gamma_{3}$ & $\gamma_{9}$ & $\gamma_{15}$ & $\gamma_{21}$ &
$\gamma_{27}$ & $\gamma_{31}$ & $\gamma_{34}$ & $\gamma_{40}$ &
$\gamma_{46}$ & $\gamma_{52}$ & $\gamma_{58}$ & $\gamma_{64}$ &
$\gamma_{70}$ & $\gamma_{74}$ & $\gamma_{75}$ & $\gamma_{78}$ &
$\gamma_{82}$ & $\gamma_{83}$ & $u_{1}$ & $u_{2}$ & $u_{3}$ \\
$[$GeV$]$ & & $[$pb/GeV$]$ & \% &\\ \hline $ 20$-$30$ & 1.53(25) &
$7.65\cdot 10^{ 6}$ & 0.40 & $\pm 11$ & $^{+ 9.0}_{-9.2}$ & $\pm 13$ &
$^{+ 11}_{ -10}$ & $\pm 4.5$ & $\pm 14$ & $\pm 2.0$ & $^{+
4.1}_{-4.2}$ & $^{+ 4.5}_{-4.4}$ & 0.0 & 0.0 & $^{+ 2.7}_{-2.8}$ & 0.0
& 0.0 & $\pm 1.3$ & $\pm 11$ & $\pm 0.6$ & $\pm 2.0$ & 1.18 & 1.00 &
0.30\\ $ 30$-$45$ & 1.26(9) & $9.38\cdot 10^{ 5}$ & 0.72 & $^{+
8.6}_{-8.0}$ & $^{+ 10}_{-9.6}$ & $^{+ 11}_{-9.8}$ & $^{+ 5.8}_{-5.3}$
& $^{+ 3.1}_{-2.8}$ & $^{+ 15}_{ -13}$ & $^{+ 2.6}_{-2.7}$ & $^{+
3.5}_{-3.4}$ & $^{+ 4.9}_{-4.7}$ & 0.0 & 0.0 & $^{+ 3.5}_{-3.4}$ &
$^{-0.2}_{+ 0.1}$ & 0.0 & $\pm 0.6$ & $\pm 7.0$ & 0.0 & $\pm 1.0$ &
0.86 & 1.00 & 0.15\\ $ 45$-$60$ & 1.15(6) & $1.59\cdot 10^{ 5}$ & 1.73
& $^{+ 2.3}_{-1.6}$ & $^{+ 9.2}_{-7.8}$ & $^{+ 7.7}_{-6.3}$ & $^{+
3.8}_{-3.1}$ & $^{+ 1.9}_{-1.3}$ & $^{+ 10}_{-8.4}$ & $^{+
3.4}_{-2.8}$ & $^{+ 2.1}_{-1.6}$ & $^{+ 5.0}_{-4.1}$ & $^{+ 0.1}_{+
0.2}$ & $^{+ 0.3}_{+ 0.0}$ & $^{+ 3.0}_{-2.4}$ & $^{+ 0.3}_{+ 0.1}$ &
0.0 & $\pm 0.7$ & $\pm 2.1$ & 0.0 & $\pm 1.0$ & 0.54 & 1.00 & 0.13\\ $
60$-$80$ & 1.10(4) & $3.78\cdot 10^{ 4}$ & 0.66 & $\mp 1.0$ & $^{+
6.9}_{-6.0}$ & $^{+ 4.3}_{-3.8}$ & $^{+ 1.9}_{-1.6}$ & $^{+
3.7}_{-3.4}$ & $^{+ 7.1}_{-6.1}$ & $^{+ 3.2}_{-2.9}$ & $^{+
1.6}_{-1.4}$ & $^{+ 5.1}_{-4.7}$ & $^{+ 0.4}_{-0.3}$ & 0.0 & $\pm 0.9$
& $^{+ 1.9}_{-1.7}$ & $\pm 0.1$ & $\pm 0.2$ & $\pm 1.6$ & 0.0 & $\pm
1.0$ & 0.50 & 1.00 & 0.10\\
\hline\hline
  \end{tabular}
  \end{tiny}\vspace*{-1.5mm}
  \caption{ Measured jet cross section for $R=0.6$, $1.2\leq|y|<2.1$.
    See Table~\ref{tab:AntiKt6_EtaLow0_Results} for description of the
    columns.  All tables are available on \hepdata~\cite{HEPDATA}.  }
  \label{tab:AntiKt6_EtaLow1.2_Results}
\vspace*{-2mm}
\end{table*}

\begin{table*}
  \begin{tiny}
  \setlength{\extrarowheight}{4pt}
  \begin{tabular}{rrrrrrrrrrrrrrrrrrrrrrrrr}\hline\hline
  $\pt$-bin & NPC & \multicolumn{1}{c}{$\sigma$} & $\delta_{\rm stat}$
& $\gamma_{4}$ & $\gamma_{10}$ & $\gamma_{16}$ & $\gamma_{22}$ &
$\gamma_{28}$ & $\gamma_{31}$ & $\gamma_{35}$ & $\gamma_{41}$ &
$\gamma_{47}$ & $\gamma_{53}$ & $\gamma_{59}$ & $\gamma_{65}$ &
$\gamma_{71}$ & $\gamma_{74}$ & $\gamma_{75}$ & $\gamma_{79}$ &
$\gamma_{82}$ & $\gamma_{84}$ & $u_{1}$ & $u_{2}$ & $u_{3}$ \\
$[$GeV$]$ & & $[$pb/GeV$]$ & \% &\\ \hline $ 20$-$30$ & 1.44(19) &
$5.82\cdot 10^{ 6}$ & 0.54 & $^{+ 13}_{ -11}$ & $^{+ 9.9}_{-9.3}$ &
$^{+ 15}_{ -13}$ & $^{+ 5.7}_{-5.4}$ & $^{+ 4.9}_{-4.8}$ & $^{+ 22}_{
-20}$ & $^{+ 2.2}_{-2.1}$ & $^{+ 4.5}_{-4.4}$ & $^{+ 4.8}_{-4.6}$ &
0.0 & 0.0 & $^{+ 3.0}_{-2.9}$ & 0.0 & 0.0 & $\pm 0.4$ & $\pm 17$ &
$\pm 0.5$ & $\pm 2.0$ & 1.42 & 1.00 & 0.02\\ $ 30$-$45$ & 1.21(8) &
$6.89\cdot 10^{ 5}$ & 1.02 & $^{+ 8.9}_{-8.0}$ & $^{+ 10}_{-9.8}$ &
$^{+ 11}_{-9.9}$ & $^{+ 5.6}_{-5.4}$ & $^{+ 3.1}_{-2.8}$ & $^{+ 23}_{
-18}$ & $^{+ 2.6}_{-2.7}$ & $\pm 3.5$ & $^{+ 5.0}_{-4.8}$ & 0.0 & 0.0
& $\pm 3.5$ & $^{-0.2}_{+ 0.1}$ & 0.0 & $\pm 0.5$ & $\pm 7.7$ & $\pm
0.2$ & $\pm 1.0$ & 0.99 & 1.00 & 0.02\\ $ 45$-$60$ & 1.12(6) &
$1.19\cdot 10^{ 5}$ & 2.32 & $^{+ 2.7}_{-2.0}$ & $^{+ 9.4}_{-8.0}$ &
$^{+ 7.9}_{-6.5}$ & $^{+ 4.1}_{-3.5}$ & $^{+ 2.1}_{-1.5}$ & $^{+ 14}_{
-11}$ & $^{+ 3.6}_{-3.0}$ & $^{+ 2.2}_{-1.7}$ & $^{+ 5.0}_{-4.4}$ &
$^{+ 0.2}_{+ 0.2}$ & $^{+ 0.4}_{+ 0.0}$ & $^{+ 3.2}_{-2.5}$ & $^{+
0.3}_{-0.1}$ & $\pm 0.2$ & $\pm 0.3$ & $\pm 2.4$ & $\pm 0.2$ & $\pm
1.0$ & 0.59 & 1.00 & 0.02\\ $ 60$-$80$ & 1.08(4) & $2.42\cdot 10^{ 4}$
& 0.87 & $^{-0.9}_{+ 0.8}$ & $^{+ 7.0}_{-5.8}$ & $^{+ 4.5}_{-3.7}$ &
$^{+ 2.4}_{-1.8}$ & $^{+ 4.1}_{-3.3}$ & $^{+ 8.5}_{-6.9}$ & $^{+
3.5}_{-2.7}$ & $^{+ 2.1}_{-1.5}$ & $^{+ 5.2}_{-4.5}$ & $^{+
0.6}_{-0.3}$ & 0.0 & $^{+ 1.2}_{-0.8}$ & $^{+ 2.1}_{-1.5}$ & $\pm 0.2$
& $\pm 0.3$ & $\pm 1.5$ & 0.0 & $\pm 1.0$ & 0.55 & 1.00 & 0.02\\
\hline\hline
  \end{tabular}
  \end{tiny}\vspace*{-1.5mm}
  \caption{ Measured jet cross section for $R=0.6$, $2.1\leq|y|<2.8$.
    See Table~\ref{tab:AntiKt6_EtaLow0_Results} for description of the
    columns.  All tables are available on \hepdata~\cite{HEPDATA}. }
  \label{tab:AntiKt6_EtaLow2.1_Results}
\vspace*{-2mm}
\end{table*}

\begin{table*}
  \begin{tiny}
  \setlength{\extrarowheight}{4pt}
  \begin{tabular}{rrrrrrrrrrrrrrrrrrrrrrrrr}\hline\hline
  $\pt$-bin & NPC & \multicolumn{1}{c}{$\sigma$} & $\delta_{\rm stat}$
& $\gamma_{5}$ & $\gamma_{11}$ & $\gamma_{17}$ & $\gamma_{23}$ &
$\gamma_{29}$ & $\gamma_{31}$ & $\gamma_{36}$ & $\gamma_{42}$ &
$\gamma_{48}$ & $\gamma_{54}$ & $\gamma_{60}$ & $\gamma_{66}$ &
$\gamma_{72}$ & $\gamma_{74}$ & $\gamma_{75}$ & $\gamma_{80}$ &
$\gamma_{82}$ & $\gamma_{85}$ & $u_{1}$ & $u_{2}$ & $u_{3}$ \\
$[$GeV$]$ & & $[$pb/GeV$]$ & \% &\\ \hline $ 20$-$30$ & 1.32(15) &
$3.90\cdot 10^{ 6}$ & 0.61 & $^{+ 14}_{ -12}$ & $^{+ 11}_{-9.8}$ &
$^{+ 16}_{ -14}$ & $^{+ 8.6}_{-7.8}$ & $^{+ 5.2}_{-4.9}$ & $^{+ 35}_{
-28}$ & $^{+ 2.3}_{-2.2}$ & $^{+ 4.9}_{-4.5}$ & $^{+ 5.2}_{-4.8}$ &
0.0 & 0.0 & $^{+ 3.3}_{-3.0}$ & 0.0 & 0.0 & $\pm 0.4$ & $\pm 16$ &
$\pm 0.1$ & $\pm 2.0$ & 4.37 & 1.00 & 0.00\\ $ 30$-$45$ & 1.15(9) &
$4.26\cdot 10^{ 5}$ & 1.42 & $^{+ 9.0}_{-7.8}$ & $^{+ 11}_{-10.0}$ &
$^{+ 11}_{-9.9}$ & $^{+ 9.4}_{-8.0}$ & $^{+ 3.1}_{-2.8}$ & $^{+ 37}_{
-26}$ & $\pm 2.8$ & $^{+ 3.5}_{-3.6}$ & $^{+ 5.3}_{-5.2}$ & 0.0 & 0.0
& $^{+ 3.6}_{-3.7}$ & $\mp 0.1$ & 0.0 & $\pm 0.8$ & $\pm 9.7$ & $\pm
0.1$ & $\pm 1.0$ & 2.99 & 1.00 & 0.00\\ $ 45$-$60$ & 1.09(6) &
$6.42\cdot 10^{ 4}$ & 3.20 & $^{+ 2.1}_{-2.4}$ & $^{+ 8.8}_{-9.4}$ &
$^{+ 7.2}_{-7.9}$ & $^{+ 7.1}_{-7.8}$ & $^{+ 2.1}_{-2.3}$ & $^{+ 24}_{
-19}$ & $^{+ 3.5}_{-3.8}$ & $^{+ 2.2}_{-2.3}$ & $^{+ 4.8}_{-5.5}$ &
$^{+ 0.1}_{+ 0.2}$ & $^{+ 0.3}_{+ 0.0}$ & $^{+ 3.0}_{-3.3}$ & $^{+
0.5}_{-0.3}$ & $\pm 0.2$ & $\pm 0.4$ & $\pm 1.3$ & 0.0 & $\pm 1.0$ &
1.83 & 1.00 & 0.00\\ $ 60$-$80$ & 1.05(5) & $1.24\cdot 10^{ 4}$ & 1.36
& $^{-1.2}_{+ 1.1}$ & $^{+ 7.5}_{-6.6}$ & $^{+ 4.4}_{-4.0}$ & $^{+
5.9}_{-5.3}$ & $^{+ 4.1}_{-3.7}$ & $^{+ 17}_{ -13}$ & $^{+
3.7}_{-3.2}$ & $^{+ 2.0}_{-1.4}$ & $^{+ 5.6}_{-4.9}$ & $^{+
0.6}_{-0.4}$ & 0.0 & $^{+ 1.0}_{-0.8}$ & $^{+ 2.5}_{-2.0}$ & $\pm 0.3$
& $\pm 0.5$ & $\pm 2.3$ & 0.0 & $\pm 1.0$ & 3.47 & 1.00 & 0.00\\
\hline\hline
  \end{tabular}
  \end{tiny}\vspace*{-1.5mm}
  \caption{ Measured jet cross section for $R=0.6$, $2.8\leq|y|<3.6$.
    See Table~\ref{tab:AntiKt6_EtaLow0_Results} for description of the
    columns.  All tables are available on \hepdata~\cite{HEPDATA}.  }
  \label{tab:AntiKt6_EtaLow2.8_Results}
\vspace*{-2mm}
\end{table*}

\begin{table*}
  \begin{tiny}
  \setlength{\extrarowheight}{4pt}
  \begin{tabular}{rrrrrrrrrrrrrrrrrrrrrrrrr}\hline\hline
  $\pt$-bin & NPC & \multicolumn{1}{c}{$\sigma$} & $\delta_{\rm stat}$
& $\gamma_{6}$ & $\gamma_{12}$ & $\gamma_{18}$ & $\gamma_{24}$ &
$\gamma_{30}$ & $\gamma_{31}$ & $\gamma_{37}$ & $\gamma_{43}$ &
$\gamma_{49}$ & $\gamma_{55}$ & $\gamma_{61}$ & $\gamma_{67}$ &
$\gamma_{73}$ & $\gamma_{74}$ & $\gamma_{75}$ & $\gamma_{81}$ &
$\gamma_{82}$ & $\gamma_{86}$ & $u_{1}$ & $u_{2}$ & $u_{3}$ \\
$[$GeV$]$ & & $[$pb/GeV$]$ & \% &\\ \hline $ 20$-$30$ & 1.19(14) &
$1.51\cdot 10^{ 6}$ & 0.81 & $^{+ 14}_{ -13}$ & $\pm 11$ & $^{+ 17}_{
-15}$ & $^{+ 3.1}_{-3.2}$ & $^{+ 5.3}_{-5.5}$ & $^{+ 60}_{ -42}$ &
$^{+ 2.4}_{-2.5}$ & $^{+ 5.0}_{-5.1}$ & $^{+ 5.3}_{-5.4}$ & 0.0 & 0.0
& $^{+ 3.2}_{-3.4}$ & 0.0 & 0.0 & $\pm 0.3$ & $\pm 13$ & $\pm 1.2$ &
$\pm 2.0$ & 5.01 & 1.00 & 0.00\\ $ 30$-$45$ & 1.07(10) & $1.38\cdot
10^{ 5}$ & 0.86 & $^{+ 10}_{-8.9}$ & $^{+ 13}_{ -11}$ & $^{+ 13}_{
-11}$ & $^{+ 3.0}_{-2.8}$ & $^{+ 3.6}_{-2.9}$ & $^{+ 65}_{ -37}$ &
$^{+ 3.3}_{-3.0}$ & $^{+ 4.1}_{-3.8}$ & $^{+ 6.3}_{-5.7}$ & 0.0 & 0.0
& $^{+ 4.1}_{-3.9}$ & 0.0 & 0.0 & $\pm 0.9$ & $\pm 6.2$ & $\pm 0.7$ &
$\pm 1.0$ & 9.22 & 1.00 & 0.00\\ $ 45$-$60$ & 1.02(8) & $1.29\cdot
10^{ 4}$ & 2.08 & $^{+ 2.5}_{-2.4}$ & $^{+ 12}_{ -11}$ & $^{+
9.8}_{-9.4}$ & $^{+ 2.0}_{-1.8}$ & $^{+ 2.9}_{-2.6}$ & $^{+ 39}_{
-26}$ & $^{+ 5.0}_{-4.7}$ & $^{+ 2.5}_{-2.6}$ & $\pm 6.8$ & $^{+
0.6}_{+ 0.2}$ & $^{+ 0.7}_{+ 0.0}$ & $^{+ 3.7}_{-3.3}$ & $^{+
1.0}_{-0.2}$ & $\pm 0.4$ & $\pm 0.5$ & $\pm 1.5$ & $\pm 0.3$ & $\pm
1.0$ & 5.64 & 1.00 & 0.00\\ $ 60$-$80$ & 1.00(7) & $1.27\cdot 10^{ 3}$
& 2.23 & $^{-1.3}_{+ 1.2}$ & $^{+ 8.9}_{ -11}$ & $^{+ 5.9}_{-7.4}$ &
$^{+ 0.8}_{-2.6}$ & $^{+ 5.0}_{-7.0}$ & $^{+ 24}_{ -19}$ & $^{+
4.4}_{-6.4}$ & $^{+ 2.0}_{-4.0}$ & $^{+ 7.7}_{-8.8}$ & $^{+
0.3}_{-1.7}$ & 0.0 & $^{+ 0.5}_{-2.2}$ & $^{+ 3.5}_{-4.9}$ & $\pm 0.3$
& $\pm 0.6$ & $\pm 4.3$ & $\pm 0.8$ & $\pm 1.0$ & 5.27 & 1.00 & 0.02\\
\hline\hline
  \end{tabular}
  \end{tiny}\vspace*{-1.5mm}
  \caption{ Measured jet cross section for $R=0.6$, $3.6\leq|y|<4.4$.
    See Table~\ref{tab:AntiKt6_EtaLow0_Results} for description of the
    columns.  All tables are available on \hepdata~\cite{HEPDATA}.}
  \label{tab:AntiKt6_EtaLow3.6_Results}
\vspace*{-2mm}
\end{table*}

\begin{table*}
\section{Dijet Tables}
\label{sec:tablesDijet}
\begin{tiny}
\setlength{\extrarowheight}{4pt}
\begin{tabular}{ccccccccccccccccccccccccc}
\hline\hline $\twomass{1}{2}$-bin & NPC & $\sigma$ & $\delta_{\rm
stat}$ & $\gamma_{1}$ & $\gamma_{7}$ & $\gamma_{13}$ & $\gamma_{19}$ &
$\gamma_{25}$ & $\gamma_{31}$ & $\gamma_{32}$ & $\gamma_{38}$ &
$\gamma_{44}$ & $\gamma_{50}$ & $\gamma_{56}$ & $\gamma_{62}$ &
$\gamma_{68}$ & $\gamma_{74}$ & $\gamma_{75}$ & $\gamma_{76}$ &
$\gamma_{82}$ & $\gamma_{83}$ & $u_{1}$ & $u_{2}$ & $u_{3}$ \\
$[$TeV$]$ & & $[$pb/TeV$]$ & \% & \\ \hline 0.07-0.11 & 1.04 ( 8 ) &
$2.4\cdot 10^{8}$ & 1.3 & $^{+4.0}_{-4.1}$ & $^{+8.2}_{-7.7}$ &
$^{+5.9}_{-5.8}$ & $^{+4.3}_{-4.2}$ & $^{+1.5}_{-1.7}$ &
$^{+9.8}_{-8.5}$ & $^{+2.0}_{-2.3}$ & $^{+2.5}_{-2.7}$ &
$^{+3.7}_{-3.9}$ & 0.0 & 0.02 & $^{+2.7}_{-2.8}$ & $^{+0.15}_{-0.14}$
& 0.0 & N/A & 2.1 & 0.01 & 2.1 & $^{+2.4}_{-2.7}$ & 1.0 & 1.0 \\
0.11-0.16 & 1.03 ( 5 ) & $3.4\cdot 10^{7}$ & 0.67 & $^{+0.95}_{-0.93}$
& $^{+6.4}_{-6.2}$ & $^{+4.9}_{-4.8}$ & 2.7 & 1.9 & $^{+5.6}_{-4.9}$ &
2.6 & $^{+1.7}_{-1.6}$ & $^{+4.1}_{-4.0}$ & $^{+0.18}_{-0.24}$ &
$^{+0.0}_{-0.03}$ & $^{+1.9}_{-1.7}$ & $^{+0.81}_{-0.87}$ & 0.04 & N/A
& 0.76 & 0.12 & 2.0 & $^{+1.6}_{-1.7}$ & 1.0 & 1.0 \\ 0.16-0.21 & 1.02
( 4 ) & $7.1\cdot 10^{6}$ & 0.52 & $^{+0.35}_{-0.33}$ &
$^{+6.0}_{-5.5}$ & $^{+3.9}_{-3.7}$ & $^{+2.2}_{-2.0}$ &
$^{+1.3}_{-1.2}$ & $^{+3.3}_{-3.0}$ & $^{+2.8}_{-2.5}$ &
$^{+1.3}_{-1.1}$ & $^{+4.5}_{-4.1}$ & $^{+0.85}_{-0.77}$ &
$^{+0.06}_{-0.04}$ & $^{+0.49}_{-0.47}$ & $^{+2.5}_{-2.4}$ & 0.04 &
N/A & 0.61 & 0.1 & 2.0 & $^{+1.2}_{-1.1}$ & 1.0 & 1.0 \\ 0.21-0.26 &
1.02 ( 4 ) & $2.0\cdot 10^{6}$ & 0.88 & 0.12 & $^{+5.7}_{-5.6}$ &
$^{+3.7}_{-4.0}$ & 2.2 & $^{+0.79}_{-0.84}$ & $^{+2.2}_{-2.3}$ & 2.6 &
$^{+0.99}_{-0.98}$ & $^{+4.7}_{-4.8}$ & $^{+1.3}_{-1.4}$ &
$^{+0.04}_{-0.07}$ & $^{+0.03}_{-0.02}$ & $^{+3.4}_{-3.7}$ & 0.2 & N/A
& 0.34 & 0.04 & 2.0 & $^{+1.0}_{-0.98}$ & 1.0 & 1.0 \\
\hline\hline
\end{tabular}
\end{tiny}\vspace*{-1.5mm}
\caption{ Measured dijet cross section for $R=0.4$ and $ y^{*} < 0.5$.
NPC stands for multiplicative non-perturbative corrections with ${\rm
error} \times 100$ in brackets, {\it i.e.} 1.25(10) means $1.25 \pm
0.10$.  $\sigma$ is the measured cross section. $\delta_{\rm stat}$ is
the statistical uncertainty. $\gamma_{i}$ and $u_{i}$ are the
correlated and uncorrelated systematic uncertainties, as described in
Sec.~\ref{sec:corr} and Table~\ref{tab:correlationsDijet}. All
uncertainties are given in $\%$.  An overall luminosity uncertainty of
$3.4\%$, which is applicable to all ATLAS data samples based on 2010
data, is not shown.  All tables are available on \hepdata~\cite{HEPDATA}.  }
\label{tab:DijetMassResults04_0}
\vspace*{-2mm}
\end{table*}


\begin{table*}
\begin{tiny}
\setlength{\extrarowheight}{4pt}
\begin{tabular}{ccccccccccccccccccccccccc}
\hline\hline $\twomass{1}{2}$-bin & NPC & $\sigma$ & $\delta_{\rm
stat}$ & $\gamma_{1}$ & $\gamma_{7}$ & $\gamma_{13}$ & $\gamma_{19}$ &
$\gamma_{25}$ & $\gamma_{31}$ & $\gamma_{32}$ & $\gamma_{38}$ &
$\gamma_{44}$ & $\gamma_{50}$ & $\gamma_{56}$ & $\gamma_{62}$ &
$\gamma_{68}$ & $\gamma_{74}$ & $\gamma_{75}$ & $\gamma_{76}$ &
$\gamma_{82}$ & $\gamma_{83}$ & $u_{1}$ & $u_{2}$ & $u_{3}$ \\
$[$TeV$]$ & & $[$pb/TeV$]$ & \% & \\ \hline 0.11-0.16 & 1.04 ( 6 ) &
$7.9\cdot 10^{7}$ & 1.2 & $^{+3.9}_{-3.5}$ & $^{+7.8}_{-7.2}$ &
$^{+6.2}_{-5.6}$ & $^{+4.2}_{-3.8}$ & $^{+1.9}_{-1.7}$ &
$^{+10}_{-8.3}$ & $^{+2.5}_{-2.3}$ & $^{+2.6}_{-2.3}$ &
$^{+4.1}_{-3.8}$ & $^{+0.02}_{-0.0}$ & $^{+0.02}_{-0.0}$ &
$^{+2.7}_{-2.5}$ & $^{+0.04}_{-0.0}$ & 0.05 & N/A & 2.1 & 0.25 & 2.1 &
$^{+2.7}_{-2.3}$ & 1.0 & 1.0 \\ 0.16-0.21 & 1.03 ( 5 ) & $1.6\cdot
10^{7}$ & 0.77 & $^{+0.48}_{-0.63}$ & 6.2 & $^{+4.4}_{-4.6}$ &
$^{+2.0}_{-2.7}$ & $^{+1.8}_{-2.5}$ & $^{+4.4}_{-4.9}$ &
$^{+2.9}_{-3.5}$ & $^{+1.2}_{-1.9}$ & $^{+4.3}_{-4.7}$ &
$^{+0.4}_{-0.93}$ & $^{+0.0}_{-0.04}$ & $^{+1.3}_{-1.9}$ &
$^{+1.4}_{-2.2}$ & 0.1 & N/A & 0.0 & 0.0 & 1.9 & $^{+1.0}_{-2.3}$ &
1.0 & 1.0 \\ 0.21-0.26 & 1.03 ( 4 ) & $4.7\cdot 10^{6}$ & 0.62 &
$^{+0.39}_{-0.45}$ & $^{+6.0}_{-5.9}$ & $^{+4.0}_{-4.1}$ &
$^{+2.2}_{-2.3}$ & $^{+1.2}_{-1.3}$ & $^{+3.4}_{-3.6}$ & 2.8 & 1.1 &
$^{+4.5}_{-4.6}$ & $^{+0.8}_{-0.87}$ & $^{+0.04}_{-0.03}$ &
$^{+0.43}_{-0.39}$ & $^{+2.5}_{-2.7}$ & 0.06 & N/A & 0.26 & 0.0 & 2.0
& $^{+1.1}_{-1.4}$ & 1.0 & 1.0 \\ 0.26-0.31 & 1.02 ( 3 ) & $1.7\cdot
10^{6}$ & 0.66 & 0.11 & $^{+6.2}_{-6.1}$ & $^{+4.2}_{-4.1}$ &
$^{+2.4}_{-2.2}$ & $^{+0.96}_{-0.67}$ & $^{+2.9}_{-3.1}$ &
$^{+2.7}_{-2.6}$ & $^{+1.1}_{-0.88}$ & 5.0 & $^{+1.3}_{-1.1}$ &
$^{+0.08}_{-0.05}$ & $^{+0.11}_{-0.0}$ & $^{+3.6}_{-3.7}$ & 0.04 & N/A
& 0.6 & 0.0 & 2.0 & $^{+1.2}_{-0.96}$ & 1.0 & 1.0 \\
\hline\hline
\end{tabular}
\end{tiny}\vspace*{-1.5mm}
\caption{ Measured dijet cross section for $R=0.4$ and $0.5 \leq y^{*}
< 1.0$.  See Table~\ref{tab:DijetMassResults04_0} for a description of
the columns.  All tables are available on \hepdata~\cite{HEPDATA}.  }
\label{tab:DijetMassResults04_1}
\vspace*{-2mm}
\end{table*}


\begin{table*}
\begin{tiny}
\setlength{\extrarowheight}{4pt}
\begin{tabular}{ccccccccccccccccccccccccc}
\hline\hline $\twomass{1}{2}$-bin & NPC & $\sigma$ & $\delta_{\rm
stat}$ & $\gamma_{2}$ & $\gamma_{8}$ & $\gamma_{14}$ & $\gamma_{20}$ &
$\gamma_{26}$ & $\gamma_{31}$ & $\gamma_{33}$ & $\gamma_{39}$ &
$\gamma_{45}$ & $\gamma_{51}$ & $\gamma_{57}$ & $\gamma_{63}$ &
$\gamma_{69}$ & $\gamma_{74}$ & $\gamma_{75}$ & $\gamma_{77}$ &
$\gamma_{82}$ & $\gamma_{83}$ & $u_{1}$ & $u_{2}$ & $u_{3}$ \\
$[$TeV$]$ & & $[$pb/TeV$]$ & \% & \\ \hline 0.16-0.21 & 1.04 ( 6 ) &
$4.3\cdot 10^{7}$ & 1.5 & $^{+4.2}_{-3.9}$ & $^{+8.3}_{-7.8}$ &
$^{+6.5}_{-6.1}$ & $^{+4.4}_{-4.1}$ & $^{+2.0}_{-1.9}$ &
$^{+13}_{-11}$ & 2.5 & $^{+2.7}_{-2.6}$ & $^{+4.2}_{-4.0}$ &
$^{+0.08}_{-0.07}$ & 0.01 & 2.8 & $^{+0.1}_{-0.04}$ & 0.04 & N/A & 2.3
& 0.2 & 2.1 & $^{+3.1}_{-2.8}$ & 2.0 & 1.0 \\ 0.21-0.26 & 1.04 ( 5 ) &
$1.3\cdot 10^{7}$ & 1.1 & $^{+0.37}_{-0.61}$ & 6.5 & $^{+5.0}_{-5.3}$
& $^{+2.1}_{-2.3}$ & $^{+1.4}_{-1.7}$ & $^{+7.9}_{-7.4}$ &
$^{+2.8}_{-2.9}$ & $^{+1.4}_{-1.5}$ & $^{+4.2}_{-4.6}$ &
$^{+0.26}_{-0.36}$ & $^{+0.03}_{-0.04}$ & 1.8 & $^{+0.71}_{-0.85}$ &
0.0 & N/A & 0.22 & 0.12 & 1.9 & $^{+1.5}_{-2.1}$ & 2.0 & 1.0 \\
0.26-0.31 & 1.03 ( 4 ) & $4.8\cdot 10^{6}$ & 1.0 & $^{+0.04}_{-0.09}$
& $^{+7.0}_{-6.4}$ & $^{+4.8}_{-4.5}$ & $^{+2.5}_{-2.2}$ &
$^{+1.8}_{-1.5}$ & $^{+6.7}_{-5.7}$ & $^{+3.0}_{-2.6}$ &
$^{+1.3}_{-1.1}$ & $^{+4.7}_{-4.4}$ & $^{+0.49}_{-0.32}$ &
$^{+0.06}_{-0.02}$ & $^{+1.1}_{-0.93}$ & $^{+1.8}_{-1.6}$ & 0.1 & N/A
& 0.76 & 0.12 & 2.0 & $^{+1.4}_{-1.5}$ & 2.0 & 1.0 \\ 0.31-0.37 & 1.03
( 4 ) & $1.9\cdot 10^{6}$ & 0.95 & 0.14 & $^{+6.1}_{-5.8}$ &
$^{+4.5}_{-4.0}$ & $^{+2.5}_{-2.2}$ & $^{+1.6}_{-1.4}$ &
$^{+5.1}_{-4.6}$ & $^{+3.1}_{-2.5}$ & $^{+1.5}_{-1.2}$ &
$^{+4.7}_{-4.4}$ & $^{+0.88}_{-0.8}$ & $^{+0.05}_{-0.03}$ &
$^{+0.57}_{-0.36}$ & $^{+2.8}_{-2.6}$ & 0.3 & N/A & 0.8 & 0.0 & 2.0 &
$^{+1.7}_{-1.4}$ & 2.0 & 1.0 \\
\hline\hline
\end{tabular}
\end{tiny}\vspace*{-1.5mm}
\caption{ Measured dijet cross section for $R=0.4$ and $1.0 \leq y^{*}
< 1.5$.  See Table~\ref{tab:DijetMassResults04_0} for a description of
the columns.  All tables are available on \hepdata~\cite{HEPDATA}.}
\label{tab:DijetMassResults04_2}
\vspace*{-2mm}
\end{table*}


\begin{table*}
\begin{tiny}
\setlength{\extrarowheight}{4pt}
\begin{tabular}{ccccccccccccccccccccccccc}
\hline\hline $\twomass{1}{2}$-bin & NPC & $\sigma$ & $\delta_{\rm
stat}$ & $\gamma_{3}$ & $\gamma_{9}$ & $\gamma_{15}$ & $\gamma_{21}$ &
$\gamma_{27}$ & $\gamma_{31}$ & $\gamma_{34}$ & $\gamma_{40}$ &
$\gamma_{46}$ & $\gamma_{52}$ & $\gamma_{58}$ & $\gamma_{64}$ &
$\gamma_{70}$ & $\gamma_{74}$ & $\gamma_{75}$ & $\gamma_{78}$ &
$\gamma_{82}$ & $\gamma_{83}$ & $u_{1}$ & $u_{2}$ & $u_{3}$ \\
$[$TeV$]$ & & $[$pb/TeV$]$ & \% & \\ \hline 0.26-0.31 & 1.04 ( 6 ) &
$1.3\cdot 10^{7}$ & 2.6 & $^{+4.2}_{-3.6}$ & $^{+8.8}_{-8.6}$ &
$^{+6.7}_{-6.8}$ & $^{+4.1}_{-3.8}$ & $^{+1.9}_{-2.1}$ &
$^{+18}_{-15}$ & $^{+2.8}_{-2.7}$ & $^{+2.8}_{-2.5}$ &
$^{+4.5}_{-4.4}$ & $^{+0.1}_{-0.01}$ & $^{+0.02}_{-0.0}$ &
$^{+3.2}_{-2.9}$ & $^{+0.14}_{-0.07}$ & 0.07 & N/A & 2.6 & 0.22 &
$^{+2.1}_{-2.0}$ & $^{+3.7}_{-3.2}$ & 2.0 & 1.0 \\ 0.31-0.37 & 1.03 (
5 ) & $5.4\cdot 10^{6}$ & 1.9 & $^{+2.0}_{-1.7}$ & $^{+8.4}_{-6.6}$ &
$^{+7.3}_{-5.6}$ & $^{+3.6}_{-2.6}$ & $^{+2.2}_{-1.7}$ &
$^{+15}_{-11}$ & $^{+3.4}_{-2.6}$ & $^{+2.2}_{-1.9}$ &
$^{+5.4}_{-4.3}$ & $^{+0.17}_{-0.13}$ & $^{+0.01}_{-0.03}$ &
$^{+2.4}_{-2.1}$ & $^{+0.62}_{-0.52}$ & 0.01 & N/A & 1.6 & 0.04 & 2.0
& $^{+3.2}_{-2.5}$ & 2.0 & 1.0 \\ 0.37-0.44 & 1.03 ( 5 ) & $2.2\cdot
10^{6}$ & 1.7 & $^{+0.92}_{-0.67}$ & $^{+8.3}_{-6.9}$ &
$^{+6.4}_{-5.3}$ & $^{+2.7}_{-2.1}$ & 2.2 & $^{+13}_{-9.6}$ &
$^{+3.4}_{-2.7}$ & $^{+1.9}_{-1.7}$ & $^{+5.6}_{-4.8}$ &
$^{+0.65}_{-0.47}$ & 0.0 & $^{+1.6}_{-1.4}$ & $^{+1.7}_{-1.2}$ & 0.06
& N/A & 1.5 & 0.38 & 2.0 & $^{+2.5}_{-2.2}$ & 2.0 & 1.0 \\ 0.44-0.51 &
1.03 ( 4 ) & $8.9\cdot 10^{5}$ & 1.7 & $^{+0.0}_{-0.36}$ &
$^{+6.8}_{-7.6}$ & $^{+4.3}_{-5.6}$ & $^{+1.7}_{-2.5}$ &
$^{+1.4}_{-2.2}$ & $^{+9.3}_{-9.4}$ & $^{+2.4}_{-3.5}$ &
$^{+0.9}_{-1.7}$ & $^{+4.1}_{-5.7}$ & $^{+0.17}_{-0.85}$ &
$^{+0.03}_{-0.05}$ & $^{+0.39}_{-1.0}$ & $^{+1.6}_{-3.1}$ & 0.2 & N/A
& 0.45 & 0.0 & 2.0 & $^{+1.5}_{-2.4}$ & 2.0 & 1.0 \\
\hline\hline
\end{tabular}
\end{tiny}\vspace*{-1.5mm}
\caption{ Measured dijet cross section for $R=0.4$ and $1.5 \leq y^{*}
< 2.0$.  See Table~\ref{tab:DijetMassResults04_0} for a description of
the columns.  All tables are available on \hepdata~\cite{HEPDATA}.}
\label{tab:DijetMassResults04_3}
\vspace*{-2mm}
\end{table*}


\begin{table*}
\begin{tiny}
\setlength{\extrarowheight}{4pt}
\begin{tabular}{ccccccccccccccccccccccccc}
\hline\hline $\twomass{1}{2}$-bin & NPC & $\sigma$ & $\delta_{\rm
stat}$ & $\gamma{4}$ & $\gamma_{10}$ & $\gamma_{16}$ & $\gamma_{22}$ &
$\gamma_{28}$ & $\gamma_{31}$ & $\gamma_{35}$ & $\gamma_{41}$ &
$\gamma_{47}$ & $\gamma_{53}$ & $\gamma_{59}$ & $\gamma_{65}$ &
$\gamma_{71}$ & $\gamma_{74}$ & $\gamma_{75}$ & $\gamma_{79}$ &
$\gamma_{82}$ & $\gamma_{84}$ & $u_{1}$ & $u_{2}$ & $u_{3}$ \\
$[$TeV$]$ & & $[$pb/TeV$]$ & \% & \\ \hline 0.37-0.44 & 1.03 ( 7 ) &
$5.8\cdot 10^{6}$ & 3.9 & $^{+5.1}_{-5.2}$ & $^{+11}_{-9.3}$ &
$^{+7.9}_{-7.5}$ & $^{+4.1}_{-4.5}$ & $^{+2.1}_{-2.7}$ &
$^{+30}_{-22}$ & $^{+2.8}_{-3.2}$ & $^{+3.0}_{-3.5}$ & 5.3 &
$^{+0.0}_{-0.08}$ & $^{+0.02}_{-0.08}$ & $^{+3.0}_{-3.7}$ &
$^{+0.12}_{-0.07}$ & 0.06 & N/A & 4.8 & 0.0 & 2.1 & $^{+5.7}_{-4.3}$ &
1.0 & $^{+1.6}_{-0.44}$ \\ 0.44-0.51 & 1.03 ( 6 ) & $2.6\cdot 10^{6}$
& 3.7 & $^{+4.7}_{-3.9}$ & $^{+11}_{-9.3}$ & $^{+8.6}_{-7.6}$ &
$^{+4.3}_{-3.8}$ & $^{+2.4}_{-2.5}$ & $^{+28}_{-20}$ &
$^{+4.0}_{-3.4}$ & $^{+3.5}_{-3.0}$ & $^{+6.0}_{-5.3}$ &
$^{+0.28}_{-0.24}$ & $^{+0.03}_{-0.02}$ & $^{+3.8}_{-3.2}$ &
$^{+0.44}_{-0.4}$ & 0.1 & N/A & 3.7 & 0.07 & 2.0 & $^{+5.2}_{-4.3}$ &
1.0 & $^{+1.1}_{-0.92}$ \\ 0.51-0.59 & 1.03 ( 6 ) & $1.1\cdot 10^{6}$
& 2.7 & $^{+2.4}_{-2.0}$ & $^{+9.4}_{-7.9}$ & $^{+7.6}_{-6.6}$ &
$^{+3.7}_{-2.6}$ & $^{+2.4}_{-2.2}$ & $^{+21}_{-16}$ &
$^{+3.5}_{-3.3}$ & $^{+2.7}_{-2.1}$ & $^{+5.9}_{-4.9}$ &
$^{+0.39}_{-0.17}$ & $^{+0.02}_{-0.0}$ & $^{+2.7}_{-2.2}$ &
$^{+0.96}_{-0.92}$ & 0.05 & N/A & 2.1 & 0.0 & 2.0 & $^{+3.6}_{-3.7}$ &
1.0 & $^{+0.9}_{-1.1}$ \\ 0.59-0.67 & 1.03 ( 5 ) & $5.2\cdot 10^{5}$ &
2.2 & $^{+0.99}_{-0.86}$ & $^{+8.3}_{-7.3}$ & $^{+6.8}_{-5.9}$ &
$^{+2.9}_{-2.5}$ & $^{+2.6}_{-2.2}$ & $^{+17}_{-13}$ &
$^{+3.4}_{-3.0}$ & $^{+2.1}_{-1.8}$ & $^{+5.4}_{-4.8}$ &
$^{+0.55}_{-0.41}$ & $^{+0.03}_{-0.0}$ & $^{+1.6}_{-1.4}$ &
$^{+1.8}_{-1.6}$ & 0.03 & N/A & 1.9 & 0.14 & 2.0 & $^{+3.1}_{-3.0}$ &
1.0 & $^{+0.96}_{-1.0}$ \\
\hline\hline
\end{tabular}
\end{tiny}\vspace*{-1.5mm}
\caption{ Measured dijet cross section for $R=0.4$ and $2.0 \leq y^{*}
< 2.5$.  See Table~\ref{tab:DijetMassResults04_0} for a description of
the columns.  All tables are available on \hepdata~\cite{HEPDATA}.}
\label{tab:DijetMassResults04_4}
\vspace*{-2mm}
\end{table*}


\begin{table*}
\begin{tiny}
\setlength{\extrarowheight}{4pt}
\begin{tabular}{ccccccccccccccccccccccccc}
\hline\hline $\twomass{1}{2}$-bin & NPC & $\sigma$ & $\delta_{\rm
stat}$ & $\gamma{4}$ & $\gamma_{10}$ & $\gamma_{16}$ & $\gamma_{22}$ &
$\gamma_{28}$ & $\gamma_{31}$ & $\gamma_{35}$ & $\gamma_{41}$ &
$\gamma_{47}$ & $\gamma_{53}$ & $\gamma_{59}$ & $\gamma_{65}$ &
$\gamma_{71}$ & $\gamma_{74}$ & $\gamma_{75}$ & $\gamma_{79}$ &
$\gamma_{82}$ & $\gamma_{84}$ & $u_{1}$ & $u_{2}$ & $u_{3}$ \\
$[$TeV$]$ & & $[$pb/TeV$]$ & \% & \\ \hline 0.67-0.76 & 1.03 ( 7 ) &
$7.3\cdot 10^{5}$ & 7.1 & $^{+4.9}_{-4.8}$ & 10 & 8.0 & 4.2 &
$^{+2.8}_{-2.6}$ & $^{+43}_{-28}$ & $^{+3.5}_{-3.3}$ &
$^{+3.5}_{-3.3}$ & $^{+5.3}_{-5.6}$ & $^{+0.19}_{-0.09}$ &
$^{+0.04}_{-0.01}$ & $^{+3.6}_{-3.4}$ & $^{+0.35}_{-0.22}$ & 0.1 & N/A
& 4.8 & 0.17 & 2.1 & $^{+5.5}_{-5.9}$ & 1.0 & $^{+0.75}_{-1.2}$ \\
0.76-0.85 & 1.03 ( 6 ) & $3.9\cdot 10^{5}$ & 8.5 & $^{+4.2}_{-3.0}$ &
$^{+11}_{-8.9}$ & $^{+9.0}_{-7.4}$ & $^{+4.2}_{-3.5}$ &
$^{+2.7}_{-2.4}$ & $^{+39}_{-23}$ & $^{+4.1}_{-3.3}$ &
$^{+3.5}_{-2.6}$ & $^{+6.7}_{-5.6}$ & $^{+0.16}_{-0.14}$ &
$^{+0.05}_{-0.0}$ & $^{+3.6}_{-2.7}$ & $^{+0.68}_{-0.5}$ & 0.2 & N/A &
4.7 & 0.4 & 2.0 & $^{+6.1}_{-4.4}$ & 1.0 & $^{+0.94}_{-1.1}$ \\
0.85-0.95 & 1.03 ( 6 ) & $1.9\cdot 10^{5}$ & 4.9 & $^{+2.5}_{-2.0}$ &
$^{+10}_{-8.5}$ & $^{+9.0}_{-6.9}$ & $^{+4.4}_{-3.1}$ & 2.6 &
$^{+32}_{-21}$ & $^{+3.9}_{-3.6}$ & $^{+2.6}_{-2.4}$ &
$^{+7.2}_{-5.4}$ & $^{+0.31}_{-0.25}$ & $^{+0.02}_{-0.0}$ &
$^{+2.5}_{-2.2}$ & $^{+1.2}_{-1.1}$ & 0.05 & N/A & 3.1 & 0.11 & 2.0 &
$^{+5.6}_{-4.0}$ & 1.0 & $^{+1.1}_{-0.95}$ \\ 0.95-1.06 & 1.02 ( 5 ) &
$9.1\cdot 10^{4}$ & 3.3 & 1.3 & $^{+9.4}_{-8.1}$ & $^{+7.7}_{-6.6}$ &
$^{+3.4}_{-3.5}$ & $^{+2.6}_{-2.7}$ & $^{+25}_{-19}$ &
$^{+3.6}_{-4.0}$ & $^{+2.0}_{-2.4}$ & $^{+6.0}_{-5.6}$ &
$^{+0.56}_{-0.53}$ & 0.0 & $^{+1.7}_{-1.9}$ & $^{+2.0}_{-2.1}$ & 0.03
& N/A & 2.0 & 0.08 & 2.0 & $^{+4.2}_{-4.5}$ & 1.0 & $^{+1.0}_{-0.97}$
\\
\hline\hline
\end{tabular}
\end{tiny}\vspace*{-1.5mm}
\caption{ Measured dijet cross section for $R=0.4$ and $2.5 \leq y^{*}
< 3.0$.  See Table~\ref{tab:DijetMassResults04_0} for a description of
the columns.  All tables are available on \hepdata~\cite{HEPDATA}.}
\label{tab:DijetMassResults04_5}
\vspace*{-2mm}
\end{table*}


\begin{table*}
\begin{tiny}
\setlength{\extrarowheight}{4pt}
\begin{tabular}{ccccccccccccccccccccccccc}
\hline\hline $\twomass{1}{2}$-bin & NPC & $\sigma$ & $\delta_{\rm
stat}$ & $\gamma{5}$ & $\gamma_{11}$ & $\gamma_{17}$ & $\gamma_{23}$ &
$\gamma_{29}$ & $\gamma_{31}$ & $\gamma_{36}$ & $\gamma_{42}$ &
$\gamma_{48}$ & $\gamma_{54}$ & $\gamma_{60}$ & $\gamma_{66}$ &
$\gamma_{72}$ & $\gamma_{74}$ & $\gamma_{75}$ & $\gamma_{80}$ &
$\gamma_{82}$ & $\gamma_{85}$ & $u_{1}$ & $u_{2}$ & $u_{3}$ \\
$[$TeV$]$ & & $[$pb/TeV$]$ & \% & \\ \hline 1.18-1.31 & 1.02 ( 7 ) &
$5.6\cdot 10^{4}$ & 6.6 & $^{+3.8}_{-4.4}$ & $^{+12}_{-11}$ &
$^{+8.9}_{-8.7}$ & $^{+4.7}_{-4.8}$ & $^{+2.3}_{-3.0}$ &
$^{+59}_{-33}$ & $^{+3.7}_{-4.2}$ & $^{+2.7}_{-3.4}$ &
$^{+6.0}_{-6.6}$ & $^{+0.17}_{-0.16}$ & 0.02 & $^{+2.9}_{-3.5}$ &
$^{+0.53}_{-0.59}$ & 0.4 & N/A & 5.8 & 0.69 & 2.0 & $^{+6.6}_{-7.1}$ &
2.0 & $^{+0.95}_{-1.1}$ \\ 1.31-1.45 & 1.02 ( 7 ) & $2.6\cdot 10^{4}$
& 6.2 & $^{+2.9}_{-3.1}$ & $^{+12}_{-10}$ & $^{+8.6}_{-8.5}$ &
$^{+5.3}_{-4.5}$ & $^{+2.4}_{-2.9}$ & $^{+44}_{-30}$ &
$^{+4.1}_{-4.5}$ & $^{+2.6}_{-2.9}$ & $^{+6.3}_{-6.7}$ &
$^{+0.26}_{-0.27}$ & $^{+0.04}_{-0.0}$ & $^{+2.8}_{-3.0}$ &
$^{+1.1}_{-1.2}$ & 0.3 & N/A & 3.0 & 0.7 & 2.0 & $^{+5.6}_{-6.8}$ &
2.0 & $^{+0.78}_{-1.2}$ \\ 1.45-1.6 & 1.01 ( 7 ) & $1.4\cdot 10^{4}$ &
7.2 & $^{+2.4}_{-2.1}$ & $^{+11}_{-10}$ & $^{+8.2}_{-8.3}$ &
$^{+5.0}_{-4.9}$ & $^{+3.2}_{-3.0}$ & $^{+40}_{-27}$ &
$^{+4.6}_{-4.3}$ & $^{+3.0}_{-2.8}$ & 6.7 & $^{+0.48}_{-0.44}$ &
$^{+0.05}_{-0.0}$ & $^{+2.9}_{-2.7}$ & 1.8 & 0.0 & N/A & 3.3 & 1.4 &
2.0 & $^{+5.6}_{-6.3}$ & 2.0 & $^{+0.88}_{-1.1}$ \\ 1.6-1.76 & 1.01 (
7 ) & $7.2\cdot 10^{3}$ & 7.3 & $^{+1.6}_{-1.3}$ & 11 & 8.4 & 5.1 &
$^{+3.6}_{-3.3}$ & $^{+39}_{-25}$ & $^{+4.9}_{-4.6}$ &
$^{+3.1}_{-2.6}$ & $^{+7.3}_{-7.2}$ & $^{+0.73}_{-0.77}$ & 0.03 &
$^{+2.2}_{-1.9}$ & 2.8 & 0.06 & N/A & 3.1 & 0.94 & 2.0 &
$^{+6.1}_{-6.6}$ & 2.0 & $^{+0.96}_{-1.0}$ \\
\hline\hline
\end{tabular}
\end{tiny}\vspace*{-1.5mm}
\caption{ Measured dijet cross section for $R=0.4$ and $3.0 \leq y^{*}
< 3.5$.  See Table~\ref{tab:DijetMassResults04_0} for a description of
the columns.  All tables are available on \hepdata~\cite{HEPDATA}.}
\label{tab:DijetMassResults04_6}
\vspace*{-2mm}
\end{table*}


\begin{table*}
\begin{tiny}
\setlength{\extrarowheight}{4pt}
\begin{tabular}{ccccccccccccccccccccccccc}
\hline\hline $\twomass{1}{2}$-bin & NPC & $\sigma$ & $\delta_{\rm
stat}$ & $\gamma{6}$ & $\gamma_{12}$ & $\gamma_{18}$ & $\gamma_{24}$ &
$\gamma_{30}$ & $\gamma_{31}$ & $\gamma_{37}$ & $\gamma_{43}$ &
$\gamma_{49}$ & $\gamma_{55}$ & $\gamma_{61}$ & $\gamma_{67}$ &
$\gamma_{73}$ & $\gamma_{74}$ & $\gamma_{75}$ & $\gamma_{81}$ &
$\gamma_{82}$ & $\gamma_{86}$ & $u_{1}$ & $u_{2}$ & $u_{3}$ \\
$[$TeV$]$ & & $[$pb/TeV$]$ & \% & \\ \hline 1.76-1.94 & 0.989 ( 10 ) &
$8.9\cdot 10^{3}$ & 13 & $^{+6.2}_{-5.2}$ & $^{+14}_{-12}$ &
$^{+10}_{-9.1}$ & 4.9 & $^{+3.2}_{-3.0}$ & $^{+100}_{-46}$ &
$^{+4.4}_{-4.1}$ & $^{+4.1}_{-3.8}$ & $^{+7.4}_{-6.7}$ &
$^{+0.07}_{-0.2}$ & 0.01 & $^{+4.3}_{-4.0}$ & $^{+0.22}_{-0.48}$ & 0.0
& N/A & 6.8 & 0.0 & 2.1 & $^{+11}_{-8.1}$ & 2.0 & $^{+2.3}_{-0.0}$ \\
1.94-2.12 & 0.989 ( 10 ) & $3.6\cdot 10^{3}$ & 14 & $^{+4.3}_{-4.4}$ &
$^{+13}_{-11}$ & $^{+9.5}_{-9.1}$ & $^{+4.1}_{-5.1}$ &
$^{+3.1}_{-2.8}$ & $^{+76}_{-37}$ & 4.1 & $^{+3.6}_{-3.3}$ &
$^{+7.0}_{-6.9}$ & $^{+0.06}_{-0.18}$ & $^{+0.0}_{-0.02}$ &
$^{+3.5}_{-3.6}$ & $^{+0.58}_{-0.79}$ & 0.5 & N/A & 2.8 & 0.0 & 2.0 &
$^{+7.3}_{-6.9}$ & 2.0 & $^{+0.72}_{-1.3}$ \\ 2.12-2.33 & 0.989 ( 10 )
& $1.3\cdot 10^{3}$ & 8.9 & $^{+1.7}_{-3.0}$ & $^{+12}_{-9.6}$ &
$^{+9.9}_{-8.4}$ & $^{+2.8}_{-3.6}$ & $^{+2.3}_{-2.0}$ &
$^{+58}_{-42}$ & $^{+3.6}_{-3.5}$ & $^{+2.2}_{-2.3}$ & 6.2 &
$^{+0.43}_{-0.32}$ & 0.02 & $^{+2.1}_{-2.3}$ & $^{+0.81}_{-0.59}$ &
0.4 & N/A & 0.0 & 0.0 & 2.0 & $^{+4.9}_{-6.3}$ & 2.0 &
$^{+0.33}_{-1.7}$ \\ 2.33-2.55 & 0.989 ( 10 ) & $6.1\cdot 10^{2}$ &
8.8 & $^{+1.9}_{-2.4}$ & $^{+18}_{-12}$ & $^{+13}_{-9.9}$ &
$^{+2.8}_{-4.1}$ & $^{+3.6}_{-3.2}$ & $^{+86}_{-42}$ &
$^{+4.3}_{-4.8}$ & $^{+3.1}_{-3.0}$ & $^{+7.6}_{-7.9}$ &
$^{+0.27}_{-0.42}$ & $^{+0.05}_{-0.02}$ & $^{+2.8}_{-2.5}$ &
$^{+1.8}_{-2.3}$ & 0.2 & N/A & 1.3 & 0.0 & 2.0 & $^{+6.5}_{-7.9}$ &
2.0 & $^{+0.68}_{-1.3}$ \\
\hline\hline
\end{tabular}
\end{tiny}\vspace*{-1.5mm}
\caption{ Measured dijet cross section for $R=0.4$ and $3.5 \leq y^{*}
< 4.0$.  See Table~\ref{tab:DijetMassResults04_0} for a description of
the columns.  All tables are available on \hepdata~\cite{HEPDATA}.}
\label{tab:DijetMassResults04_7}
\vspace*{-2mm}
\end{table*}


\begin{table*}
\begin{tiny}
\setlength{\extrarowheight}{4pt}
\begin{tabular}{ccccccccccccccccccccccccc}
\hline\hline $\twomass{1}{2}$-bin & NPC & $\sigma$ & $\delta_{\rm
stat}$ & $\gamma{6}$ & $\gamma_{12}$ & $\gamma_{18}$ & $\gamma_{24}$ &
$\gamma_{30}$ & $\gamma_{31}$ & $\gamma_{37}$ & $\gamma_{43}$ &
$\gamma_{49}$ & $\gamma_{55}$ & $\gamma_{61}$ & $\gamma_{67}$ &
$\gamma_{73}$ & $\gamma_{74}$ & $\gamma_{75}$ & $\gamma_{81}$ &
$\gamma_{82}$ & $\gamma_{86}$ & $u_{1}$ & $u_{2}$ & $u_{3}$ \\
$[$TeV$]$ & & $[$pb/TeV$]$ & \% & \\ \hline 2.55-3.04 & 0.937 ( 15 ) &
$2.6\cdot 10^{2}$ & 21 & $^{+5.8}_{-6.2}$ & $^{+16}_{-14}$ & 12 &
$^{+2.0}_{-2.4}$ & $^{+2.8}_{-3.1}$ & $^{+120}_{-59}$ &
$^{+4.5}_{-4.7}$ & $^{+3.9}_{-4.0}$ & $^{+7.7}_{-8.4}$ &
$^{+0.08}_{-0.29}$ & 0.0 & $^{+4.1}_{-4.5}$ & $^{+0.42}_{-0.87}$ & 0.3
& N/A & 7.2 & 3.2 & $^{+2.1}_{-2.0}$ & $^{+8.9}_{-9.0}$ & 2.0 & 1.0 \\
3.04-4.27 & 0.937 ( 15 ) & $1.8\cdot 10^{1}$ & 24 & $^{+3.9}_{-2.8}$ &
$^{+18}_{-15}$ & $^{+14}_{-13}$ & $^{+4.2}_{-2.6}$ & $^{+5.8}_{-4.3}$
& $^{+110}_{-50}$ & 7.7 & $^{+5.4}_{-3.1}$ & 11 & $^{+2.1}_{-1.1}$ &
$^{+0.0}_{-0.28}$ & $^{+5.1}_{-3.0}$ & $^{+4.1}_{-2.5}$ & 3.0 & N/A &
2.9 & 4.1 & 2.0 & $^{+7.9}_{-9.3}$ & 2.0 & 1.0 \\ \hline\hline
\end{tabular}
\end{tiny}\vspace*{-1.5mm}
\caption{ Measured dijet cross section for $R=0.4$ and $4.0 \leq y^{*}
< 4.4$.  See Table~\ref{tab:DijetMassResults04_0} for a description of
the columns.  All tables are available on \hepdata~\cite{HEPDATA}.}
\label{tab:DijetMassResults04_8}
\vspace*{-2mm}
\end{table*}

\begin{table*}
\begin{tiny}
\setlength{\extrarowheight}{4pt}
\begin{tabular}{ccccccccccccccccccccccccc}
\hline\hline $\twomass{1}{2}$-bin & NPC & $\sigma$ & $\delta_{\rm
stat}$ & $\gamma_{1}$ & $\gamma_{7}$ & $\gamma_{13}$ & $\gamma_{19}$ &
$\gamma_{25}$ & $\gamma_{31}$ & $\gamma_{32}$ & $\gamma_{38}$ &
$\gamma_{44}$ & $\gamma_{50}$ & $\gamma_{56}$ & $\gamma_{62}$ &
$\gamma_{68}$ & $\gamma_{74}$ & $\gamma_{75}$ & $\gamma_{76}$ &
$\gamma_{82}$ & $\gamma_{83}$ & $u_{1}$ & $u_{2}$ & $u_{3}$ \\
$[$TeV$]$ & & $[$pb/TeV$]$ & \% & \\ \hline 0.07-0.11 & 1.35 ( 9 ) &
$4.6\cdot 10^{8}$ & 0.97 & $^{+6.9}_{-6.2}$ & $^{+9.3}_{-8.5}$ &
$^{+9.6}_{-8.6}$ & $^{+5.6}_{-5.1}$ & $^{+2.4}_{-2.3}$ &
$^{+9.6}_{-7.5}$ & $^{+2.7}_{-2.6}$ & $^{+3.1}_{-2.9}$ &
$^{+4.5}_{-4.3}$ & $^{+0.03}_{-0.02}$ & 0.0 & $^{+3.3}_{-3.1}$ &
$^{+0.11}_{-0.12}$ & 0.0 & N/A & 2.8 & 0.45 & 2.1 & $^{+5.5}_{-4.9}$ &
1.0 & 1.0 \\ 0.11-0.16 & 1.22 ( 5 ) & $5.6\cdot 10^{7}$ & 0.59 &
$^{+0.91}_{-0.87}$ & $^{+6.9}_{-6.5}$ & $^{+5.2}_{-4.8}$ &
$^{+3.2}_{-2.9}$ & $^{+2.5}_{-2.4}$ & $^{+4.9}_{-4.2}$ &
$^{+2.8}_{-2.6}$ & 1.7 & $^{+4.3}_{-4.1}$ & $^{+0.22}_{-0.25}$ &
$^{+0.04}_{-0.03}$ & $^{+2.0}_{-1.9}$ & $^{+0.74}_{-0.76}$ & 0.03 &
N/A & 0.89 & 0.17 & 2.0 & $^{+2.8}_{-2.9}$ & 1.0 & 1.0 \\ 0.16-0.21 &
1.15 ( 3 ) & $1.1\cdot 10^{7}$ & 0.44 & $^{+0.57}_{-0.6}$ &
$^{+6.0}_{-5.6}$ & $^{+2.9}_{-2.7}$ & 2.1 & $^{+2.5}_{-2.3}$ &
$^{+3.1}_{-2.9}$ & $^{+3.0}_{-2.9}$ & 1.3 & $^{+4.8}_{-4.5}$ &
$^{+0.92}_{-0.86}$ & $^{+0.05}_{-0.03}$ & $^{+0.47}_{-0.49}$ &
$^{+2.9}_{-2.7}$ & 0.02 & N/A & 0.58 & 0.15 & 2.0 & $^{+1.9}_{-2.2}$ &
1.0 & 1.0 \\ 0.21-0.26 & 1.12 ( 3 ) & $2.9\cdot 10^{6}$ & 0.6 &
$^{+0.14}_{-0.15}$ & $^{+5.7}_{-5.3}$ & $^{+2.8}_{-2.9}$ &
$^{+1.9}_{-2.0}$ & 1.1 & 2.3 & $^{+2.6}_{-2.7}$ & $^{+0.99}_{-1.1}$ &
$^{+5.0}_{-4.7}$ & $^{+1.3}_{-1.4}$ & $^{+0.03}_{-0.07}$ &
$^{+0.04}_{-0.03}$ & $^{+3.7}_{-3.6}$ & 0.0 & N/A & 0.51 & 0.05 & 2.0
& $^{+1.6}_{-1.7}$ & 1.0 & 1.0 \\
\hline\hline
\end{tabular}
\end{tiny}\vspace*{-1.5mm}
\caption{ Measured dijet cross section for $R=0.6$ and $ y^{*} < 0.5$.
See Table~\ref{tab:DijetMassResults04_0} for a description of the
columns.  All tables are available on \hepdata~\cite{HEPDATA}.}
\label{tab:DijetMassResults06_0}
\vspace*{-2mm}
\end{table*}


\begin{table*}
\begin{tiny}
\setlength{\extrarowheight}{4pt}
\begin{tabular}{ccccccccccccccccccccccccc}
\hline\hline $\twomass{1}{2}$-bin & NPC & $\sigma$ & $\delta_{\rm
stat}$ & $\gamma_{1}$ & $\gamma_{7}$ & $\gamma_{13}$ & $\gamma_{19}$ &
$\gamma_{25}$ & $\gamma_{31}$ & $\gamma_{32}$ & $\gamma_{38}$ &
$\gamma_{44}$ & $\gamma_{50}$ & $\gamma_{56}$ & $\gamma_{62}$ &
$\gamma_{68}$ & $\gamma_{74}$ & $\gamma_{75}$ & $\gamma_{76}$ &
$\gamma_{82}$ & $\gamma_{83}$ & $u_{1}$ & $u_{2}$ & $u_{3}$ \\
$[$TeV$]$ & & $[$pb/TeV$]$ & \% & \\ \hline 0.11-0.16 & 1.28 ( 6 ) &
$1.4\cdot 10^{8}$ & 0.91 & $^{+5.3}_{-4.9}$ & $^{+8.8}_{-8.1}$ &
$^{+8.8}_{-8.0}$ & $^{+4.9}_{-4.6}$ & $^{+2.2}_{-2.3}$ &
$^{+9.1}_{-7.8}$ & $^{+2.5}_{-2.6}$ & 2.7 & $^{+4.4}_{-4.2}$ &
$^{+0.07}_{-0.03}$ & $^{+0.0}_{-0.01}$ & 2.8 & $^{+0.07}_{-0.06}$ &
0.07 & N/A & 2.3 & 0.28 & 2.1 & $^{+5.3}_{-4.6}$ & 1.0 & 1.0 \\
0.16-0.21 & 1.2 ( 4 ) & $2.5\cdot 10^{7}$ & 0.69 & 1.2 &
$^{+6.4}_{-6.0}$ & $^{+3.3}_{-3.1}$ & 2.0 & 2.4 & $^{+4.0}_{-3.6}$ &
$^{+3.0}_{-2.9}$ & $^{+1.2}_{-1.1}$ & $^{+4.4}_{-4.3}$ &
$^{+0.43}_{-0.38}$ & 0.03 & 1.2 & 1.4 & 0.1 & N/A & 0.07 & 0.0 & 1.9 &
$^{+1.8}_{-2.6}$ & 1.0 & 1.0 \\ 0.21-0.26 & 1.15 ( 3 ) & $7.1\cdot
10^{6}$ & 0.59 & $^{+0.58}_{-0.59}$ & $^{+5.8}_{-5.4}$ &
$^{+2.8}_{-2.5}$ & $^{+2.0}_{-1.9}$ & $^{+2.4}_{-2.1}$ &
$^{+3.1}_{-2.9}$ & $^{+2.8}_{-2.6}$ & $^{+1.2}_{-1.1}$ &
$^{+4.4}_{-4.3}$ & $^{+0.82}_{-0.83}$ & $^{+0.06}_{-0.04}$ & 0.43 &
$^{+2.7}_{-2.5}$ & 0.0 & N/A & 0.63 & 0.0 & 2.0 & $^{+2.0}_{-2.1}$ &
1.0 & 1.0 \\ 0.26-0.31 & 1.12 ( 3 ) & $2.5\cdot 10^{6}$ & 0.49 &
$^{+0.12}_{-0.13}$ & $^{+6.0}_{-5.9}$ & $^{+3.1}_{-3.0}$ & 2.1 &
$^{+1.5}_{-1.3}$ & $^{+3.1}_{-3.0}$ & $^{+2.8}_{-2.7}$ &
$^{+1.2}_{-1.1}$ & $^{+5.3}_{-5.1}$ & 1.3 & 0.04 & $^{+0.22}_{-0.04}$
& $^{+3.9}_{-3.8}$ & 0.02 & N/A & 0.61 & 0.0 & 2.0 & 2.1 & 1.0 & 1.0
\\
\hline\hline
\end{tabular}%
\end{tiny}\vspace*{-1.5mm}
\caption{ Measured dijet cross section for $R=0.6$ and $4.0 \leq y^{*}
< 4.4$.  See Table~\ref{tab:DijetMassResults04_0} for a description of
the columns.  All tables are available on \hepdata~\cite{HEPDATA}.}
\label{tab:DijetMassResults06_8}
\vspace*{-2mm}
\end{table*}

\clearpage
\onecolumngrid
\input{atlas_authlist} 

\end{document}

%% file: atlas_authlist.tex
\begin{flushleft}
{\Large The ATLAS Collaboration}

\bigskip

G.~Aad$^{\rm 48}$,
B.~Abbott$^{\rm 110}$,
J.~Abdallah$^{\rm 11}$,
A.A.~Abdelalim$^{\rm 49}$,
A.~Abdesselam$^{\rm 117}$,
O.~Abdinov$^{\rm 10}$,
B.~Abi$^{\rm 111}$,
M.~Abolins$^{\rm 87}$,
H.~Abramowicz$^{\rm 152}$,
H.~Abreu$^{\rm 114}$
% ...
\bigskip

% ...
$^{10}$ Institute of Physics, Azerbaijan Academy of Sciences, Baku, Azerbaijan\\
$^{11}$ Institut de F\'isica d'Altes Energies and Departament de F\'isica de la Universitat Aut\`onoma  de Barcelona and ICREA, Barcelona, Spain\\
% ...
$^{48}$ Fakult\"{a}t f\"{u}r Mathematik und Physik, Albert-Ludwigs-Universit\"{a}t, Freiburg i.Br., Germany\\
$^{49}$ Section de Physique, Universit\'e de Gen\`eve, Geneva, Switzerland\\
% ...
$^{87}$ Department of Physics and Astronomy, Michigan State University, East Lansing MI, United States of America\\
% ...
$^{110}$ Homer L. Dodge Department of Physics and Astronomy, University of Oklahoma, Norman OK, United States of America\\
$^{111}$ Department of Physics, Oklahoma State University, Stillwater OK, United States of America\\
% ...
$^{114}$ LAL, Univ. Paris-Sud and CNRS/IN2P3, Orsay, France\\
% ...
$^{117}$ Department of Physics, Oxford University, Oxford, United Kingdom\\
% ...
$^{152}$ Raymond and Beverly Sackler School of Physics and Astronomy, Tel Aviv University, Tel Aviv, Israel\\
% ...
\end{flushleft}
